\documentclass[pra,aps,superscriptaddress,notitlepage,longbibliography,twocolumn,nofootinbib,superscriptaddress,10pt]{revtex4-2}
\usepackage{amsmath}
\usepackage{amsfonts}
\usepackage{graphicx}
\usepackage{bm}
\usepackage{bbold}
\usepackage{color}
\usepackage{braket}
\usepackage{comment}
\usepackage{multirow}

\usepackage[dvipsnames]{xcolor}

\usepackage[colorlinks]{hyperref}

\hypersetup{
    colorlinks=true,
    linkcolor=blue,
    filecolor=magenta,      
    urlcolor=magenta,
    citecolor={blue},
    }

\DeclareMathAlphabet{\mathcallc}{U}{dutchcal}{m}{n}
\SetMathAlphabet{\mathcallc}{bold}{U}{dutchcal}{b}{n}
\DeclareMathAlphabet{\mathbcallc}{U}{dutchcal}{b}{n}

\usepackage[english]{babel}

\begin{document}

\abovedisplayskip=6pt
\abovedisplayshortskip=6pt
\belowdisplayskip=6pt
\belowdisplayshortskip=6pt

\title{Dissipation in the Broadband and Ultrastrong Coupling Regimes of Cavity Quantum Electrodynamics: An \emph{Ab Initio} Quantized Quasinormal Mode Approach}
\author{Chris Gustin}
\email{cgustin@stanford.edu}
\affiliation{E.\,L.\,Ginzton Laboratory, Stanford University, Stanford, California 94305, USA}
\author{Juanjuan Ren}
\affiliation{\hspace{0pt}Department of Physics, Engineering Physics, and Astronomy, Queen's University, Kingston, Ontario K7L 3N6, Canada\hspace{0pt}}
\author{Sebastian Franke}
\affiliation{Technische Universit\"at Berlin, Institut f\"ur Theoretische Physik,
Nichtlineare Optik und Quantenelektronik, Hardenbergstra{\ss}e 36, 10623 Berlin, Germany}
\affiliation{\hspace{0pt}Department of Physics, Engineering Physics, and Astronomy, Queen's University, Kingston, Ontario K7L 3N6, Canada\hspace{0pt}}
\author{Stephen Hughes}
\affiliation{\hspace{0pt}Department of Physics, Engineering Physics, and Astronomy, Queen's University, Kingston, Ontario K7L 3N6, Canada\hspace{0pt}}

\begin{abstract}
Phenomenological approaches to photon loss have long been the workhorse of cavity-QED, but prove inadequate in the presence of sufficiently broadband light-matter interactions.
We present a rigorous and \emph{ab initio} derivation of a quantum master equation for a quantized optical cavity mode coupled to a dipole, using a quasinormal mode (QNM) quantization procedure for
plasmonic and dielectric open-system cavity-QED, which is valid in broadband light-matter interaction regimes, including ultrastrong coupling (USC).
The theory supports general three-dimensional resonators with arbitrary dispersion and loss, and thus can be applied to a wide range of 
open cavities.
Our \emph{ab initio} and gauge-invariant approach fully recovers the recent result of Phys. Rev. Lett. 134, 123601 (2025) for the spectral density of a quantized cavity with a single dipole, exhibits a dissipative classical-quantum correspondence for bosonic Hopfield model systems, and reveals important departures from previous heuristic assumptions about system-bath coupling. 
We identify a new criterion for what we term the ``broadband'' dissipative regime of cavity-QED, where phenomenological models require corrections in accordance with the intrinsic and spatially-dependent complex phase of the QNM, and also shed light on \emph{fundamental limits} to single-mode models in extreme coupling regimes. Using plasmonic and dielectric cavity examples, we show validity ranges of our QNM master equation and spectral USC calculations, and discuss prospects for near-term experimental observation of broadband dissipative effects.
\end{abstract}

\maketitle
%\begin{comment}
\section{Introduction}
Over the past decades, cavity quantum electrodynamics (cavity-QED) has become a ubiquitous platform both for fundamental research and quantum information processing (QIP) technologies~\cite{Mabuchi2002Nov,Browne2017,doi:10.1021/acs.chemrev.2c00855,van_enk_quantum_2004}. Photons are ideal carriers of quantum information, traveling at the speed of light with minimal decoherence at room temperature, and naturally transmittable along classical channels. Interfacing with atoms---or more precisely, an entire broad range of material degrees of freedom which interact with light---allows for the implementation of nonlinear and often deterministic quantum operations, while engineered photonic structures can mold the flow of light and resonances to enable both a greater degree of unitary control over the system, as well as system-bath engineering~\cite{PhysRevX.5.021025,verstraete_quantum_2009,gonzalez-tudela_lightmatter_2024}.

Over the decades, a set of canonical theoretical tools has been built up alongside experimental and technological progress in cavity-QED which enables intuitive, computationally inexpensive, and precise modeling of the quantum dynamics~\cite{gardiner_quantum_2004,carmichael2013statistical,carmichael2009statistical,breuer2002theory}. Such theoretical progress has undoubtedly contributed to the growth of the field, providing researchers with effective tools with which to predict new physical effects, develop new QIP protocols, and understand experimental results. Milestones along this path range from the Jaynes-Cummings~\cite{larson_jaynescummings_2024} and Dicke~\cite{PhysRev.93.99,https://doi.org/10.1002/qute.201800043} models (unitary) to system-reservoir and input-output theory~\cite{Gardiner1,gardiner_quantum_2004,combes_slh_2017} (dissipative).

In particular, the latter development of open quantum system theory has allowed for realistic and concise treatment of photon loss, intrinsic in any physical cavity-QED system. A particular powerful approach has been the Lindblad master equation (and related techniques---e.g., the stochastic Schr{\" o}dinger equation), which, for example, can describe Markovian photon loss from quantized cavity modes (a reservoir---or interchangeably, bath) without the need to include a continuum Hilbert space in final calculations, and while retaining all of the information associated with the discrete electromagnetic resonances that are physically present in a given system~\cite{lindblad_generators_1976,breuer2002theory}.

Such \emph{system-reservoir} theories (and their associated input-output relations) have proven extremely effective in connecting purely unitary quantum dynamics to the realistic dissipative behavior seen in reality. However, while the general theory of open quantum systems can generically treat couplings between a system and its environment by means of a \emph{spectral density} function $\Lambda^2(\omega)$~\cite{breuer2002theory}, when it comes to photon loss, such system-reservoir theories have almost invariably to date relied on \emph{phenomenological} assumptions on the form of the interaction Hamiltonian which couples an 
electromagnetic system (e.g., modes of a quantized cavity or waveguide) to its reservoir. For a long time, such an approach proved sufficient, as the highly Markovian nature of photon decay (occurring over timescales many orders of magnitude larger than the photon cycle time) meant that the only relevant parameter to the process (putting aside photonic Lamb shifts which can be absorbed into the observed system frequencies) was a single value $\Lambda^2(\omega_c)$, where $\omega_c$ is the frequency of photonic resonance, which is proportional to the linewidth of the resonance~\cite{Gustin2025}.

Such phenomenological models are now being challenged in a variety of emerging regimes of light-matter interaction, where increased strength of the system coupling parameters and tightly-confined mode volumes probe system-environment interactions over a large spectral range, which we refer to as \emph{broadband} dissipative cavity-QED.

One striking and often-studied example of broadband dissipative cavity-QED is the case of ultrastrong coupling (USC)~\cite{forn-diaz_ultrastrong_2019,RevModPhys.91.025005}.
Regimes of quantum light-matter interaction 
in cavity-QED
are typically characterized as {\it weak}, {\it strong}, and {\it ultrastrong}
~\cite{frisk_kockum_ultrastrong_2019,forn-diaz_ultrastrong_2019,LeBoite2020Jul}.
Weak to intermediate coupling gives rise to modified spontaneous emission rates through the Purcell
effect~\cite{purcell1946resonance},
when the cavity-emitter coupling rate 
$g$ is less than the dissipation rates of the emitter or cavity ($\gamma,\kappa$).
The
strong coupling 
occurs when 
$g>\gamma,\kappa$, which can manifest in
vacuum Rabi oscillations and signatures
of anharmonic cavity-QED with a pump field.
The USC regime
occurs when a rotating wave approximation (RWA) is no longer valid, roughly
when $g \geq 0.1\omega_c$ (where $\omega_c$ is the cavity frequency), and
the ground state properties of the hybridized states become an entangled state of photons and matter.
%A fourth regime occurs
%when $g>\omega_0$, yielding the so-called
%deep USC regime.
This latter regime is not only driven by fundamental considerations, but by emerging experiments, for example, including  intersubband polaritons~\cite{ciuti_quantum_2005,anappara_signatures_2009,zaks_thz-driven_2011}, 
circuit-QED \cite{frisk_kockum_ultrastrong_2019,forn-diaz_ultrastrong_2019,PhysRevX.12.031004}, Landau polaritons~\cite{scalari_ultrastrong_2012, rajabali_ultrastrongly_2022, tay_multimode_2025, li_vacuum_2018}, and plasmonic resonators~\cite{mueller_deep_2020,hu_robust_2024,chang_ultrastrong_2025,yoo_ultrastrong_2021}.

In such extreme light-matter coupling regimes, beyond the breakdown of the RWA [which can be addressed by generalizing the JC model to the Quantum Rabi Model (QRM)], another issue in the system dynamics is that naive truncation of the matter degrees of freedom to a two-level system (TLS), as is commonly done, can introduce spurious gauge-dependence.
This particular problem has been addressed by introducing a generalized minimal coupling replacement~\cite{DiStefano2019Aug}, which also is necessary for dissipative models~\cite{Salmon2022Mar}.

While much progress has been made in the generalization of light-matter models on the \emph{system} level, a rigorous and efficient treatment of loss and dissipation in such coupling regimes is still an open question. Previous works have typically employed a normal mode expansion with real eigenfrequencies, and the photon decay is added phenomenologically as a Lindblad dissipator or by coupling the normal modes to supplementary photon baths {\it by hand} via the aforementioned system-reservoir theory. 
Moreover, the rising importance of metallic resonators for USC physics demands a more generalized quantum theory that rigorously takes into account the absorptive and dispersive nature of the scattering structures, which is also 
%completely 
lacking in usual phenomenological normal mode theories. While some progress has been made toward an \emph{ab initio} approach in 1D~\cite{Dutra2000Oct,PhysRevA.88.043819,khanbekyan} and dispersionless~\cite{Viviescas2003Jan,PhysRevX.10.011008,PhysRevLett.133.203604} systems, a tractable approach to general 3D cavity structures including potential material dispersion and absorption that remains valid in USC has not yet been found.

A critical problem lies in the fact that reliance on phenomenology has been shown recently to be insufficient to accurately predict spectral observables from systems in the USC regime~\cite{Salmon2022Mar,Hughes2024Jun,PhysRevLett.132.106902}. The need for \emph{ab initio} dissipative few-mode models has also been highlighted recently in other contexts in quantum optics~\cite{PhysRevResearch.2.023396,PhysRevX.10.011008}. We have recently made a significant advance in this problem by identifying, using a semi-phenomenological approach, the correct form of the spectral density for a single lossy quantized cavity mode~\cite{Gustin2025}, which can be used to \emph{accurately} calculate  spectral observables in broadband coupling regimes, including USC, when the system is described by a sufficiently isolated single optical resonance.

The result in Ref.~\cite{Gustin2025} employs a quasinormal mode (QNM) approach
\cite{2ndquant2,NormKristHughes,Lalanne_review,Kristensen:20}. The QNMs $\tilde{\mathbf{f}}_{\mu}(\mathbf{r})$ are \emph{complex} solutions to the Helmholtz equation, 
\begin{equation}\label{eq:QNM_def}
\boldsymbol{\nabla}\times\boldsymbol{\nabla}\times\tilde{\mathbf{f}}_\mu(\mathbf{r})-\frac{\tilde{\omega}_\mu^2}{c^2}\epsilon(\mathbf{r},\tilde{\omega}_\mu)\tilde{\mathbf{f}}_\mu(\mathbf{r})=0,
\end{equation}
which is  solved with suitable radiation conditions. As a consequence of the open boundary conditions and possibly also material absorption, the corresponding QNM eigenfrequencies $\tilde{\omega}_{\mu}$ are complex-valued. While the real part describes the resonant oscillation of the QNMs, the (negative) imaginary part reflects the width of the resonance (dissipation).
These ``modes'' are well behaved within and near the scattering geometry, but spatially diverge outside.
However, 
one can use methods to regularize the QNMs which are then well behaved in the far field, and can be used in both classical~\cite{GeNJP2014}
and quantum light-matter regimes~\cite{PhysRevLett.122.213901}.

Notably, the central result
in Ref.~\cite{Gustin2025} finds that the phase of the QNM at the dipole location in cavity-QED setups, $\text{arg}\{\tilde{\mathbf{f}}\}$, is crucial to the underlying spectral density of the cavity-reservoir interaction. Furthermore, we find (elaborated more precisely later in the text), that when the product of the cavity quality (Q) factor and the QNM phase at the dipole location becomes sufficiently large, the dissipative cavity-QED system enters a \emph{broadband} dissipative regime, where phenomenological models begin to partially break down, and our approach allows for accurate quantitative modeling of spectral observables. The broadband dissipative regime includes not only USC, but also weak and strong coupling regimes with sufficiently large detunings between the cavity and dipole~\cite{Gustin2025}, and, even on resonance, can often be reached with coupling strengths orders of magnitude below USC. The product of the QNM phase and Q-factor is also intrinsically related to the limitations of single-mode models, and thus, our approach allows for important insights into fundamental limits to single-mode models in broadband dissipative regimes. The role of the QNM phase on frequency-dependent dissipation had previously been noted in instances where the QNM phase is particularly large~\cite{PhysRevB.92.205420}, but in fact is essential to consider even when very small (i.e., at a modal antinode) in the context of broadband dissipation~\cite{Gustin2025}.

In this work, we extend the recent results of Ref.~\cite{Gustin2025} to derive a fully \emph{ab initio} single-mode and single-dipole theory of dissipative cavity-QED, using the formalism of quantized QNMs, which can be used to accurately model the USC and broadband dissipative regimes and predict substantial corrections beyond existing phenomenological approaches.   The theory of quantized QNMs was developed recently, and 
allows one to maintain the idea of a mode decomposition for absorptive/open cavities by utilizing a symmetrization transformation to construct proper QNM photon Fock states~\cite{PhysRevLett.122.213901,meschede2025quantumquasinormalmodetheory}. The theory has previously been applied to the weak and intermediate to strong light-matter coupling regime for simulating quantum-optical processes, such as cavity-enhanced spontaneous emission~\cite{PhysRevLett.122.213901}, single-photon emission~\cite{Hughes_SPS_2019}, multi-photon statistics~\cite{franke2020quantized}, as well as for exploring phenomena close to an exceptional point for coupled resonators~\cite{Ren2022,Ren2022b} including amplifiers \cite{PhysRevLett.127.013602,PhysRevX.11.041020,PhysRevA.105.023702,VanDrunen:24,rrwd-47xr}.

Our quantized QNM approach allows, for the first time to our knowledge, an \emph{ab initio} quantum theory of dissipative cavity-QED which is capable of recovering our recent results for the spectral density of a lossy cavity mode~\cite{Gustin2025}, while still retaining the full convenience of a single-mode master equation approach which most effectively captures the physics associated with isolated discrete photonic resonances (in contrast to, e.g., pseudomode approaches~\cite{PhysRevLett.126.093601,PhysRevLett.132.106902, sanchez-barquilla_few-mode_2022} which, while powerful and \emph{ab initio}, often require the introduction of many more modes than physically present resonances in the system). 

We furthermore advance previously-existing theories of quantized QNMs by presenting them as a direct partitioning of the Hilbert space of the entire complex medium (resonator structure + environment), using the rigorous theory of gauge-invariant macroscopic QED~\cite{Gustin2023Jan}, and identify a new ``spatially specified'' quantization representation of the QNMs, which we find necessary to recover known results, including semiclassical limits where appropriate. This work {\it significantly generalizes all previous quantum models in the USC regime for open cavity systems}, in that it allows one to access to the coupling constants from the QNM and the geometry properties of the scattering structure {\it without any phenomenological fitting parameters}. Our results are relevant to emerging experiments and we predict they should in principle be observable in the near future.

The structure of the rest of the paper is as follows: in Sec.~\ref{sec:theory1}, we describe the foundational theoretical components of our \emph{ab initio} approach to broadband dissipative cavity-QED, including the macrocopic QED framework for quantization of electromagnetic fields in a arbitrary dispersive and absorbing medium, the classical theory of QNMs, and our projection approach to QNM quantization and its ``spatially specified'' representation in the single-mode regime.  In Sec.~\ref{Sec: Light-Matter}, we introduce light-matter coupling between the quantized QNM cavity system and a single dipole which respects the gauge principal for arbitrary coupling strengths.

In Sec.~\ref{sec:QNM_me}, we use the system-reservoir Hamiltonian for the dissipative cavity-QED system to derive a single-mode master equation valid in broadband dissipative regimes, including USC. We then use this master equation theory to introduce and define the broadband dissipative regime of cavity-QED in terms of the dimensionless parameter, $\tilde{\Omega}_{\rm BB}$, which is the value that a system dynamical rate (e.g., the cavity-dipole coupling strength) divided by the cavity frequency must take to enter the regime where broadband dissipative corrections become significant. We also discuss our near-field photodetection model.

In Sec.~\ref{sec:classical_results}, we perform detailed electromagnetic simulations for a variety of dielectric and plasmonic cavity designs, verifying our quantized QNM theory for the weak dipole-cavity coupling regime where comparison with perturbative results is possible~\cite{Gustin2025}, show how broadband dissipative corrections can occur in many dielectric cavity designs for coupling strengths orders-of-magnitude below the usual threshold for USC, and estimate corresponding limits to validity of single-mode dissipative models. In Sec.~\ref{sec:quantum_sim}, we then perform quantum master equation simulations in the USC regime, showing the strong impact of our \emph{ab initio} theory's corrections when compared with heuristic models, and showing how a complete classical-quantum correspondence can be obtained in the thermodynamic limit of many-dipole couplings, which we previously showed for phenomenological dissipation models~\cite{Hughes2024Jun}. 

In Sec.~\ref{sec:exp}, we connect to current and emerging experiments involving cavity designs with ultrasmall mode volumes, both dielectric and plasmonic, and discuss prospects for near-term experimental observations of the broadband dissipative dynamics we predict in this paper. Finally, in Sec.~\ref{sec:conclusions}, we conclude and summarize the main findings of the work. We also include five appendices, which give further technical details on our QNM quantization scheme and master equation, and show additional cavity resonator designs and their associated QNM and broadband dissipative regime parameters.

\section{Theoretical background}\label{sec:theory1}

\subsection{Quantization in spatially-dependent absorptive and dispersing media}
\label{Subsec: MacroQED}

First we briefly describe the theory of macroscopic QED, which is a quantization scheme for arbitrary spatial-inhomogeneous and absorptive media~\cite{Dung,grunwel,Suttorp,Gustin2023Jan}. 
We concentrate on the electromagnetic part of the Hamiltonian in passive 
materials and do not introduce any active emitter contributions yet, as this is later added via gauge-invariant coupling to a TLS. 

The Hamiltonian of the combined medium (e.g., cavity resonator + environment) and photon field
can be written in the general form~\cite{Dung,vogel2006},
\begin{equation}
    \hat{H}_{\rm em}=\int_0^\infty{ d}\omega_{\rm m}\int{ d}^3 r~\hbar\omega_{\rm m}\hat{\mathbf{b}}^\dagger(\mathbf{r},\omega_{\rm m})\cdot\hat{\mathbf{b}}(\mathbf{r},\omega_{\rm m}),\label{eq: H_em}
\end{equation}
where the vector-valued and spatially-dependent bosonic operators $\hat{\mathbf{b}}(\mathbf{r},\omega_{\rm m})$ represent the fundamental variables of the polariton (medium-photon) degrees of freedom with mode indices $\omega_{\rm m}$ and $\mathbf{r}$, where $[\hat{b}_i(\mathbf{r},\omega_{\rm m}), \hat{b}^{\dagger}_j(\mathbf{r'},\omega_{\rm m}'] = \delta_{ij}\delta(\mathbf{r}-\mathbf{r'})\delta(\omega_{\rm m} - \omega_{\rm m}')$. The associated medium-assisted electric field operator $\hat{\mathbf{E}}_{\rm F}(\mathbf{r},\omega_{\rm m})$ fulfills the (quantum) Helmholtz equation:
\begin{equation}
    \left[\boldsymbol{\nabla}\times\boldsymbol{\nabla}\times-\frac{\omega^2_{\rm m}}{c^2}\epsilon(\mathbf{r},\omega_{\rm m})\right]\hat{\mathbf{E}}_{\rm F}(\mathbf{r},\omega_{\rm m})=i\omega_{\rm m}\mu_0\hat{\mathbf{j}}_{\rm N}(\mathbf{r},\omega_{\rm m}),\label{eq: Helmholtz_E}
\end{equation}
where $\epsilon(\mathbf{r},\omega_{\rm m})=\epsilon_R(\mathbf{r},\omega_{\rm m})+i\epsilon_I(\mathbf{r},\omega_{\rm m})$ is the complex permittivity function (or dielectric constant), whose real and imaginary part describes the dispersion and absorption of the electromagnetic environment, respectively.
Note that when coupling to matter degrees of freedom (the TLS) is introduced later, the \emph{full} electric field operator $\hat{\mathbf{E}}(\mathbf{r},\omega_{\rm m})$ also contains a contribution from these material components; the subscript F refers to the portion of the field which can be expressed in terms of the bosonic polariton operators, which can be found from a Fano diagonalization of a model of the passive medium as an oscillator bath~\cite{Suttorp,philbin2010canonical,Gustin2023Jan}.

The current source operator, 
\begin{equation}
\hat{\mathbf{j}}_{\rm N}(\mathbf{r},\omega_{\rm m})=\omega_{\rm m}\sqrt{\frac{\hbar\epsilon_0\epsilon_I(\mathbf{r},\omega_{\rm m})}{\pi}}\,\hat{\mathbf{b}}(\mathbf{r},\omega_{\rm m}),
\end{equation}
is a 
noise current density, which preserves the fundamental field commutation relations.
In the following, we will concentrate on the case of an absorptive scattering structure, that is embedded in a spatial-homogeneous and non-absorptive 
background medium with $\epsilon(\mathbf{r},\omega_{\rm m})=n_{\rm B}^2$.
%However, 
We emphasize that the formalism %is not restricted to spatial-homogeneous systems and 
can also take into account more non-trivial and practical photonic background media, such as photonic crystals or layered structures involving different substrates and waveguides, 
while fully accounting for dispersive properties.

The quantum Helmholtz equation~\eqref{eq: Helmholtz_E} has the formal solution: 
\begin{equation}
    \hat{\mathbf{E}}_{\rm F}(\mathbf{r},\omega_{\rm m})=\frac{i}{\epsilon_0\omega_{\rm m}}\int{d}^3s\, \mathbf{G}(\mathbf{r},\mathbf{s},\omega_{\rm m})\cdot \hat{\mathbf{j}}_{\rm N}(\mathbf{s},\omega_{\rm m}),\label{eq: E_mode}
\end{equation}
where $\mathbf{G}(\mathbf{r},\mathbf{s},\omega_{\rm m})$ is the (classical) photonic Green function, defined from 
\begin{equation}
    \left[\boldsymbol{\nabla}\times\boldsymbol{\nabla}\times-\frac{\omega^2_{\rm m}}{c^2}\epsilon(\mathbf{r},\omega_{\rm m})\right]\mathbf{G}(\mathbf{r},\mathbf{s},\omega_{\rm m})=\frac{\omega_{\rm m}^2}{c^2}\mathbf{I}\delta(\mathbf{r-\mathbf{s}}),\label{eq: GreenHelm}
\end{equation}
together with suitable radiation conditions, i.e.,  the Silver-Müller radiation conditions:
\begin{equation}
    \frac{\mathbf{r}}{|\mathbf{r}|}\times\boldsymbol{\nabla}\times\mathbf{G}(\mathbf{r},\mathbf{s},\omega_{\rm m})\rightarrow in_{\rm B}\frac{\omega_{\rm m}}{c}\mathbf{G}(\mathbf{r},\mathbf{s},\omega_{\rm m}) \label{eq: SM_Cond_GF}.
\end{equation}
We highlight that $\omega_{\rm m}$ is not a Fourier variable of time $t$, but a continuous {\it modal} index. 
Thus, for example, in the Heisenberg picture, one has
$\hat{\bf E}({\bf r}, \omega_{\rm m}, t)$.
The electric field operator, $\hat{\mathbf{E}}_{\rm F}(\mathbf{r})$ is the sum over all mode indices $\omega_{\rm m}$, so that
\begin{equation}\label{eq:EF_def}
\hat{\mathbf{E}}_{\rm F}(\mathbf{r})=\int_0^\infty{d}\omega_{\rm m}\hat{\mathbf{E}}_{\rm F}(\mathbf{r},\omega_{\rm m})+{\rm H.c.}
\end{equation}

As we show in Sec.~\ref{Sec: Light-Matter}, the transverse part of the vector potential (which is manifestly gauge-invariant) $\hat{\mathbf{A}}_{\perp}(\mathbf{r})$ plays a fundamental role in the introduction of gauge-invariant coupling to a TLS, and is given by 
\begin{equation}
\hat{\mathbf{A}}_{\perp}(\mathbf{r})=\frac{1}{\epsilon_0}\int \frac {d\omega_{\rm m}}{\omega_{\rm m}^2}\int{ d}^3s\,  \mathbf{G}^{\perp}(\mathbf{r},\mathbf{s},\omega_{\rm m})\cdot \hat{\mathbf{j}}_{\rm N}(\mathbf{s},\omega_{\rm m}),\label{eq: Avec_components}
\end{equation}
where
\begin{equation}
    \mathbf{G}^{\perp}(\mathbf{r},\mathbf{s},\omega_{\rm m})=\int{ d}^3r' \boldsymbol{\delta}^{\perp}(\mathbf{r}-\mathbf{r}')\cdot\mathbf{G}(\mathbf{r}',\mathbf{s},\omega_{\rm m})
\end{equation}
is the (left-sided) transverse part of the Green tensor, with $\boldsymbol{\delta}^{\perp}(\mathbf{r}-\mathbf{r}')$ being the transverse part of the Dirac delta distribution. Thus, the electromagnetic field operators are  determined by the classical Green function, while the quantum character is fully encoded in the noise operators.

The macroscopic QED formalism allows for rigorous quantization in arbitrary media, but, as it expresses the electromagnetic fields in terms of continuous frequency and spatially-dependent operators, it alone does not yet take advantage of the discrete nature of photonic resonances that occur in (e.g.) cavity structures.
However, using QNM theory ~\cite{PhysRevLett.122.213901}, one can indeed construct an efficient and concise mode decomposition of the field operators, which will be explained in more detail below. In addition to being valid for arbitrary dispersive and absorptive media,  the QNM approach also recovers the important and practical limit of a non-absorptive dielectric medium~\cite{franke2020fluctuation}, and is far more practical than a continuum approach for resonant structures that are well defined in terms of a few transverse modes of interest.

\subsection{Classical field quasinormal mode theory}\label{subsec:classic}

Next we present some basic aspects of (classical) QNM theory, that are needed for their quantization. such as the Green function expansion and mode regularization.

We first introduce the concept of open cavity modes, which we refer to as QNMs~\cite{2ndquant2,NormKristHughes,Lalanne_review,Kristensen:20}.  
Formally, the vector-valued QNM eigenfunctions, $\tilde{\mathbf{f}}_\mu({\bf r})$, of a dispersive and three-dimensional scattering structure are solutions to the classical Helmholtz equation, Eq.~\eqref{eq:QNM_def},
which is solved with 
the Silver-Müller radiation conditions:
\begin{equation}
    \frac{\mathbf{r}}{|\mathbf{r}|}\times\boldsymbol{\nabla}\times\tilde{\mathbf{f}}_\mu(\mathbf{r})\rightarrow in_{\rm B}\frac{\tilde{\omega}_\mu}{c}\tilde{\mathbf{f}}_\mu(\mathbf{r}) \label{eq: SM_Cond_QNM}.
\end{equation}

As a consequence of the radiation conditions, the QNM eigenfrequencies $\tilde{\omega}_\mu=\omega_\mu - i\gamma_\mu$ are complex numbers, where $\omega_\mu$ and  $\gamma_\mu$ describe the resonance frequency and half width at half maximum of the associated QNM resonance, respectively. In addition, $\epsilon(\mathbf{r},\tilde{\omega}_\mu)$ is the analytical continuation of the permittivity function into the complex plane. Note, 
in the following, we will assume that the QNMs are purely transverse (thus, we neglect any static/longitudinal QNM parts), so that the modal part of the Green function expansion is fully encoded in its transverse part
which is found to completely dominate for a wide range of dielectric and plasmonic cavities~\cite{Dezfouli2019,KamandarDezfouli2017b,franke2020quantized,PhysRevA.98.043806,PhysRevLett.126.257401}, including dimer plasmonic gaps, with gaps less than 1-nm~\cite{Dezfouli2019}; indeed, this is also found to be the case with non-local effects included~\cite{KamandarDezfouli2017b,PhysRevLett.126.257401}.

Assuming completeness for the QNMs, and also based
on various rigorous numerical checks for accurate QNM expansions for a wide range of cavity modes, 
the transverse part of the Green function for positions $\mathbf{r}_0$, $\mathbf{r}$ inside or near the resonator can be expanded in terms of the (transverse) QNMs~\cite{lee_dyadic_1999,LeungSP1D,GeNJP2014,Kristensen:20},
\begin{equation}
\mathbf{G}^{\perp}\left(\mathbf{r}_0,\mathbf{r},\omega_{\rm m}\right)= \sum_{\mu} A_{\mu}\left(\omega_{\rm m}\right)\,\tilde{\mathbf{f}}_{\mu}\left({\bf r}_0\right)\tilde{\mathbf{f}}_{\mu}\left({\bf r}\right),\label{eq: GFwithSUM}
\end{equation}
where $A_{\mu}(\omega_{\rm m})$ is a frequency-dependent QNM expansion coefficient, 
\begin{equation}\label{eq:A_coeff}
         A_{\mu}(\omega_{\rm m}) = \frac{\omega_{\rm m}}{2(\tilde{\omega}_{\mu}-\omega_{\rm m})},
     \end{equation}
     and the QNMs are appropriately normalized. 
    The
     QNM expansion uses an unconjugated product of vector fields, which is essential to capture important QNM phase effects.

     Formally speaking, the expansion in terms of QNMs in Eq.~\eqref{eq: GFwithSUM} should only be used \emph{inside} of the scatterer (resonator) geometry. However, when the QNMs dominate the electromagnetic response for a given frequency bandwidth, the use of a strictly QNM expansion is also highly accurate at spatial locations in the near-field of the resonator (i.e., where a material dipole emitter could be placed in a cavity-QED setting). One consequence of this approximation is that ``background'' terms, which arise from the continuous background modes and are required to solve Maxwell's equations at locations far from the resonator geometry, are neglected~\cite{GeNJP2014}. These terms are necessary to capture background field couplings to material emitter excitations, which give rise to the usual spontaneous emission into non-cavity modes. In USC, such terms are often very small, and thus we neglect them going forward.

In the following, any additional longitudinal parts, $\mathbf{G}^{\parallel}\left(\mathbf{r}_0,\mathbf{r},\omega_{\rm m}\right)$, are formally included as weakly contributing parts to the total Green function.
In many QNM calculations for resonant 
cavity systems including metallic dimers or hybrid dielectric-metal resonators, the response is typically
completely described by one or 
just several  QNMs~\cite{Ge2014,KamandarDezfouli2017,Dezfouli2019,carlson2019dissipative,Kristensen:20} (excluding transverse background terms, which as mentioned above are often small in USC), which is one of the main advantages of using a 
{\it discrete} mode quantization scheme.

\begin{figure}[ht]
    \centering
    \includegraphics[width=0.99\columnwidth]{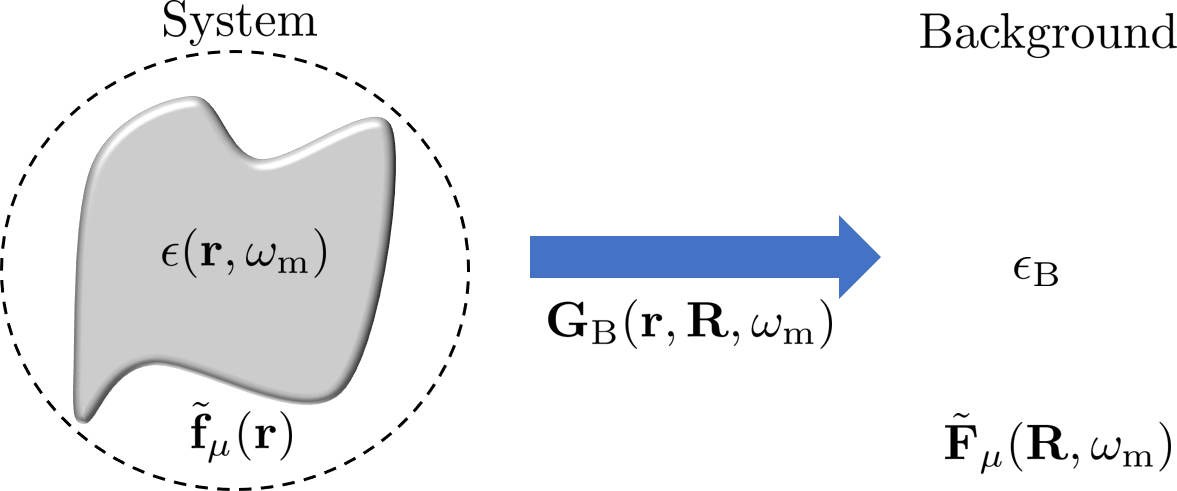}
    \caption{Visualization of the field expansion in terms of QNMs and background contributions. The {\it system} region inside or close to the scattering geometry (here sketched by the exemplary region enclosed by a circle) is represented by the discrete set of QNMs $\tilde{\mathbf{f}}_\mu(\mathbf{r})$. The {\it background} region is described by
    regularized QNMs $\tilde{\mathbf{F}}_\mu(\mathbf{R},\omega_{\rm m})$. The field operators, e.g., $\hat{\mathbf{A}}_{\perp}(\mathbf{r})$, are formed by an integration of the total photon Green function together with the polariton noise operators over all real space, $\mathbb{R}^{3}$.
    }\label{fig: SchematicQNM}
\end{figure}

We emphasize that the expansion in Eq.~\eqref{eq: GFwithSUM} is assumed to be performed at spatial positions within or near the cavity region, which we refer to as the ``system'' (cf.~Fig.~\ref{fig: SchematicQNM}). The actual bounds of the system region are not important, and do not enter into the theory, provided that coupling to active material emitters occurs within the system region, and that the far-field is contained in the background region. 

To fully determine the electromagnetic field operators in the system region, such as the vector potential in Eq.~\eqref{eq: Avec_components}, a Green function expansion for its second spatial entry outside the system is also required. 
At locations far outside the system, practically  the QNMs
must be {\it regularized} as they no longer directly form a good basis for the Green function~\cite{Ge2014,ren_near-field_2020,PhysRevA.108.043502}. The regularizations we use are based on fundamental Green identities and require either: (i) the QNMs {\it inside} the resonator through a form of the Dyson equation~\cite{Ge2014}, or (ii) the QNMs at a boundary enclosing the resonator region through a near-field to far-field transformation~\cite{PhysRevB.101.205402}, which exploits the field equivalence principle~\cite{schelkunoff1936some}. Utilizing these regularizations, one can formulate the transverse Green function for positions (i.e., the second spatial argument of the Green function) outside the system as 
\begin{align}
    \mathbf{G}^{\perp}(\mathbf{r},\mathbf{R},\omega_{\rm m})
    &=
    \sum_\mu A_\mu(\omega_{\rm m})\tilde{\mathbf{f}}_\mu(\mathbf{r})\tilde{\mathbf{F}}_\mu(\mathbf{R},\omega_{\rm m}),\label{eq: RegGreen_plus_Background}
\end{align}
where $\tilde{\mathbf{F}}_\mu(\mathbf{R},\omega_{\rm m})$ represents the {\it regularized QNM} functions, which can be obtained from the Dyson equation approach (which we use going forward) to yield~\cite{GeNJP2014}:
\begin{equation}
    \tilde{\mathbf{F}}_\mu(\mathbf{R},\omega_{\rm m})=\int {d}^3s\Delta\epsilon(\mathbf{s},\omega_{\rm m})\mathbf{G}_{\rm B}(\mathbf{R},\mathbf{s},\omega_{\rm m})\cdot\tilde{\mathbf{f}}_\mu(\mathbf{s}),\label{eq: DysEqBigF} 
    \end{equation}
and
 $\Delta\epsilon(\mathbf{s},\omega_{\rm m})=\epsilon(\mathbf{r},\omega_{\rm m})-n_{\rm B}^2$ is the permittivity difference (defining the cavity) and $\mathbf{G}_{\rm B}(\mathbf{R},\mathbf{s},\omega_{\rm m})$ is the background Green function, which solves Eq.~\eqref{eq: GreenHelm} with $\epsilon(\mathbf{r},\omega)=n_{\rm B}^2$. 
 Here, in line with our earlier discussion, we have again neglected non-modal contributions to the Green function coming from background contributions~\cite{GeNJP2014}.
 
 We remark that there are different forms of the QNM expansion coefficient $A_{\mu}(\omega_{\rm m})$ that one could choose, and the difference in choosing which form to use corresponds to a different set of ways to make a \emph{few-mode approximation}; 
 however, when all modes are included, either choice should lead to equivalent results. When considering the single-mode limit, by use of a Green function identity, either form leads to equivalent results~\cite{Gustin2025}. More specifically, we will see that the use of this identity leads to two different representations of our theory: a \emph{spatially unspecified} form%\sh{for the ...}
, and a \emph{spatially specified} form (defined later in the text) which picks out a specific location in space, and thus depends explicitly on the phase of the QNM at the specified location. For the spatially unspecified form, the choice of $A_{\mu}(\omega_{\rm m})$ does 
 in fact lead to different results, but we show that the spatially specified form is generally needed to obtain the correct result. For very large light-matter couplings (i.e., the deep-strong coupling regime, where the coupling $g$ is nearly equal to the bare frequencies of the system), different coefficient forms can lead to different results even in the spatially specified representation~\cite{Gustin2025}. Here, other issues can occur (e.g., quasistatic effects) as well, so we preclude these extreme regimes from consideration in the (ultimately single-mode) quantized QNM model we present in this work.

For our considered case of a homogeneous background, there is an
explicit analytical form for 
$\mathbf{G}_{\rm B}(\mathbf{r},\mathbf{r}',\omega_{\rm m})$,  which 
is given through~\cite{Scheel}
\begin{align}\label{eq:G_transverse} {\mathbf{G}}_{\mathrm{\rm B}}
=&  - \frac{\delta(\mathbf{X})}{3 n^2_{\rm B}}{\mathbf{I}}+  \frac{{ k_{0}^{2}} \exp \left(i k_{\rm B} X\right)}{4 \pi X} \Bigg[\left(1+\frac{i k_{\rm B} X \! - \! 1}{k^{2}_{\rm B} X^{2}}\right) {\mathbf{I}}  \nonumber\\
&
+ \left(\frac{3-3 i k_{\rm B} X-k^{2}_{\rm B} X^{2}}{k^{2}_{\rm B} X^{2}}\right)\mathbf{n}_{\mathbf{X}}\mathbf{n}_{\mathbf{X}}\Bigg],
\end{align} 
with $\mathbf{X}=\mathbf{r}-\mathbf{r}'=\mathbf{n}_{\mathbf{X}}X$, $X=|\mathbf{X}|$, $k_0=\omega_{\rm m}/c$ and $k_{\rm B}=n_{\rm B}k_0$.
The presented theoretical framework also covers other background structures, e.g. a 
%two-dimensional 
photonic crystal slab structure with an embedded waveguide. In that case, most of the output from the cavity (usually realized by missing holes in the photonic crystal) will be directed into the direction of the (quasi) one-dimensional waveguide, where the respective Green function $\mathbf{G}_{\rm wg}$ is known analytically \cite{PhysRevB.75.205437}.

Summarizing the different contributions to the total Green function, we have 
\begin{align}\label{eq:GF_inside}
    \mathbf{G}(\mathbf{r}_0,\mathbf{r},\omega_{\rm m})=&\sum_\mu A_\mu(\omega_{\rm m})\tilde{\mathbf{f}}_\mu(\mathbf{r}_0)\tilde{\mathbf{f}}_\mu(\mathbf{r})\nonumber\\
    &+\mathbf{G}^{\parallel}(\mathbf{r}_0,\mathbf{r},\omega_{\rm m}),
\end{align}
for $\mathbf{r}$ (or $\mathbf{r}_0$) at positions {\it inside} the system region and 
\begin{align}\label{eq:GF_outside}
    \mathbf{G}(\mathbf{r}_0,\mathbf{R},\omega_{\rm m})=&\sum_\mu A_\mu(\omega_{\rm m})\tilde{\mathbf{f}}_\mu(\mathbf{r}_0)\tilde{\mathbf{F}}_\mu(\mathbf{R},\omega_{\rm m})\nonumber\\
&+\mathbf{G}^{\parallel}(\mathbf{r}_0,\mathbf{R},\omega_{\rm m}),
\end{align}
for $\mathbf{R}$ at positions {\it outside} the system region.

\subsection{Quantized quasinormal modes and reservoir modes\label{subsec: QNM_reservoir}}
We next summarize and  extend previous theoretical work on the expansion of the electromagnetic field operators and Hamiltonian in terms of QNMs
(see Refs.~\cite{PhysRevLett.122.213901,franke2020quantized,franke2020fluctuation,Gustin2023Jan}), which is an essential step for the derivation of the TLS-field interaction in the USC regime described in Sec.~\ref{Sec: Light-Matter}. 

In particular, we will concentrate on the transverse part of the vector potential $\hat{\mathbf{A}}_{\perp}(\mathbf{r})$ as well as the longitudinal part of the electric field operator $\hat{\mathbf{E}}_{\rm F}^{\parallel}(\mathbf{r})$, as these operators are the ones which appear in the minimal coupling Hamiltonian which couples the fields to the TLS~\cite{Gustin2023Jan}.  
While one may naively expect the non-radiative processes to be entirely captured by the longitudinal part of the electromagnetic fields, there are also significant non-radiative decay processes associated with the (transverse) QNM part, since the coupling of the electromagnetic field and the medium is present in both the transverse and longitudinal fields (in principle). Indeed, all plasmonic QNMs
have non-radiative contributions through Ohmic heating,
which can be used to quantitatively describe the 
non-radiative beta factor (quenching), as well as the
radiative beta factor (or quantum efficiency)~\cite{Ge:14Dimer}. This is often poorly understood in the plasmonics community, but both modal contributions (radiative and nonradiative) scale precisely with the usual Purcell factor mode scaling, $\propto Q/V_{\rm eff}$.

In previous works involving quantized QNMs, a \emph{spatially unspecified} scheme was used to construct the QNMs, where the quantization involves integration over all spatial regions, and does not pick out a specific location. Additionally in this work, by use of a Green's function identity, we can move to a \emph{spatially specified} representation by picking out a specific location
of interest (e.g., the location of the dipole), which we will show is necessary to identify the correct frequency-dependence of the master equation decay rates of the cavity in the USC and dissipative broadband regimes.  This is discussed in Sec.~\ref{sec:spatial_spec}.

\subsubsection{Projection of discrete oscillator subspace from continuum polariton excitations of electromagnetic fields and the medium}
 
To obtain a representation of the electromagnetic fields in terms of quantized QNMs and residual reservoir components, we employ a recently proposed projection approach~\cite{Gustin2023Jan}, adapted from previous QNM quantization schemes~\cite{PhysRevLett.122.213901}. Specifically, we write the medium polariton operators as 
\begin{equation}\label{eq:projection}
    \hat{\mathbf{b}}(\mathbf{r},\omega_{\rm m}) = \sum_{\mu} \mathbf{L}^*_{\mu}(\mathbf{r},\omega_{\rm m})\hat{a}_{\mu} + \hat{\mathbf{c}}(\mathbf{r},\omega_{\rm m}),
\end{equation}
where we have defined quantized discrete operators
\begin{equation}
    \hat{a}_{\mu} = \int d^3r\int d\omega_{\rm m } \mathbf{L}_{\mu}(\mathbf{r},\omega_{\rm m}) \cdot \hat{\mathbf{b}}(\mathbf{r},\omega_{\rm m}),
\end{equation}
and $\hat{\mathbf{c}}(\mathbf{r},\omega_{\rm m})$ is a residual reservoir operator, defined implicitly by Eq.~\eqref{eq:projection}. For the discrete operators to correspond to distinct (quasi-)modes, we require $[\hat{a}_{\mu},\hat{a}_{\nu}]=\delta_{\mu\nu}$, which implies
\begin{equation}\label{eq:orthonormal}
    \int d^3r \int d\omega_{\rm m} \mathbf{L}_{\mu}(\mathbf{r},\omega_{\rm m}) \cdot \mathbf{L}_{\nu}^*(\mathbf{r},\omega_{\rm m}) = \delta_{\mu \nu}.
\end{equation}

Importantly, by construction, the discrete operators $\hat{a}_{\mu}^{(\dagger)}$ commute with the continuum operators $\hat{\mathbf{c}}^{(\dagger)}(\mathbf{r},\omega_{\rm m})$. As such, we can then decompose the field Hamiltonian $\hat{H}_{\rm em}$ into discrete, continuum, and interaction parts as follows:
\begin{subequations}\label{eq:disc_cont}
    \begin{equation}\label{eq:Hem_sys}
        \hat{H}^{\rm discr}_{\rm em} = \sum_{\mu \nu} \chi_{\mu \nu} \hat{a}^{\dagger}_{\mu}\hat{a}_{\nu},
    \end{equation}
    \begin{equation}
        \hat{H}^{\rm cont}_{\rm em} = \int d^3r \int d\omega_{\rm m} \hbar\omega_{\rm m} \hat{\mathbf{c}}^{\dagger}(\mathbf{r},\omega_{\rm m}) \cdot \hat{\mathbf{c}}(\mathbf{r},\omega_{\rm m}),
    \end{equation}
    \begin{equation}
        \hat{H}^{\rm int}_{\rm em} = \sum_{\mu}\int d^3r \int d\omega_{\rm m} \hbar\omega_{\rm m} \mathbf{L}_{\mu}(\mathbf{r},\omega_{\rm m}) \cdot \hat{\mathbf{c}}(\mathbf{r},\omega_{\rm m}) \hat{a}^{\dagger}_{\mu} + \text{H.c.},
    \end{equation}
    where 
    \begin{equation}\label{eq:chi}
        \chi_{\mu \nu} = \int d^3r \int d\omega_{\rm m } \omega_{\rm m} \mathbf{L}_{\mu}(\mathbf{r},\omega_{\rm m}) \cdot \mathbf{L}^*_{\nu}(\mathbf{r},\omega_{\rm m}).
    \end{equation}
\end{subequations}
Note that the reservoir continuum operators in general have non-local commutation relations:
\begin{align}\label{eq:commutator}
[\hat{\mathbf{c}}(\mathbf{r},\omega_{\rm m}), & \ \hat{\mathbf{c}}^{\dagger}(\mathbf{r}',\omega_{\rm m}')] = \mathbf{I}\delta(\mathbf{r}-\mathbf{r}')\delta(\omega_{\rm m }-\omega'_{\rm m}) \nonumber \\ 
& - \sum_{\mu} \mathbf{L}^*_{\mu}(\mathbf{r},\omega_{\rm m})\mathbf{L}_{\mu}(\mathbf{r}',\omega_{\rm m}'),
\end{align}
though for the case of quantized QNMs, we show in Appendix~\ref{app:boson} that the second term of Eq.~\eqref{eq:commutator} can be neglected when deriving the quantum master equation, and the reservoir operators are effectively local under the Markov approximation.

Up to now, this construction is completely general. Additionally, to obtain a mode expansion for the transverse vector potential of the form $\hat{\mathbf{A}}_{\perp}(\mathbf{r}) = \hat{\mathbf{A}}^{\perp}_{\rm discr}(\mathbf{r}) + \hat{\mathbf{A}}^{\perp}_{\rm B}(\mathbf{r})$, where
\begin{equation}
   \hat{\mathbf{A}}^{\perp}_{\rm discr} = \sum_{\mu} \sqrt{\frac{\hbar}{2\epsilon_0 \chi_{\mu \mu}}}\tilde{\mathbf{f}}_{\mu}^{\rm s}(\mathbf{r})\hat{a}_{\mu} + \text{H.c.},
\end{equation}
and
    \begin{align}\label{eq:AB}
\hat{\mathbf{A}}^{\perp}_{\rm B}&(\mathbf{r})=\int  {\rm d}^3\!r'   \int   \frac{d \omega_{\rm m}}{\omega_{\rm m}}  \sqrt{\frac{\hbar \epsilon_I(\mathbf{r}',\omega_{\rm m})}{\epsilon_0 \pi}}  \nonumber \\ & \times \mathbf{G}^{\perp}(\mathbf{r},\mathbf{r}',\omega_{\rm m})\cdot \hat{\mathbf{c}}(\mathbf{r}',\omega_{\rm m}) + \text{H.c.},
\end{align}
which, in terms of $\hat{\mathbf{A}}_{\rm discr}^{\perp}(\mathbf{r})$, is similar in form to more usual normal mode expansions, we define a corresponding (quasi-)mode profile:
\begin{align}\label{eq:qnm_sym}
&\tilde{\mathbf{f}}_{\mu}^{\rm s}(\mathbf{r}) = \nonumber \\ &\sqrt{\frac{2\chi_{\mu \mu}}{\pi}} \!\! \int \! \! d^3r' \! \!\! \int \! \! \frac{ d\omega_{\rm m}}{\omega_{\rm m}}\sqrt{\epsilon_I(\mathbf{r}',\omega_{\rm m})} \mathbf{G}^{\perp}(\mathbf{r},\mathbf{r}',\omega_{\rm m})\!\cdot\! \mathbf{L}_{\mu}^*(\mathbf{r}',\omega_{\rm m}).
\end{align}

\subsubsection{Construction of quantized quasinormal modes}

We now move from this general construction of a discrete oscillator basis to a description in terms of \emph{quantized quasinormal modes}. To do so, we choose  the projection functions to satisfy~\cite{Gustin2023Jan}:
\begin{align}\label{eq:projection_fn}
&\mathbf{L}_{\mu}(\mathbf{r},\omega_{\rm m}) = \nonumber \\ &\sum_{\nu} \left[S^{-\frac{1}{2}}\right]_{\mu \nu} \sqrt{\frac{2\omega_{\nu}\epsilon_I(\mathbf{r},\omega_{\rm m})}{\pi}}\frac{A_{\nu}(\omega_{\rm m})}{\omega_{\rm m}} \tilde{\mathbf{F}}'_{\nu}(\mathbf{r},\omega_{\rm m}),
\end{align}
where $\tilde{\mathbf{F}}'_{\nu}(\mathbf{r},\omega_{\rm m}) = \tilde{\mathbf{f}}_{\nu}(\mathbf{r})$ for $\mathbf{r}$ inside the system region (i.e., the QNM mode profile), and $\tilde{\mathbf{F}}'_{\nu}(\mathbf{R},\omega_{\rm m}) = \tilde{\mathbf{F}}_{\nu}(\mathbf{R},\omega_{\rm m})$ for $\mathbf{R}$ outside the system region (i.e., the regularized QNM).
The former $\tilde{\mathbf{f}}_{\nu}(\mathbf{r})$ are frequency independent and spatially divergent, and the latter
$\tilde{\mathbf{F}}_{\nu}(\mathbf{R},\omega_{\rm m})$ are frequency-dependent and spatially convergent (i.e., well behaved in the far field).

The factor $S_{\mu\nu}$ is a symmetrizing matrix (closely connected to the radiative and non-radiative loss of the QNMs), which is found by enforcing Eq.~\eqref{eq:orthonormal}, 
\begin{align}\label{eq:S}
    S_{\mu \nu}& = \frac{2\sqrt{\omega_{\mu}\omega_{\nu}}}{\pi} \int \frac{d\omega_{\rm m}}{\omega_{\rm m}^2} \int d^3r  \nonumber \\ 
    & \epsilon_I(\mathbf{r},\omega_{\rm m})A_{\rm \mu}(\omega_{\rm m}) A^*_{\rm \nu}(\omega_{\rm m})\tilde{\mathbf{F}}'_{\mu}(\mathbf{r},\omega_{\rm m}) \cdot \tilde{\mathbf{F}}'^*_{\nu}(\mathbf{r},\omega_{\rm m}).
\end{align}
We refer to this form of the symmetrizing matrix as the {\it spatially unspecified form}, as it does not pick out any specific position where the QNMs are evaluated, in contrast to the spatially specified form we identify in Sec.~\ref{sec:spatial_spec} for the single-mode limit. 

Then, inserting the definition of the projection functions in Eq.~\eqref{eq:projection_fn} into the mode expansion coefficients from Eq.~\eqref{eq:qnm_sym}, we see that the relationship between the field expansion functions $\tilde{\mathbf{f}}^{\rm s}_{\mu}(\mathbf{r})$ and the QNM mode profiles is
\begin{equation}
\tilde{\mathbf{f}}^{\rm s}_{\mu}(\mathbf{r}) = \sum_{\nu} \tilde{\mathbf{f}}_{\nu}(\mathbf{r})\left[S^{\frac{1}{2}}\right]_{\nu \mu}\sqrt{\frac{\chi_{\mu \mu}}{\omega_{\nu}}},
\end{equation}
and they take the form of symmetrized QNM mode profile functions, recovering previous work~\cite{PhysRevLett.122.213901}. As such, we can identify $\hat{\mathbf{A}}_{\rm discr}^{\perp} = \hat{\mathbf{A}}^{\perp}_{\rm QNM}$, in terms of quantized QNMs. Explicitly, for $\mathbf{r}$ within the system region,
\begin{equation}\label{eq:QNM_expansion}
\hat{\mathbf{A}}^{\perp}_{\rm QNM}(\mathbf{r}) = \sum_{\mu} \sqrt{\frac{\hbar}{2\epsilon_0 \chi_{\mu \mu}}}\tilde{\mathbf{f}}_{\mu}^{\rm s}(\mathbf{r})\hat{a}_{\mu} + \text{H.c.},
\end{equation}
and, in fact, due to our definition of the QNM projection functions $\mathbf{L}_{\mu}(\mathbf{r},\omega_{\rm m})$, we have simply $\hat{\mathbf{A}}_{\rm B}^{\perp}(\mathbf{r})=0$ for the case of QNMs (see Appendix~\ref{app:background_A} for a derivation). Note that the vanishing of the background component is unique to the transverse potential---for example, $\hat{\mathbf{E}}_{\rm F, B}^{\perp} \neq 0$. Moreover, it is a consequence of our approximation of neglecting non-modal transverse background contributions to the Green function expansion, which, as previously mentioned, would be necessary to include to model things like nonresonant background radiation loss of emitters to non-QNM modes on an \emph{ab initio} basis, if in a regime where such terms are substantial.

The difference between the quantization of QNMs here and previous work~\cite{PhysRevLett.122.213901,franke2020fluctuation,PhysRevLett.127.013602,PhysRevA.105.023702} is twofold: (i) We define the quantized QNM expansion in terms of the transverse vector potential (as opposed to the electric field operator), as the vector potential plays the more fundamental role 
in the gauge-invariant theory of macroscopic QED, as it is used to define the Hamiltonian which remains gauge-invariant upon truncation of material levels to a finite basis, which is especially important to ensure gauge-invariant results in the USC coupling regime~\cite{Gustin2023Jan,Savasta2021May}. This definition is more convenient, as, for this choice, $\hat{\mathbf{A}}_{\rm B}^{\perp} = 0$, which allows for straightforward transformations between gauges in terms of the quantized QNM operators. As certain gauges are more convenient in specific situations (for example, the Coulomb gauge is more convenient in the case of time-dependent couplings to matter~\cite{PhysRevResearch.3.023079,Gustin2023Jan}), the ability to transform between gauges with ease is a substantial benefit of our scheme; (ii) We define the mode expansion in terms of the $\chi_{\mu\mu}$ parameters, as opposed to the bare QNM frequency $\omega_{\mu}$, as in the single mode limit, $\chi_{\mu\mu}$ is the coefficient of the bare cavity QNM resonance that occurs in the Hamiltonian [see Eq.~\eqref{eq:Hem_sys}]. Note that for high-Q resonances (and in practice, even modest Q factors  $\sim 10$ encountered in plasmonics), the poles in the projection functions $\mathbf{L}_{\mu}(\mathbf{r},\omega)$ centered around $\omega_\mu$ allow for the $\omega_{\rm m}$ factor in Eq.~\eqref{eq:chi} to be approximated as $\omega_{\mu}$, recovering $\chi_{\mu\mu} \approx \omega_{\mu}$. In practice, we do this for all our calculations later in the paper.

Despite these differences, the fundamental approach of our quantization scheme here and the associated numerical techniques for calculating the associated parameters, including the symmetrizing matrix $S_{\mu\nu}$ (i.e., in the spatially unspecified representation) and its positive definiteness, remain the same as in previous works. Additionally, we note that the choice of defining the QNM expansion in terms of the transverse vector potential is not essential, and one can also use a definition in terms of the transverse electric field (as in previous works), and we do this in Appendix~\ref{app:E_field_QNM}. A formulation with respect to the electric field is more convenient in particular if using the dipole gauge, which can be sometimes be preferable when calculating TLS observables~\cite{Lamb1987Sep} and formulating input-output theories~\cite{Milonni1989Oct}.

Note that, in the general case of lossy materials, and in the spatially unspecified form,
we have separate  contributions
from radiative and non-radiative losses.
The term 
\begin{align}\label{eq:s_nrad}
   S_{\mu\nu}^{\rm nrad}= \frac{2}{\pi} \int d\omega_{\rm m}\frac{\sqrt{\omega_{\mu}\omega_{\nu}}}{\omega_{\rm m}^2} A_{\mu}(\omega_{\rm m}) A^*_{\nu}(\omega_{\rm m})S_{\mu\nu}^{\rm nrad}(\omega_{\rm m}),
\end{align}
where
\begin{align}
   &S_{\mu\nu}^{\rm nrad}(\omega_{\rm m})= \int_{\rm V_{in}}{d}^3r \epsilon_I(\mathbf{r},\omega_{\rm m})\tilde{\mathbf{f}}_\mu(\mathbf{r})\cdot\tilde{\mathbf{f}}_\nu^*(\mathbf{r}),
\end{align}
reflects the non-radiative loss contribution through a QNM overlap integral over the absorptive region (the scattering geometry ${\rm V_{in}}$), and
\begin{align}\label{eq:s_rad}
   S_{\mu\nu}^{\rm rad}= \frac{2}{\pi} \int d\omega_{\rm m}\frac{\sqrt{\omega_{\mu}\omega_{\nu}}}{\omega_{\rm m}^2} A_{\mu}(\omega_{\rm m}) A^*_{\nu}(\omega_{\rm m})   S_{\mu\nu}^{\rm rad}(\omega_{\rm m}),
\end{align}
where
\begin{align}
   S_{\mu\nu}^{\rm rad}(\omega_{\rm m})= \frac{n_{\rm B} c}{\omega_{\rm m}}\oint_{\mathcal{S}_\infty} {d}A_\mathbf{s}\tilde{\mathbf{F}}_{\mu}(\mathbf{s},\omega_{\rm m})\cdot\tilde{\mathbf{F}}_{\nu}^*(\mathbf{s},\omega_{\rm m})
\end{align}
describes the radiative loss processes through a power flow expression at a far-field surface $S_{\infty}$; here  the Silver-Müller radiation conditions are approximately valid (compared to the strict relation at $|\mathbf{s}|\rightarrow\infty$). As mentioned before, the QNM part contains both non-radiative and radiative loss processes. Obviously for a lossless material system, then we only have radiative contributions, a limit that is fully recovered for any finite-size
real dielectric resonator~\cite{franke2020fluctuation}.
From a practical viewpoint, one can approximate the radiative and non-radiative symmetrization factors through a resonant pole approximation, around the peaks of the appearing Lorentzian functions (due to the pole in the QNM expansion coefficients)~\cite{ren_near-field_2020}.

Moving on to the longitudinal electric field operator, the QNM-basis contribution becomes
\begin{align}
    \hat{\mathbf{E}}^{\parallel}_{\rm F, QNM}(\mathbf{r})= i\sum_{\mu}\sqrt{\frac{\hbar\chi_{\mu\mu}}{2\epsilon_0}}\tilde{\mathbf{f}}_{\mu}^{\parallel}(\mathbf{r})\hat{a}_{\mu} + \text{H.c.},
\end{align}
where we have defined the projected longitudinal ``mode'' profiles as 
\begin{align}
\tilde{\mathbf{f}}_{\mu}^{\parallel}(\mathbf{r}) = &\int d^3r' \int d\omega_{\rm m}  \nonumber \\ & \times \sqrt{\frac{2\epsilon_I(\mathbf{r}',\omega_{\rm m})}{\pi \chi_{\mu\mu}}} \mathbf{G}^{\parallel}(\mathbf{r},\mathbf{r}',\omega_{\rm m}) \cdot \mathbf{L}_{\mu}^*(\mathbf{r}',\omega_{\rm m}), 
\end{align}
and the reservoir contribution is 
\begin{align}
    \hat{\mathbf{E}}^{\parallel}_{\rm F, B}(\mathbf{r})=&i\sqrt{\frac{\hbar}{\pi\epsilon_0}}\int_0^\infty{d}\omega_{\rm m}\int{d}^3s \sqrt{\epsilon_I(\mathbf{s},\omega_{\rm m})}\nonumber\\
    &\times \mathbf{G}^{\parallel}(\mathbf{r},\mathbf{s},\omega_{\rm m})\cdot \hat{\mathbf{c}}(\mathbf{s},\omega_{\rm m})+{\rm H.c.}
\end{align}
Note that $\tilde{\mathbf{f}}^{\parallel}_{\mu}(\mathbf{r})$ are not mode profiles in a typical sense, but rather spatially-dependent and longitudinal expansion coefficient associated with the quantized QNMs when the full longitudinal electric field is decomposed in the QNM-reservoir basis. Although the QNMs we use 
%themselves 
are transverse, the mode projection functions $\mathbf{L}_{\mu}(\mathbf{r},\omega_{\mu})$ are not strictly transverse over all space, and as such, retain small longitudinal components.
In a regime where the transverse QNMs dominate over the total electromagnetic fields, we expect the contribution of these longitudinal components to be negligible (as can easily be confirmed numerically).

In addition to the field decompositions, we also note that the discrete-continuum Hamiltonian in Eq.~\eqref{eq:disc_cont} can also be understood in terms of this QNM-reservoir separation, with $\hat{H}^{\rm em}_{\rm discr} = \hat{H}^{\rm em}_{\rm QNM}$, $\hat{H}^{\rm em}_{\rm cont} = \hat{H}^{\rm em}_{\rm R}$, and $\hat{H}^{\rm em}_{\rm int} = \hat{H}^{\rm em}_{\rm QNM-R}$. The last interaction term can also be written as (cf.~Ref.~\onlinecite{franke2020quantized} for details)
\begin{align}
    H_{\rm QNM-R} &= \hbar\sum_{\mu}\int_0^\infty \! { d}\omega_{\rm m}\int \! { d}^3r\,  \mathbf{g}_\mu(\mathbf{r},\omega_{\rm m})\cdot\hat{\mathbf{c}}(\mathbf{r},\omega_{\rm m})\hat{a}^{\dagger}_\mu \nonumber\\
    &+ {\rm H.c.},
\end{align}
where
\begin{align}\label{eq:projection_fn_g}
&\mathbf{g}_{\mu}(\mathbf{r},\omega_{\rm m}) = \nonumber \\ &\sum_{\nu} \left[S^{-\frac{1}{2}}\right]_{\mu \nu} \sqrt{\frac{2\omega_{\nu}\epsilon_I(\mathbf{r},\omega_{\rm m})}{\pi}}\frac{B_{\nu}(\omega_{\rm m})}{\omega_{\rm m}} \tilde{\mathbf{F}}'_{\nu}(\mathbf{r},\omega_{\rm m});
\end{align}
this has the same form as the projection function $\mathbf{L}_{\mu}(\mathbf{r},\omega_{\rm m})$, but with
\begin{equation}
    B_{\nu}(\omega_{\rm m}) = -(\tilde{\omega}_{\nu} - \omega_{\rm m})A_{\nu}(\omega_{\rm m})
\end{equation}
replacing $A_{\nu}(\omega_{\rm m})$; i.e., with the simple pole at the complex QNM frequency removed. For the choice of coefficient we use in this work, Eq.~\eqref{eq:A_coeff}, one simply has $B_{\nu}(\omega_{\rm m}) = -\omega_{\rm m}/2$ for all $\nu$.

Our model can be regarded as an \emph{ab initio} system-bath theory, where the operators $\hat{\mathbf{c}}(\mathbf{r},\omega_{\rm m})$ and associated states take on the role of the (residual) reservoir modes and $\hat{a}_\mu$ are the system QNM modes. The inherent QNM decay rate $\gamma_\mu$ is now implicitly included in the more general coupling $\mathbf{g}_\mu(\mathbf{r},\omega_{\rm m})$ through evaluation for frequencies around $\omega_{\mu}$, i.e., for a single QNM $\mu$ and no active emitter coupling, $\int{\rm d}^3r\,  |\mathbf{g}_\mu(\mathbf{r},\omega_{\mu})|^2\approx \gamma_{\mu}/\pi$. This function allows us to calculate frequency-dependent decay rates corresponding to optically active transitions of the light-matter Hamiltonian in the USC regime, without any phenomenological assumptions on the form of the system-reservoir interaction. This will be discussed in more detail in Sec.~\ref{subsec: NM_Limits}.

\subsubsection{Spatially specified form in the single mode approximation }\label{sec:spatial_spec}

In the single-mode case (taking $\mu = c$), 
we can also move to a spatially specified form of the $S_c = S_{cc}$ quantization parameter, which, as we show later, is necessary to obtain the correct frequency-dependent modifications to the master equation in the presence of coupling to active dipole TLS coupling~\cite{Gustin2025}. This can be done by making use of the fundamental Green function relation~\cite{Dung} (suppressing the frequency index):
\begin{equation}\label{eq:gf_id}
    \int d^3r \epsilon_I(\mathbf{r})\mathbf{G}^{\perp}(\mathbf{r}_1,\mathbf{r}) \cdot \mathbf{G}^{\perp *}(\mathbf{r},\mathbf{r}_2) = \text{Im}\{\mathbf{G}^{\perp}(\mathbf{r}_1,\mathbf{r}_2)\}
\end{equation}
for any $\mathbf{r}_1$ and $\mathbf{r}_2$. Thus, taking Eq.~\eqref{eq:GF_inside} and Eq.~\eqref{eq:GF_outside} in the single-mode limit, letting $\mathbf{r}_1=\mathbf{r}_2$, and taking the dot product on the right and left with an arbitrary real unit vector $\mathbf{n}$, we obtain the relation
\begin{align}\label{eq:GF_id_1mode}
    &\int d^3r \epsilon_I(\mathbf{r},\omega_{\rm m}) |A_c(\omega_{\rm m})\tilde{\mathbf{F}}'_c(\mathbf{r},\omega_{\rm m})|^2 = \text{Im}\{A_c(\omega_{\rm m})e^{i2\phi_1}\},
\end{align}
which is clearly dependent on the QNM phase at the specified location, projected onto the arbitrary vector $\phi_1 = \text{arg}\{\mathbf{ n} \cdot \mathbf{\tilde{f}}_c(\mathbf{r}_1)\}$.
We then have 
\begin{equation}
    \text{Im}\{A_c(\omega_{\rm m})e^{i2\phi_1}\} = \frac{\omega_{\rm m}\gamma_c \cos{(2\phi_1)}}{2\left[(\omega_{\rm m}-\omega_c)^2 + \gamma_c^2\right]}\zeta_c(\phi_1,\omega_{\rm m}),
\end{equation}
where
\begin{equation}\label{eq:zetadef}
    \zeta_c(\phi_1,\omega_{\rm m}) = 1-2Q_c\tan{\left(2\phi_1\right)}\left(
    \frac{\omega_{\rm m}}{\omega_c} - 1\right)
\end{equation}
is a factor which depends on the intrinsic QNM phase at the chosen location, and determines the deviation from a Lorentzian lineshape (modulated by a linear frequency factor) of the spontaneous emission rate of a dipole with resonant frequency $\omega_0 = \omega_{\rm m}$ weakly coupled to a single cavity mode~\cite{Gustin2025}.
Now taking the single-mode limit of Eq.~\eqref{eq:S}, we find, in the spatially specified representation,
\begin{equation}
    S_c(\mathbf{n},\mathbf{r}_1) = \frac{\gamma_c \cos{(2\phi_1)}}{\pi} \int \frac{d\omega_{\rm m}}{\omega_{\rm m}} \frac{\zeta_c(\phi_1,\omega_{\rm m})}{(\omega_{\rm m}-\omega_c)^2 + \gamma_c^2}.
\end{equation}

Assuming the integral to be sharply peaked around the resonance $\omega_{\rm m} = \omega_c$, we can make a pole approximation by letting $\omega_{\rm m} \approx \omega_c$ in the factor that appears in the denominator (consistent with previous evaluations of the $S$ quantization matrix~\cite{franke2020quantized}), and extend the limits of the frequency integral to $\pm \infty$, obtaining:
\begin{equation}
    S_c(\mathbf{n},\mathbf{r}_1) \approx \cos{(2\phi_1)}.
\end{equation}
We remark that this approximation breaks down when $|\tan{(2\phi_1)}| \gg 1$; thus, we should exclude this representation in any regions where $\phi_1 \approx \pi/4$ (or more generally $\pm \pi/4 (2m+1)$,
with $m=0,1,2,\cdots$). Moreover, when considering broadband coupling to matter within the Markov approximation, the resultant master equation can (as we will see later) have unphysical negative decay rates for any QNM phase that is not sufficiently small, so more broadly we restrict ourselves to $|\phi_1| \ll 1$.

It is important to note that, in this representation, $S_c({\bf n}, \mathbf{r}_1)$ {\it depends explicitly on the position of the QNM that has been specified}, and as such, will generally give different results that the spatially unspecified form given in the previous subsection. Additionally, this form does not allow one to identify radiative and non-radiative contributions separately as a function of position (i.e., of a placed dipole).
Nonetheless, we will show in Sec.~\ref{subsec: NM_Limits} that the spatially specified representation is required to identify the correct frequency-dependent decay rates in the USC regime, as well as more generally at spatial positions where the QNM phase is substantial, even with weak coupling.

\section{Quasinormal mode light-matter interaction in the ultrastrong coupling regime\label{Sec: Light-Matter}}
In this section, we present the light-matter interaction Hamiltonian, valid for arbitrary light-matter coupling strengths, and we will assume a single effective matter particle (TLS with a dipole approximation).

\begin{figure}[ht]
    \centering
    \includegraphics[width=0.7\columnwidth]{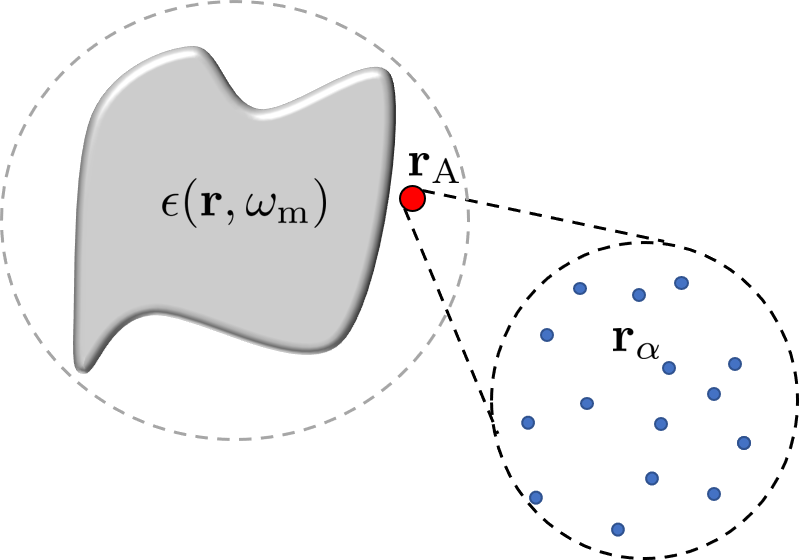}
    \caption{Schematic of an arbitrary shaped scattering structure with (complex) permittivity $\epsilon(\mathbf{r},\omega_{\rm m})$, which is located near a cloud of charged particles with positions $\mathbf{r}_\alpha$ (blue dots). The charged particles form an effective particle at the center of mass position $\mathbf{r}_{\rm A}$ (red dot). The grey dashed circle reflects the region of the ``system'', where the QNMs form a good basis for the field expansion (cf.~Fig.~\ref{fig: SchematicQNM}). Note that in this work we let $\mathbf{r}_{\rm A} = \mathbf{0}$.
}\label{fig: Schematic}
\end{figure}

\subsection{Generalized minimal coupling replacement}\label{sec:min_coupling}
In recent years, it has become well-established that the usual approach to truncate the minimal coupling Hamiltonian giving light-matter interactions can lead to inconsistent gauge-dependent predictions when in the USC regime, and that gauge-invariance can be restored by means of an appropriate generalization of the minimal coupling procedure~\cite{DeBernardis2018Nov,DiStefano2019Aug,Taylor2020Sep,Salmon2022Mar,Gustin2023Jan,Taylor2022Mar}. In Ref.~\cite{Gustin2023Jan} we have recently shown how to derive arbitrary-gauge Hamiltonians consistent with gauge-invariance in the truncated space for both material and mode truncation. Here we briefly show how to write down the correctly-truncated Coulomb and dipole gauge Hamiltonians under the case of material truncation; since our separation of QNM and reservoir parts of the electromagnetic fields preserves the entire spatio-frequency structure of their expansions, mode truncation concerns are not directly applicable here.

Before introducing interactions with the active matter degrees of freedom and the electromagnetic fields, we assume an initial {\it uncoupled} (or bare) Hamiltonian of light and matter, which takes the form:
\begin{equation}
    \hat{H}_{\rm bare} = \hat{H}_0 + \hat{H}_{\rm em},
\end{equation}
where $\hat{H}_{\rm em}$ is defined 
in Eq.~\eqref{eq: H_em} 
and $\hat{H}_0$ is an essentially arbitrary \emph{effective matter} Hamiltonian, but ultimately should have an energy spectrum with an optical transition dipole moment between two states which are well-separated in energy from all other transitions of its spectrum to allow for a TLS approximation; $H_0$ thus contains any Coulomb interactions between the constituent particles of the matter system. However, we stress that 
interactions with the medium-assisted field, transverse or longitudinal, have not yet been included. It should be noted that medium-assisted longitudinal fields can play a role in modifying the bare ($\hat{H_0}$) material particle system energy, in a way not easily captured by our formalism~\cite{PhysRevA.68.013822}. Following the standard approach, we assume this effect to be already incorporated in the definition of $\hat{H}_0$, and we include the longitudinal medium-assisted terms later for completeness.

We then introduce the projector operator for a single TLS:
\begin{equation}
    \hat{P} = \ket{e}\bra{e} + \ket{g}\bra{g},
\end{equation}
where $\ket{e}$ and $\ket{g}$ are the two states (``excited'' and ``ground'', respectively) which constitute the two-level approximation for the matter Hamiltonian; these have energy separation $\hbar\omega_0$, such that 
\begin{equation}
    \hat{P}\hat{H}_0\hat{P} = \frac{\hbar\omega_0}{2}\hat{\sigma}_z.
\end{equation}

We assume that the only information about the TLS which is required, in addition to its energy level separation, is its dipole moment, defined from:
\begin{subequations}\label{eq: d_def}
\begin{equation}\label{eq:dia}
    \bra{e}\hat{\mathbf{d}}\ket{e} = \bra{g}\hat{\mathbf{d}}\ket{g}=0,
\end{equation}
\begin{equation}\label{eq:off}
    \bra{e}\hat{\mathbf{d}}\ket{g} = \bra{g}\hat{\mathbf{d}}\ket{e} = \mathbf{d}, 
\end{equation}
\end{subequations}
where $\hat{\mathbf{d}}=\sum_{\alpha} q_{\alpha}\hat{\mathbf{r}}_{\alpha} = \mathbf{d}\hat{\sigma}_x$ is the dipole operator and $\mathbf{d}$ is the dipole moment (assumed real). 
Since Eqs.~\eqref{eq:dia} and~\eqref{eq:off} 
are satisfied so long as the TLS has parity symmetry,  our results are general, and thus apply to  any dipole emitter system. Figure~\ref{fig: Schematic} shows a schematic of this setup.

Including dipole-field interactions, 
it has been shown recently that the Hamiltonian which preserves gauge-invariance under material truncation within the framework of macroscopic QED and quantized QNM theory is~\cite{Gustin2023Jan}:
\begin{equation}
    \hat{\mathcal{H}} = \hat{\mathcal{V}}\hat{H}_{\rm em}\hat{\mathcal{V}}^{\dagger} + \hat{\mathcal{U}}\hat{\mathcal{H}}_0\hat{\mathcal{U}}^{\dagger} + \hat{\mathcal{H}}_{\parallel},
\end{equation}
where we use calligraphic letters to denote variables truncated with respect to the material degrees of freedom. The choice of gauge under which $\hat{\mathcal{H}}$ is to be realized is determined by the unitary operators $\hat{\mathcal{V}}^{(\dagger)}$ and $\hat{\mathcal{U}}^{(\dagger)}$. As $\hat{\mathcal{U}}^{(\dagger)}$ in particular is expressed directly as a function of the truncated dipole operator (proportional to the position operator), the minimal-coupling replacement is self-consistently implemented in the two-level basis, preserving gauge-invariance.

Here,  $\hat{\mathcal{H}}_{\parallel} = -\mathbf{d} \cdot \hat{\mathbf{E}}^{\parallel}_{\rm F}(\mathbf{0}) \hat{\sigma}_x$ is the manifestly gauge-invariant term corresponding to longitudinal electric field coupling to the dipole. The coupling between the transverse electromagnetic fields and the dipole is given by the unitary operators $\hat{\mathcal{V}}^{(\dagger)}$ and $\hat{\mathcal{U}}^{(\dagger)}$, which are expressed as functions of the \emph{truncated} dipole operator:
\begin{equation}
    \hat{\mathcal{V}} = \exp{\left[-\frac{i}{\hbar}\int d^3r \hat{\bm{\mathcal{P}}}_{\perp}(\mathbf{r}) \cdot \mathbf{A}_{\perp}(\mathbf{r})\right]}
\end{equation}
\begin{equation}
    \hat{\mathcal{U}} = \exp{\left[\frac{i}{\hbar}\int d^3r \left(\hat{\bm{\mathcal{P}}}^{\rm d}_{\perp}(\mathbf{r})  - \hat{\bm{\mathcal{P}}}_{\perp}(\mathbf{r})\right) \cdot \mathbf{A}_{\perp}(\mathbf{r})\right]},
\end{equation}
where $\hat{\bm{\mathcal{P}}}_{\perp}$ is the arbitrary-gauge transverse polarization expressed in terms of the truncated dipole operator; note these functions can also be expressed in terms of the arbitrary-gauge vector potential through its longitudinal field component within the quantization scheme we employ here~\cite{Gustin2023Jan}. 

In this work we will focus on two well used gauges in quantum optics\footnote{Similar quantization schemes can be performed for superconducting QED systems, e.g., using the flux gauge and the charge gauge~\cite{PhysRevResearch.3.023079}.}: the dipole (d) gauge and Coulomb (C) gauge. Within the quantization function scheme of Ref.~\cite{Gustin2023Jan}, the transverse polarization corresponding to the Coulomb gauge is $\hat{\bm{\mathcal{P}}}_{\perp}^{\rm C}(\mathbf{r}) = 0$, and the dipole gauge $\hat{\bm{\mathcal{P}}}_{\perp}^{\rm d} = \hat{\mathbf{d}} \cdot {\bm \delta}^{\perp}(\mathbf{r})$ (that is, the multipolar gauge polarization within the dipole approximation). 
Considering these two gauges, we find immediately then that $\hat{\mathcal{U}}_{\rm d} = \hat{\mathcal{V}}_{\rm C} = 1$, and $\hat{\mathcal{U}}_{\rm C} = \hat{\mathcal{V}}_{\rm d}^{\dagger}$. Thus, we can define $\hat{\mathcal{W}} \equiv \hat{\mathcal{U}}_{\rm C}$, and write
\begin{equation}\label{eq:H_coul}
    \hat{\mathcal{H}}_{\rm C} = \hat{H}_{\rm em} + \hat{\mathcal{W}}\hat{\mathcal{H}}_0\hat{\mathcal{W}}^{\dagger}+ \hat{\mathcal{H}}_{\parallel}
\end{equation}
for the Coulomb gauge Hamiltonian, and 
\begin{equation}\label{eq:H_dip}
    \hat{\mathcal{H}}_{\rm d} = \hat{\mathcal{W}}^{\dagger}\hat{H}_{\rm em}\hat{\mathcal{W}} + \hat{\mathcal{H}}_0 + \hat{\mathcal{H}}_{\parallel}
\end{equation}
for the dipole gauge. Explicitly, we have
\begin{equation}
\hat{\mathcal{W}} = \text{exp}\left[\frac{i}{\hbar} \mathbf{d} \cdot \hat{\mathbf{A}}_{\perp}(\mathbf{0}) \hat{\sigma}_x\right],
\end{equation}
and we see explicitly the advantage of having defined our quantized QNMs in terms of the vector potential; the operator used to transform between gauges $\hat{\mathcal{W}}$ (which preserves the correct gauge-invariant results) is naturally expressed in terms of purely quantized QNM operators. We also show a definition in terms of the transverse electric field operators in Appendix~\ref{app:E_field_QNM}, which is what has been used in previous work, but is much less convenient when transforming between gauges, as one must take extra care in calculating the transformation $\hat{\mathcal{W}}$, which becomes expressed in this case in terms of both quantized QNM operators and residual bath operators. In particular, one must be careful to include the non-local commutators of the reservoir operators to obtain the correct transformation.

\subsection{System-reservoir Hamiltonian}\label{subsec:sys-res}

We now discuss how to split up the entire Coulomb and dipole gauge Hamiltonians into system (QNM), reservoir, and system-reservoir parts, in preparation for the derivation of a master equation.

By calculating the unitary transformation in Eq.~\eqref{eq:H_coul}, we find the general form of the full Coulomb gauge Hamiltonian,
\begin{equation}
    \hat{\mathcal{H}}_{\rm C} = \hat{H}_{\rm em} + \frac{\hbar \omega_0}{2}\left[\cos{(\hat{\Phi})}\hat{\sigma}_z + \sin{(\hat{\Phi})}\hat{\sigma}_y\right] + \hat{\mathcal{H}}_{\parallel},
\end{equation}
where
\begin{align}\hat{\Phi} &= \frac{2}{\hbar} \mathbf{d} \cdot \hat{\mathbf{A}}^{\perp}_{\rm QNM}(\mathbf{0}) \nonumber \\ 
& = 2\sum_{\mu} \left[\eta_{\mu}\hat{a}_{\mu} + \eta_{\mu}^* \hat{a}^{\dagger}_{\mu}\right],
\end{align}
with the complex coupling constants 
\begin{equation}\label{eq:eta_def}
\eta_{\mu} = \frac{\mathbf{d} \cdot \ \tilde{\mathbf{f}}^{{\rm 
 s}}_\mu(\mathbf{0})}{\sqrt{2\epsilon_0 \hbar\chi_{\mu \mu}}}.
 \end{equation}
For the system component, we have, straightforwardly, 
\begin{equation}\label{eq:HC_sys}
    \hat{\mathcal{H}}_{\rm C}^{\rm S} = \hat{H}^{\rm em}_{\rm QNM} + \frac{\hbar \omega_0}{2}\Big[ \cos{(\hat{\Phi}_{\rm })}\hat{\sigma}_z+\sin{(\hat{\Phi}_{\rm })}\hat{\sigma}_y\Big] + \hat{\mathcal{H}}_{\parallel}^{\rm S},
\end{equation}
 and the longitudinal term is
\begin{equation} \hat{\mathcal{H}}_{\parallel}^{\rm S} = -i\sum_{\mu} \hbar\chi_{\mu\mu}\left(\eta_{\mu}^\parallel \hat{a}_{\mu} - \eta_{\mu}^{\parallel *}\hat{a}_{\mu}^{\dagger}\right)\hat{\sigma}_x,
\end{equation}
where $\eta^{\parallel}_{\mu} = \mathbf{d} \cdot \ \tilde{\mathbf{f}}^{{\parallel}}_\mu(\mathbf{0})/\sqrt{2\epsilon_0 \hbar\chi_{\mu \mu}}$.

The system-reservoir coupling Hamiltonian is 
\begin{equation}\label{eq:sr_coul}
    \hat{\mathcal{H}}_{\rm C}^{\rm SR} = \hat{H}^{\rm em}_{\rm QNM-R} + \hat{B}_{\parallel}\hat{\sigma}_x,
\end{equation}
where $\hat{B}_{\parallel} = -\mathbf{d} \cdot \hat{\mathbf{E}}_{\rm F, B}^{\parallel}(\mathbf{0})$.
The total Coulomb gauge Hamiltonian is then $\hat{\mathcal{H}}_{\rm C} = \hat{\mathcal{H}}_{\rm C}^{\rm S} + \hat{\mathcal{H}}_{\rm C}^{\rm SR} + \hat{H}^{\rm em}_{\rm R}$.

In the dipole gauge, we have, performing a similar separation of $\hat{\mathcal{H}}_{\rm d} = \hat{\mathcal{H}}_{\rm d}^{\rm S} + \hat{\mathcal{H}}_{\rm d}^{\rm SR} + \hat{H}^{\rm em}_{\rm R}$,
\begin{subequations}\label{eq:H_cavdip}
\begin{equation}
    \hat{\mathcal{H}}_{\rm d}^{\rm S} = \hat{\mathcal{H}}'^{\rm em}_{\rm QNM} + \frac{\hbar\omega_0}{2}\hat{\sigma}_z+\hat{\mathcal{H}}_{\parallel}^{\rm S},
\end{equation}
\begin{equation}
    \hat{\mathcal{H}}_{\rm d}^{\rm SR}= \hat{\mathcal{H}}'^{\rm em}_{\rm QNM-R} +  \hat{B}_{\parallel}\hat{\sigma}_x,
\end{equation}
\end{subequations}
where we use primes to indicate that $\hat{\mathcal{H}}'^{\rm em}_{\rm QNM}$ and $\hat{\mathcal{H}}'^{\rm em}_{\rm QNM-R}$ are
expressed in terms of $\hat{a}_{\mu}'$ and $\hat{a}_{\mu}^{{\dagger}'}$, where $\hat{a}_{\mu} = \hat{W}_{\rm QNM}^{\dagger}\hat{a}_{\mu} \hat{W}_{\rm QNM} = \hat{a}_{\mu} + i\eta_{\mu}^* \hat{\sigma}_x$. 

Dropping a term proportional to the TLS subspace identity operator, one can put the dipole gauge system Hamiltonian in the form of the canonical multimode quantum Rabi model:
\begin{equation}\label{eq:dipolegauge}
    \hat{\mathcal{H}}_{\rm d}^{\rm S} = \hat{H}_{\rm QNM} + \frac{\hbar \omega_0}{2}\hat{\sigma}_z + \sum_{\mu}\hbar\left[g^{\rm d}_{\mu}\hat{a}_\mu + g^{{\rm d}*}_{\mu}\hat{a}_{\mu}^{\dagger}\right]\hat{\sigma}_x,
\end{equation}
where the dipole gauge QNM coupling constant is $g^{\rm d}_{\mu} = -i\sum_{\eta} \chi_{\eta \mu}\left(\eta_{\eta}+\eta^{\parallel}_{\eta}\right)$, which is generally complex.

We note that these Hamiltonians easily recover previously known results; in the single mode limit, where $\chi_{cc} \approx \omega_{\rm c}$, and in the non-USC regime where we can expand $\eta_\mu$ to leading order (and neglecting $\eta_{c}^{\parallel}$), both gauges have the same system-reservoir and reservoir Hamiltonians, and the system Hamiltonians are $\hat{\mathcal{H}}_{\rm C}^{\rm C/d} \approx \omega_c a_c^{\dagger} a_c + \frac{\hbar\omega_0}{2}\hat{\sigma}_z + \hat{\mathcal{H}}_{\rm int}^{\rm C/d}$, where $\hat{\mathcal{H}}_{\rm int}^{\rm C/d} = \hbar \left[g_{c}^{\rm C/d}\hat{a}_{c} + g_{c}^{{\rm C/d}*}\hat{a}_{c}^{\dagger}\right]\hat{\sigma}_x$, and $g_c^{\rm C} = \eta_c \omega_0$, such that the ratio of Coulomb gauge to dipole gauge cavity coupling strengths is $|g_c^{\rm C}/g_c^{\rm d}| = \omega_0/\omega_c$, recovering  well-known results
 (e.g., see Refs.~\onlinecite{DeBernardis2018Nov,DiStefano2019Aug}).

\section{Quasinormal mode master equation in the ultrastrong coupling regime}\label{sec:QNM_me}
In the previous section, we derived the light-matter interaction Hamiltonian valid for arbitrary coupling regimes in the Coulomb as well as dipole gauge. In the following, we will use this theory to setup a QNM master equation. We first derive these results in the spatially unspecified representation, and then show how to transform the resultant master equation to the spatially specified representation, as required to obtain the correct QNM phase-dependent decay rates of the lossy dipole-cavity system~\cite{Gustin2025}.

\subsection{Dressed-state basis}
To start, we first introduce the dressed-state basis $\{|j\rangle\}$ in which the TLS-QNM system Hamiltonian $\hat{\mathcal{H}}_{\rm C}^{\rm S}$ takes on the diagonal form,
\begin{equation}
    \hat{\mathcal{H}}_{\rm C}^{\rm S}=\hbar\sum_{j} \omega_{j} |j\rangle\langle j|,
\end{equation}
where $\hbar\omega_j$ are the associated eigenenergies of the dressed states, and we write the system-reservoir Hamiltonian as
\begin{equation}\label{eq:HSR_app1}
    \hat{\mathcal{H}}_{\rm C}^{\rm SR} = \hat{B}_{\parallel}\hat{S}_{\parallel}+ \left[\sum_{\mu}\hat{C}_{\mu}^{\dagger}\hat{a}_{\mu} + \text{H.c.}\right],
\end{equation}
where we have defined $\hat{S}_{\parallel} = \hat{\sigma}_x$, and $\hat{C}_\mu=\hbar\int_0^\infty{ d}\omega_{\rm m}\int{d}^3r\mathbf{g}_{\mu}(\mathbf{r},\omega_{\rm m})\cdot\mathbf{c}(\mathbf{r},\omega_{\rm m})$. We work in the Coulomb gauge for this derivation, but the final master equation we derive has a gauge-invariant form. The first term of Eq.~\eqref{eq:HSR_app1} can be associated with the TLS coupling to background longitudinal modes, respectively (and we reiterate we have neglected background transverse couplings), and the term in square brackets gives the coupling of the QNM with the reservoir. Going forward from here, we will neglect the longitudinal contribution for simplicity, assuming the transverse QNM response to dominate, which is appropriate for realistic separations of the dipole to the plasmonic resonator. However, this term could be retained if desired, and the potential effect on the decay rates through $\mathbf{G}^{\parallel}(\mathbf{r_0},\mathbf{r},\omega_{\rm m})$ can be calculated using the same method as those we derive below. 

Next, to facilitate a derivation in the USC regime, where light and matter degrees of freedom strongly hybridize, we move into the interaction picture. Denoting the interaction picture with a time argument, we find
\begin{equation}
    \hat{C}_\mu(t)=\hbar\int_0^\infty \! \! { d}\omega_{\rm m}\! \! \int \! \! {d}^3r\mathbf{g}_{\mu}(\mathbf{r},\omega_{\rm m})\cdot\mathbf{c}(\mathbf{r},\omega_{\rm m})e^{-i\omega_{\rm m}t},
    \end{equation}
where we have assumed local bosonic commutation relations for the reservoir operators $\hat{\mathbf{c}}(\mathbf{r},\omega)$ and $\hat{\mathbf{c}}^{\dagger}(\mathbf{r},\omega)$, as justified in Appendix~\ref{app:boson}.  Similarly, we can decompose the system operators by writing them in the dressed-state basis
\begin{equation}\label{eq:amu_intP}
    \hat{a}_{\mu}(t) = \sum_{\alpha}c_{\alpha}^{\mu}\hat{\sigma}_{\alpha}e^{-i\omega_{\alpha} t} + \text{H.c.}
\end{equation}
Here, we have introduced the abbreviated notation in which $\alpha$ is an index that refers to the pair of combined photon-matter eigenstate indices $(j,k)$, such that the energy of the state $k$ is greater than that of the state $j$, and $c_{\alpha}^{\perp} = \bra{j}\hat{S}^{\perp}\ket{k}$, $c_{\alpha}^{\mu} = \bra{j}\hat{a}_{\mu}\ket{k}$,  $\hat{\sigma}_{\alpha} = \ket{j}\bra{k}$, and $\omega_{\alpha} = \omega_k - \omega_j$. In short, $\alpha$ runs over the transitions of the dressed states. Note that we are able to neglect diagonal matrix elements in Eq.~\eqref{eq:amu_intP}, as $\hat{a}_{\mu}$ is odd under parity transformations, and parity is a symmetry of the system Hamiltonian $\hat{\mathcal{H}}_{\rm C}^{\rm S}$.

At this stage, we make a RWA 
for the bath coupling, by dropping all terms in $\hat{\mathcal{H}}_{\rm SR}(t)$ that oscillate at a sum of a bath frequency $\omega_{\rm m}$ and transition frequency $\omega_{\rm \alpha}$~\cite{Settineri2018Nov,Salmon2022Mar,Bamba2014Feb}. These terms are always very rapidly-oscillating, and can be expected to contribute contributions to the master equation only beyond the 2nd-order perturbative validity of the Born-Markov approximation~\cite{breuer2002theory} (and we have verified numerically they do not notably modify the spectra in our simulations). Making the bath-coupling RWA, we obtain 
\begin{equation}
\hat{\mathcal{H}}_{\rm C}^{\rm SR}(t) = \sum_{\alpha} \sum_{\mu} \hat{C}_{\mu}^{\dagger}(t)c_{\alpha}^{\mu} \hat{\sigma}_{\alpha}e^{-i\omega_{\alpha} t} + \text{H.c.}
\end{equation}
\subsection{Second-order Born-Markov master equation}\label{subsec:BM}
To derive the desired QNM master equation, we perform a standard second-order Born-Markov approximation~\cite{breuer2002theory}:
\begin{align}
     \frac{d}{dt}\rho_I(&t)= \nonumber \\
     &  -\frac{1}{\hbar^2}\int_0^\infty{\rm tr}_{\rm R}\left [\hat{\mathcal{H}}_{\rm C}^{\rm SR}(t),[\hat{\mathcal{H}}_{\rm C}^{\rm SR}(t-\tau),\rho_I(t)\rho_{\rm R}]\right]{\rm d}\tau,\label{eq: 2ndOrderBorn}
\end{align}
where we have traced over the reservoir degrees of freedom, and used $\rho_I$ to denote the reduced system interaction picture density matrix. We let $\rho_{\rm R}$ correspond to the reservoir density matrix, which we choose to be the vacuum state, although other choices are possible (e.g., thermal, coherent states for coherent driving~\cite{Salmon2022Mar}, or a squeezed vacuum~\cite{PhysRevX.6.031004,PhysRevLett.58.2539}).

Under the assumption of a vacuum state reservoir (zero temperature), and bosonic commutator relations, we have the result that ${\rm tr}_{\rm R}[\hat{\mathbf{c}}(\mathbf{r},\omega_{\rm m})\hat{\mathbf{c}}^\dagger(\mathbf{r}',\omega_{\rm m}')]=\mathbf{I}\delta(\mathbf{r}-\mathbf{r}')\delta(\omega_{\rm m}-\omega_{\rm m}')$, and all other combinations of two reservoir operators vanish.

Under these assumptions, we move back to the Schr{\"o}dinger picture and simplify Eq.~\eqref{eq: 2ndOrderBorn} to find
\begin{align}\label{eq:me1}
    \frac{d}{dt}\rho&= -\frac{i}{\hbar}[\hat{\mathcal{H}}_{\rm S},\rho] \nonumber \\ &+ \sum_{\alpha \alpha}F_{\alpha' \alpha} \left[\hat{\sigma}_{\alpha} \rho \hat{\sigma}_{\alpha'}^{\dagger} - \hat{\sigma}_{\alpha'}^{\dagger}\hat{\sigma}_{\alpha}\rho\right] + \text{H.c.},
\end{align}
where
\begin{equation}\label{eq:F_full}
    F_{\alpha'\alpha}^{\rm QNM}  =\frac{1}{\hbar^2} \sum_{\mu\eta}c_{\alpha'}^{\mu *}c_{\alpha}^{\eta} \int_0^{\infty} \! \! d\tau\langle \hat{C}_{\mu}\hat{C}^{\dagger}_{\eta}(-\tau)\rangle e^{i\omega_{\alpha} \tau}.
\end{equation}

At this stage, one could make a secular approximation by dropping all terms where $\alpha \neq \alpha'$. The justification for this is that, upon moving back to the interaction picture, the second line of Eq.~\eqref{eq:me1} would pick up an oscillatory factor of $e^{-i (\omega_{\alpha}-\omega_{\alpha'})t}$. If $\omega_{\alpha} - \omega_{\alpha'}$ were much greater than the greatest values of $F_{\alpha \alpha'}$, these terms would oscillate fast in the interaction picture and average out to provide no contribution to the system dynamics. In this case, the secular criterion is typically satisfied for strong coupling of the dipole-cavity system (coupling strength exceeding the decay rates), such that the anharmonic quantum Rabi model spectrum gives well-separated emission lines for all driven transitions~\cite{Mercurio2022Apr}.
Moving forward, we will assume strong coupling,
 and thus make the secular approximation for the sake of simplifying the master equation presentation. In our numerical results, we keep the full non-secular master equation for potentially more precise results.

By integrating over $\tau$ in Eq.~\eqref{eq:F_full}, we find $F_{\alpha\alpha} = \frac{1}{2}\Gamma_{\alpha}^{\rm R} + i\delta_{\alpha}^{\rm V}$, where
$\Gamma_{\alpha}^{\rm R}$ and $\delta_{\alpha}^{\rm V}$ are real parameters, and we can write the QNM master equation as 
\begin{equation}\label{eq:QNM_ME}
    \dot{\rho} = -\frac{i}{\hbar}[\hat{\mathcal{H}}_{\rm S} ,\rho] +\sum_{\alpha}\frac{\Gamma_{\alpha}}{2}\mathcal{L}[\hat{\sigma}_{\alpha}]\rho,
\end{equation}
where  $\Gamma_{\alpha}$ is the decay rate corresponding to loss of polariton excitations from the dipole-cavity system for the transition between eigenstates of the system Hamiltonian indexed by $\alpha$.
In general, there should also be a Hamiltonian term $\hat{\mathcal{H}}_{\rm V} = \sum_{\alpha} \hbar \delta_{ \alpha}^{\rm V}\hat{\sigma}_{\alpha}^{\dagger}\hat{\sigma}_{\alpha}$ present in Eq.~\eqref{eq:QNM_ME}. This is the (bath-enabled) Lamb shift Hamiltonian, which is responsible for small bath-induced shifts of the system energy levels due to virtual polariton-polariton exchange with the reservoir. We shall neglect these shifts here and focus solely on the dissipator terms, as is the common approach in cavity-QED.

The total decay rate in the spatially unspecified representation for transition indexed by $\alpha$ is given by $\Gamma_{\alpha}$, where
\begin{equation}\label{eq:g_a_qnm}
    \Gamma_{\alpha} = 2\pi\sum_{\mu\eta}c_{\alpha}^{\mu *}c_{\alpha}^{\eta}g_{\mu\eta}(\omega_{\alpha}),
\end{equation}
where
\begin{align}\label{eq:g_def}
    &g_{\mu\eta}(\omega_{\rm m}) = \int d^3r \mathbf{g}_{\mu}(\mathbf{r},\omega_{\rm m}) \cdot \mathbf{g}^*_{\eta}(\mathbf{r},\omega_{\rm m}) \nonumber \\ & = \frac{2}{\pi}\sum_{\nu\nu'}\left[S^{-\frac{1}{2}}\right]_{\mu \nu}\left[S^{-\frac{1}{2}}\right]_{\nu' \eta}\frac{\sqrt{\omega_{\nu}\omega_{\nu'}}}{\omega_{\rm m}^2}\Bigg[ \nonumber \\ & B_{\nu}(\omega_{\rm m})B_{\nu '}^{ *}(\omega_{\rm m}) S_{\nu \nu'}^{\rm nrad}(\omega_{\rm m}) 
   + B_{\nu}(\omega_{\rm m})B_{\nu '}^{ *}(\omega_{\rm m}) S_{\nu \nu'}^{\rm rad}(\omega_{\rm m}) \Bigg]. 
\end{align}
As we will discuss in the following subsection, the decay rates in Eq.~\eqref{eq:g_a_qnm} are in the spatially unspecified representation, and do not accurately reflect frequency-dependent corrections beyond phenomenological models in the single-mode approximation, due to their resulting independence of the QNM phase at the dipole location.

Note that although we have derived the QNM master equation [Eq.~\eqref{eq:QNM_ME}] in the Coulomb gauge, as it is expressed entirely in terms of operators of the dressed-state basis of the system, it is clearly gauge-invariant. To calculate the decay rates $\Gamma_{\alpha}$, the matrix elements $c^{\mu}_{\alpha} = \bra{j}\hat{a}_{\mu}\ket{k}$ were defined using their Coulomb gauge representations, but one could as easily calculate these using eigenstates calculated from the dipole gauge representation by instead considering matrix elements of the operators $\hat{W}^{\dagger}\hat{a}_{\mu}\hat{W} = \hat{a}_{\mu}' = \hat{a}_{\mu} + i \eta_{\mu}^*\hat{\sigma}_x$. This \emph{manifest} gauge invariance is a consequence of our formulation of the quantized QNMs in terms of the transverse vector potential (and neglect of background terms), which leads to an entirely system-level gauge transformation. Nonetheless, even with a gauge transformation which involves the reservoir, we have shown recently that one can expect gauge invariance of the Born-Markova master equation to be retained generally wherever the Markov approximation holds~\cite{gustin2025gaugeinvariancenaturallineshape}.
To be clear, the eigenstates, photon-matter entanglement and other state properties are gauge-dependent (especially in the USC regime), but eigenenergies and physical observables
obtained from either master equation are not.

\subsection{Single mode limit and spatially specified representation\label{subsec: NM_Limits}}
In this subsection, we consider the single-mode limit of the quantized QNM master equation derived in the previous subsection. To check the accuracy of our approach, which calculates frequency-dependent cavity decay rates (i.e., using an ab initio formulation of the spectral density), we compare with the result we recently derived for the spectral density of a lossy quantized cavity where a single dipole is weakly coupled to a single cavity mode with index $\mu = c$~\cite{Gustin2025} (within the secular approximation considered here):
\begin{equation}\label{eq:spectral_density}
\Lambda_c(\omega_{\alpha}) = \sqrt{\frac{\gamma_c}{\pi}\frac{\omega_c}{\omega_{\alpha}}\zeta_c(\phi_0,\omega_{\alpha})} \, ,
\end{equation}
where $2\gamma_c$ is the empty-cavity photon decay rate, 
and $\phi_0 = \text{arg}\{\mathbf{n}_{\mathbf{d}} \cdot \tilde{\mathbf{f}}_c(\mathbf{0})\}$ 
is the QNM phase at the dipole location projected onto the dipole moment unit vector $\mathbf{n}_{\mathbf{d}}$.  Crucially, frequency-dependent corrections to system decay rates beyond those of phenomenological (e.g., flat spectral density) models of lossy quantized cavities are highly dependent on the QNM phase at the dipole location for realistic cavities~\cite{Gustin2025}.

However, the decay rates calculated in Eq.~\eqref{eq:g_a_qnm} \emph{do not}, in the spatially unspecified form written, depend directly on the QNM phase at the dipole location. As such, when making a single (or few) mode approximation, they are not fully accurate in accounting for corrections beyond phenomenological lossy cavity models. Quasi-static plasmon-based approaches which lead similarly to phase-independent frequency-dependent decay rates~\cite{PhysRevB.103.045421,PhysRevB.105.245411} also fail to capture the accurate frequency dependence for the dominant transverse QNMs studied here.  We stress that the spatially unspecified representation has previously been used to successfully explain many important quantum optical effects beyond phenomenological models, such as large Fano-like resonances, and coupled gain-loss systems,  but in that case the dominant quantized QNM effects were captured from inter-QNM interference effects, beyond the single QNM  approximation~\cite{franke2020quantized,Ren2022,PhysRevA.105.023702}. By moving to the spatially specified representation in the single QNM approximation, we can correctly identify additional phase effects that were less important in previous works, where the dominant effects were captured by the quantization approach in the spatially unspecified representation, and USC, and more generally, the broadband dissipative regime, was not considered.

To fix this situation, we move to the spatially specified form discussed in Sec.~\ref{sec:spatial_spec} in the context of the quantization parameter $S_c$. By employing Eq.~\eqref{eq:gf_id}
 in the single-mode limit, letting $\mathbf{r}_1=\mathbf{r}_2=\mathbf{0}$, and taking the dot product on the right and left with $\mathbf{n_d}$, we obtain the result from Eq.~\eqref{eq:GF_id_1mode}, with $\phi_0$.

Next, we note that the QNM decay rate $\Gamma_{\alpha}$ in the single-mode approximation can be written as (with the implied frequency argument $\omega_{\alpha}$):
\begin{align}
    \Gamma_{\alpha} &= 4\omega_c|c_{\alpha}^{c}|^2\frac{(\omega_{\alpha} - \omega_c)^2 \! + \! \gamma_c^2}{S_c\omega^2_\alpha} \! \int \! d^3r \epsilon_I(\mathbf{r}) |A_c\tilde{\mathbf{F}}'_c(\mathbf{r})|^2 \nonumber \\
    & =4\omega_c|c_{\alpha}^{c}|^2\frac{(\omega_{\alpha} - \omega_c)^2 + \gamma_c^2}{S_c\omega^2_\alpha} \text{Im}\{A_c e^{i2\phi_0}\} \nonumber \\ 
    &= |c_{\alpha}^c|^2\frac{2\gamma_c\cos{(2\phi_0)}\omega_c}{S_c\omega_\alpha}\zeta_c(\phi_0,\omega_{\alpha}),
\end{align}
where in the second line we have used Eq.~\eqref{eq:GF_id_1mode}, and $\zeta_c(\phi_0,\omega_{\alpha})$ was defined in Eq.~\eqref{eq:zetadef}

Clearly, this form of the quantized cavity decay rate as a function of the hybridized TLS-cavity resonances $\omega_{\alpha}$, now depends explicitly on the QNM phase at the dipole location $\phi_0$, which is known to have a significant impact for realistic cavities. By identifying the empty-cavity decay rate as $\kappa_c = 2\gamma_c$ (consistent with previous work on QNMs~\cite{PhysRevLett.122.213901,franke2020quantized}), and noting that, within the dipole spatial location representation, $S_c \approx \cos{2\phi_0}$, we see that the empty-cavity result $\Gamma = \kappa_c$ is recovered. 
In summary, we have
\begin{equation}\label{eq:QNM_rate}
\Gamma_{\alpha} = \kappa_c|c_{\alpha}^c|^2\frac{\omega_c}{\omega_{\alpha}}\zeta_c(\phi_0,\omega_{\alpha}).
\end{equation}

Finally, identifying $\Gamma_{\alpha} = 2\pi |c_\alpha^c|^2\Lambda_c^2(\omega_{\alpha})$, we see that the \emph{ab initio} quantized QNM theory within the spatially specified representation fully recovers the correct form of the spectral density [Eq.~\eqref{eq:spectral_density}] needed to see frequency-dependent corrections beyond phenomenological models. We also see explicitly that for a model of cavity-reservoir coupling where the cavity operators $\hat{a}_c$, $\hat{a}^{\dagger}_c$ couple directly to the reservoir \emph{in the Coulomb gauge}, the correct spectral density for this coupling scales as $1/\omega_{\rm m}$, confirming our findings in Ref.~\cite{Gustin2025} from a fully \emph{ab initio} construction.

To the best of our knowledge, this is the \emph{first time} the correct form of the cavity system-reservoir coupling has been derived from an \emph{ab initio} perspective valid for general 3D resonators.

Finally, we note that for the single-mode approximation to give physical results, we require $\Gamma_{\alpha} >0$ for all active transitions, and thus $\zeta_c(\phi_0,\omega_{\alpha})>0$. Using a simple on-resonance model where $\omega_0 = \omega_c$, and assuming $\omega_{\alpha} \sim \omega_c(1+|\eta_c|)$ (i.e., to leading order in $|\eta_c|$), we find, for the single-mode approximation to remain accurate, as an \emph{upper bound}, $|\eta_c| < |\eta_c^{(1)}|$, where
\begin{equation}\label{eq:heuristic}
|\eta_c^{(1)}| = \frac{1}{| 2Q_c\tan{\left(2\phi_0\right)}|}.
\end{equation}
Of course this is only a heuristic expression, and becomes less quantitatively accurate when higher-order (in $\eta_c$) contributions to the transition $\omega_{\alpha}$ are significant, as well as when the cavity and TLS are detuned, but nonetheless provides a useful rough analytic heuristic to determine approximate rough bounds of single-modedness. It also highlights an intrinsic function of the QNM phase as a limiting factor in the approximation of single-mode models under broadband coupling regimes. Indeed, in Sec.~\ref{sec:classical_results}, we find that in practice the single-mode QNM theory breaks down (in terms of providing accurate corrections to phenomenological models) well below this threshold.

\subsection{Definition of dissipative broadband regime of cavity-QED}\label{sec:broadband_regime}
We are now in a position to posit a definition of the broadband regime of dissipative cavity-QED.
We have stressed that even \emph{outside of the USC regime}, corrections to phenomenological lossy cavity master equation models based on our \emph{ab initio} approach would be highly perceptible for cavities operating in a regime with sufficiently large values of $2Q_c\tan{2\phi_0}$. For a system with a characteristic dynamical rate $\Omega$ which satisfies $\Omega/\omega_c \ll 1$, this ``broadband'' regime emerges when $\Omega/\omega_c$ becomes substantial relative to $[2Q_c\tan{2\phi_0}]^{-1}$. To put this more precisely, we can use the analytical solution for the spectral linewidths of the dominant two peaks of a resonantly-coupled ($\omega_0=\omega_c$) TLS-cavity system for weak $|\eta_c|$ (derived in Appendix~\ref{app:BS}): 
\begin{equation}\label{eq:lws}
    \Gamma_{\pm} = \frac{\kappa_c}{2}\left[1 \pm \frac{|\eta_c|}{2}\left(1 - 4Q_c \tan{(2\phi_0)}\right)\right],
    \end{equation}
    where `$+$' denotes the higher-energy (blue) peak and `$-$' denotes the lower-energy (red) peak. 
In analogy with the typical definition of the USC regime as $|\eta_c| >0.1$, we thus define the ``broadband'' regime of (single-mode) lossy cavity-QED as $|\eta_c| \geq \tilde{\Omega}_{\rm BB}$, where $\tilde{\Omega}_{\rm BB}$ is defined as
\begin{equation}\label{eq:BB}
\tilde{\Omega}_{\rm BB} = 0.1 \times \text{min}\{1, |1 - 4Q_c \tan{(2\phi_0)}|^{-1}\}.
\end{equation}
In more general terms, this criterion should also hold whenever any characteristic dynamical rate of the system divided by $\omega_c$ approaches  $\tilde{\Omega}_{\rm BB}$.

It should be noted that this definition is only a rough estimate for when broadband dissipative effects become substantial, and perceptible modifications to (e.g.,) cavity spectra can appear below this threshold as well, which we indeed show in Sec.~\ref{sec:quantum_sim}.
\subsection{Output observables and near-field detection}\label{sec:photodet}
To relate the dynamics of the ultrastrongly-coupled cavity-QED system to observables which can be probed in photodetection experiments (that is, normally-ordered correlation functions of the electromagnetic fields), we will assume near-field detection, where the field can be described by the QNM expansion of the field. The more common scenario of far-field detection requires an input-output model, of which formulating a self-consistent theory in the USC regime using QNMs is a subtle and more involved process than the usual formulation (and can involve further gauge considerations~\cite{Milonni1989Oct,gustin2025gaugeinvariancenaturallineshape}), and so this problem 
will be 
addressed in future work.

In Ref.~\cite{Gustin2023Jan}, we showed that the correct operator to use to model photodetection near the resonator (where the field can be well represented as a sum of QNM functions) is $ - \mathbf{d}_{\rm det} \cdot \hat{\bm{\mathcal{E}}}_{\perp}(\mathbf{r}_{\rm det})$, where $\mathbf{d}_{\rm det}$ is an effective dipole moment of the detector system located at $\mathbf{r}_{\rm det}$, and, in the Coulomb gauge,
\begin{equation}\label{eq:e_trunc}
\hat{\bm{\mathcal{E}}}_{\perp}(\mathbf{r}_{\rm det}) = i\sum_{\mu} \sqrt{\frac{\hbar\omega_{\mu}}{2\epsilon_0}} \mathbf{\tilde{f}}_{\mu}^{\text{s}(\mathbf{E})}(\mathbf{r}_{\rm det}) \hat{a}_{\mu} + \text{H.a.},
\end{equation}
where
\begin{equation}  \mathbf{\tilde{f}}_{\mu}^{\text{s}(\mathbf{E})}(\mathbf{r}_{\rm det}) = \sum_\nu\frac{\chi_{\nu \mu}}{\sqrt{\omega_{\mu}\omega_{\nu}}}\mathbf{\tilde{f}}_{\nu}^{\rm s}(\mathbf{r}_{\rm det}),
\end{equation}
and $\hat{\bm{\mathcal{E}}}_{\perp}$ is the ``correct'' (in terms of photodetection) form of the part of the transverse electric field that can be expressed in terms of QNM operators, when the QNM operators are defined with respect to a modal expansion in terms of the transverse vector potential, as we show in Ref.~\cite{Gustin2023Jan}. Note in Ref.~\cite{Gustin2023Jan}, Eq.~\eqref{eq:e_trunc} is referred to as the ``correctly-truncated form of the electric field'' in the sense of a mode truncation of the field expansion. In this case, it is more correct to say this is the correct form of the electric field operator when decomposing the total field into QNM and reservoir subspaces, but the procedure and result for the QNM (system) part is the same. As an aside, and in contrast, it can be shown that directly substituting the projection functions $\mathbf{L}_{\mu}(\mathbf{r},\omega_{\rm m})$ into the definition of $\hat{\mathbf{E}}_{\rm F}^{\perp}$ gives $
\hat{\mathbf{E}}^{\perp}_{\rm F, QNM}(\mathbf{0}) = -i\sum_{\mu} \hbar g^{\rm d}_{\mu}\hat{a}_{\mu} + \text{H.c.},$
where $g^{\rm d}_{\mu}$ is defined in Sec.~\ref{sec:QNM_me}. Under the approximation $\chi_{\mu \mu} \approx \omega_{\mu}$, we recover $\hat{\mathbf{E}}_{\rm F, QNM}^{\perp}(\mathbf{0}) = \hat{\bm{\mathcal{E}}}_{\perp}(\mathbf{0})$. 

In the dipole gauge, the only difference is that $\hat{\bm{\mathcal{E}}}_{\perp}$ should be expanded in terms of the $\hat{a}_{\mu}'$ and $\hat{a}_{\mu}^{'{\dagger}}$ operators. Note that this procedure is consistent with previous calculations using phenomenological loss models~\cite{Salmon2022Mar,Mercurio2022Apr}.

In the single-mode approximation, we have
\begin{equation}  \mathbf{\tilde{f}}_{c}^{\text{s}(\mathbf{E})}(\mathbf{r}_{\rm det}) = \frac{\chi_c S_c}{\omega_c}\mathbf{\tilde{f}}_{c}(\mathbf{r}_{\rm det}),
\end{equation}
or, using the spatially specified representation and approximating $\chi_c \approx \omega_c$, 
\begin{equation}  \mathbf{\tilde{f}}_{c}^{\text{s}(\mathbf{E})}(\mathbf{r}_{\rm det}) \approx \sqrt{\cos{(2\phi_0)}}\mathbf{\tilde{f}}_{c}(\mathbf{r}_{\rm det}).
\end{equation}
In either case, the detected observables can be calculated by using the $\hat{a}_c$ and $\hat{a}_c^{\dagger}$ operators (or with primes if using the dipole gauge), up to a proportionality constant. In the USC regime, we further must move to the dressed state basis to ensure observables correspond to real (and not virtual) excitations~\cite{Settineri2018Nov}, which is done in Sec.~\ref{sec:quantum_sim}.

\section{Results and Simulations for Various Resonator Designs Under Weak-Coupling}\label{sec:classical_results}
In this section, we show specific examples using both plasmonic and dielectric cavity designs. In addition to performing modal simulations to identify dominant (and higher-order) QNM contributions to the transverse coupling of the dipole, we also compare with full non-modal Maxwell simulations to verify the accuracy of our theory.

\begin{figure*}[htb]
    \centering
    \includegraphics[width = 1.69\columnwidth]{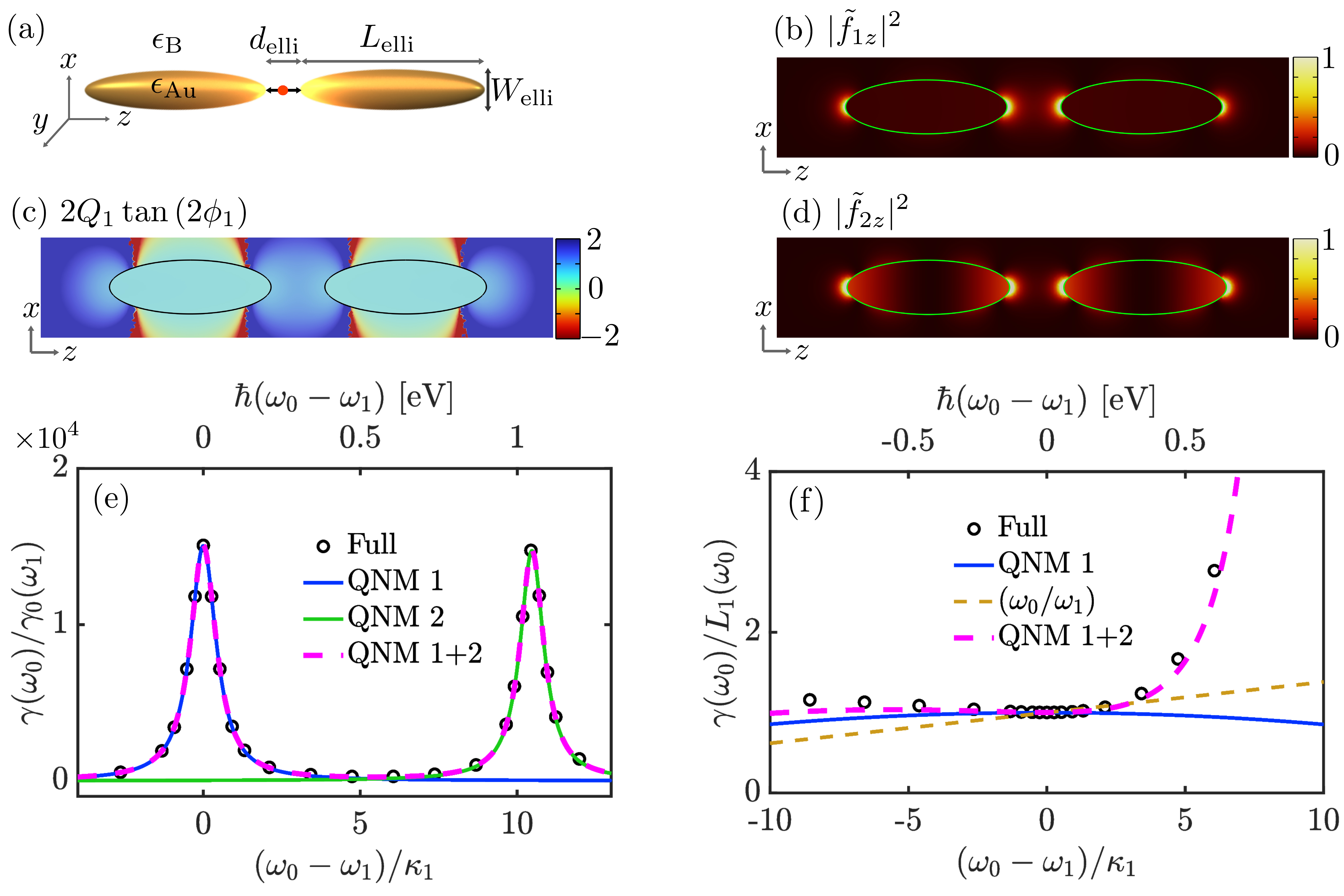}   
    \caption{(a) Schematic of gold ellipsoid dimer in free space ($\epsilon_{\rm B}=1.0$) with dimensions $W_{\rm elli}=10~$nm, $L_{\rm elli}=30~$nm, $d_{\rm elli}=10~$nm. The dielectric function of the gold nanorods is governed by the Drude model Eq.~\eqref{eq: Drude}. QNM profile for (b) dominant QNM (1) and (d) second QNM (2) (arb. units). (c) Phase distribution of $2Q_{\rm 1}\tan{(2\phi_1)}$ for dominant QNM, where the phase is defined as $\tilde{f}_{1z}(\mathbf{r})=|\tilde{f}_{1z}(\mathbf{r})|e^{i\phi_{1}(\mathbf{r})}$. (e) Purcell factors for a $z-$polarized emitter at the gap center, normalized by the free-space rate $\gamma_0(\omega)  =|\mathbf{d}|^2\omega^3/(3\pi \epsilon_0 \hbar c^3)$. (f) The decay rate normalized by  Lorentzian function $L_{1} = L_c$. Here, we plot the full dipole decay rate with the background $\gamma_0(\omega_0)$ subtracted out. The dominant QNM (1) parameters are found in Table~\ref{tab:tab}, and QNM 2 has (complex) eigenfrequency $\hbar\tilde{\omega}_2 = (3.667 - 0.04602i)$ eV and projected phase at the dipole location $\phi_2(\mathbf{r}_0) = 0.00616$.
    }
    \label{fig: metal_dimer_elli_W10L30d10}
\end{figure*}

As a primary observable to assess the validity of our approach, here we use the Purcell-enhanced decay rate of a \emph{weakly-coupled} single dipole, similar to the approach of Ref.~\cite{Gustin2025}. The advantage of this approach is that we can employ the well-known perturbative result (which can be obtained using macroscopic QED) for the dipole decay rate 
\begin{equation}\label{eq:gamma_GF}
\gamma(\omega_0) = \frac{2}{\epsilon_0}\mathbf{d} \cdot \text{Im}\{\mathbf{G}_{\perp}(\mathbf{0},\mathbf{0},\omega_0)\} \cdot \mathbf{d},
\end{equation}
expressed as a function of the TLS frequency. 
Under a single-QNM expansion for the transverse Green's function, this rate becomes
\begin{equation}\label{eq:gamma_qnm}
    \gamma_{\rm QNM}(\omega_0) = \frac{4|\tilde{g}_{\rm c}^{\rm d}|^{2}}{\kappa_c} \frac{\omega_0}{\omega_c}\frac{\kappa_c^2/4}{\kappa_c^2/4  + (\omega_0-\omega_c)^2} \zeta_c(\phi_0,\omega_0),
\end{equation}
expressed in terms of the dipole gauge TLS-cavity coupling $\tilde{g}_{c}^{\rm d}$. Equation~\eqref{eq:gamma_qnm} can also be derived from the quantized QNM theory we present in this work, which gives the spectral density of Eq.~\eqref{eq:spectral_density}, which was shown in Ref.~\cite{Gustin2025} to yield the result of Eq.~\eqref{eq:gamma_qnm} when using the Coulomb gauge. After confirming the validity of our quantized QNM theory in the weak-coupling regime, we are free to use it to perform quantum simulations in the USC regime, as we do in Sec.~\ref{sec:quantum_sim}.

\begin{table*}
    \centering
    \begin{tabular}{|l||c|c|c|c|} \hline 
         & Ellipsoid Dimer & Bowtie & 2D PC & WGM \\ \hline 
          $\hbar\tilde{\omega}_c$ (eV) & $ 2.620 - 0.04996i$ & $0.7987 - 0.002507i$ & $1.957 - 6.319 \times 10^{-4}i$ & $0.8337 - 4.120 \times 10^{-6} i$  \\ \hline 
          $Q_c$ & 26.22 & 159.3 & 1549 & $1.012 \times 10^5$  \\ \hline 
          \multirow{2}{*}{$\tilde{f}_z$ } & $2.670\times 10^{11}$ & $2.235 \times 10^9$ & $3.302 \times 10^6$  & 86750 \\
           & $+ 2.553\times 10^{9}i$ m$^{-\frac{3}{2}}$& $+7.915 \times 10^6 i$ m$^{-\frac{3}{2}}$ & $-7.178 \times 10^3i$ m$^{-1}$ & $-2.440i$ m$^{-1}$  \\ \hline 
          $\tan{(2\phi_0)} \approx  2\phi_0$ & 0.0191 & 0.00708 & $-0.00435$ & $-5.63 \times 10^{-5}$  \\ \hline
      $|\eta_c^{(1)}|$ & 1.0 & 0.44 & 0.074 & 0.088  \\ \hline $\tilde{\Omega}_{\rm BB}$ & 0.1 & 0.028 & 0.0036 & 0.0042 \\ \hline
    \end{tabular}
\caption{Parameters of dominant QNM for multiple cavity structures. $|\eta_c^{(1)}|$ [see Eq.~\eqref{eq:heuristic}] gives an upper bound and first order approximation of the maximum normalized $|\eta_c|$ attainable before the single-mode model breaks down, and $\tilde{\Omega}_{\rm BB}$ [see Eq.~\eqref{eq:BB}] gives the value of a system characteristic dynamical coupling rate (e.g., $|\eta_c|$) divided by the cavity frequency, above which broadband corrections as derived by our \emph{ab initio} quantized QNM theory in the spatially specified representation become significant. The QNM profile $\tilde{f}_z$ values are the dominant components of the vector $\mathbf{\tilde{f}}_c(\mathbf{0})$.
}
\label{tab:tab}
\end{table*}

To compare the result of Eq.~\eqref{eq:gamma_qnm} for a variety of cavity designs with the general expression in Eq.~\eqref{eq:gamma_GF} with the full non-modal Green's function found from solving Maxwell's equations, it is useful to normalize our results to the Lorentzian function
\begin{equation}\label{eq:lorentzian}
L_c(\omega) = \gamma_{\rm QNM}(\omega_c)\frac{\kappa_c^2/4}{\kappa_c^2/4 + (\omega-\omega_c)^2}.
\end{equation}
The frequency dependence of $L_c(\omega_0)$ is precisely what is obtained by phenomenological models of loss in quantized cavity systems (i.e., with a simple Lindblad term with rate $\kappa_c$ and collapse operator $\hat{a}_c$), and as such, substantial deviations in $\gamma(\omega_0)/L_c(\omega_0)$ from unity are indicative of the broadband dissipative regime of cavity-QED. For full dipole simulation results, we subtract off a slowly-varying background component, proportional to the free-space decay rate scaled by a fitting factor constant, to isolate the QNM contribution from the decay from the TLS directly into non-resonant modes, which has been neglected in this work for simplicity of presentation. Note that for some resonators (particularly those supporting sufficiently large Purcell factors) this subtraction is not needed to see good agreement due to the domination of the QNM decay over the residual background coupling (e.g., those in Ref.~\cite{Gustin2025}, where no subtraction was implemented).

We summarize the relevant QNM parameters for each of the dominant modes of the cavity designs we study in Table~\ref{tab:tab}, where we also calculate the approximate criterion from Eq.~\eqref{eq:heuristic} for limits of the single-mode model, by means of the parameter $|\eta_c^{\rm (1)}| = [2Q_c \tan{2\phi_0}|^{-1}$, which gives the first-order estimate of the maximum $|\eta_c|$ before the single-mode approximation leads to unphysical predictions, and is an upper-bound on the validity of such an approximation. We also show the parameter $\tilde{\Omega}_{\rm BB}$ in Table~\ref{tab:tab} as an estimate of the value of $|\eta_c|$ (or more broadly any dynamical rate scaled by the QNM frequency) above which broadband dissipative effects giving corrections beyond the usual phenomenological models become substantial, as discussed in Sec.~\ref{sec:broadband_regime}. 

In all of the examples, we consider the resonator structure to be placed in free space ($n_{\rm B} = 1)$. We also in some instances investigate the contribution of higher modes by means of a multimode generalization of Eq.~\eqref{eq:gamma_qnm}, where a sum runs over the QNM mode indices~\cite{Gustin2025}. In this case, $\mu = c = 1$ corresponds to the dominant QNM, with $\mu > 1$ corresponding to subdominant modes. 

The results of the following subsections suggest five important findings which we posit as \emph{potential} general rules-of-thumb for cavity structures which support a dominant mode in a relevant optical frequency window of interest: (i) for plasmonic resonators, the USC and broadband dissipative regimes typically coincide, and our \emph{ab initio} dissipative single-mode QNM theory can accurately be used to model dissipative dynamics beyond the predictions of phenomenological models. (ii) For dielectric resonator designs (excluding 1D or quasi-1D cavities), the broadband dissipative regime is reached with dynamical rates orders of magnitude below the USC threshold, and our approach can sometimes be used to identify single-mode corrections beyond phenomenological models, depending on the cavity design in question. (iii) For dielectric cavities, USC dynamics span a bandwidth far beyond the regime where the single-QNM calculation agrees with full Maxwell simulations, and thus we do not expect single-mode models to be accurate in this regime. (iv) In 1D or quasi-1D cavity designs, the quasi-harmonic distribution of longitudinal modes likely precludes the possibility of accurate single-mode calculations in the dissipative broadband regime. (v) In all cases, full Maxwell simulations begin to deviate with the QNM theory predictions at a bandwidth smaller than that suggested by $|\eta_{c}^{(1)}|$, highlighting its role as an \emph{upper bound} of validity of single-mode models.

\subsection{Plasmonic dimer resonator}

\begin{figure*}[htb]
    \centering
    \includegraphics[width = 1.46\columnwidth]{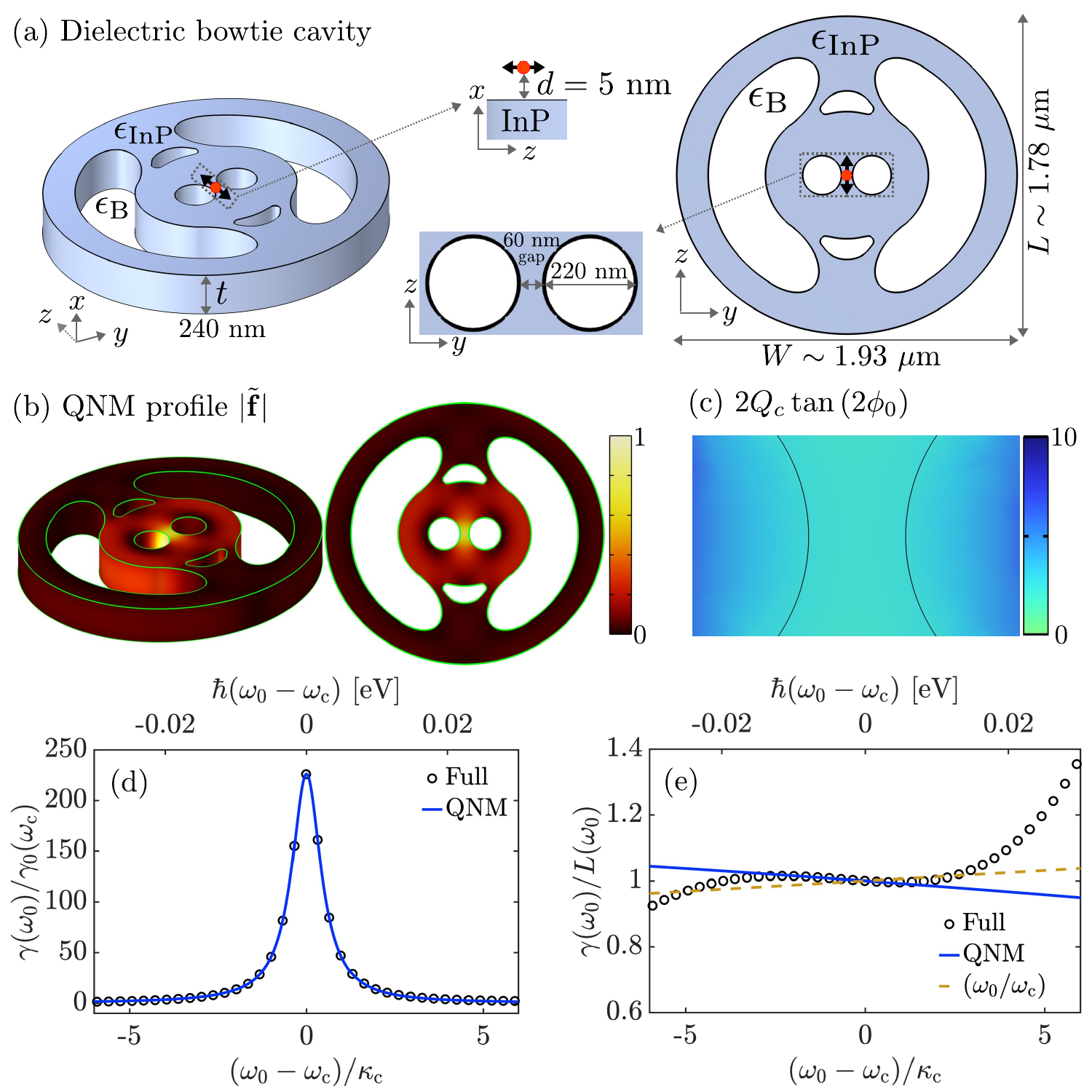}   
    \caption{(a) Schematic of 3D dielectric bowtie resonator in free space ($\epsilon_{\rm B}=1.0$). 
    (b) Mode profile of QNM $|\tilde{\mathbf{f}}|$ (including all components) (arb. units). (c) Phase distribution of $2Q_{\rm c}\tan{(2\phi_0)}$ for $z-$component $\tilde{f}_z$, where the phase is defined as $\tilde{f}_{z}(\mathbf{r})=|\tilde{f}_{z}(\mathbf{r})|e^{i\phi_{0}(\mathbf{r})}$. (d) Purcell factors for a $z-$polarized emitter $5$ nm above the bridge gap center, normalized by the free-space rate $\gamma_0(\omega)  =|\mathbf{d}|^2\omega^3/(3\pi \epsilon_0 \hbar c^3)$. (e) Decay rate normalized by Lorentzian function $L_{c}(\omega_0)$. Here, we plot the full dipole decay rate with an approximate background $1.77\gamma_0(\omega_0)$ subtracted out. The dominant QNM parameters are found in Table~\ref{tab:tab}. 
    }
    \label{fig: bowtie_20250507}
\end{figure*}

We first consider a gold dimer [Fig.~\ref{fig: metal_dimer_elli_W10L30d10} (a)] consisting of two identical ellipsoidal nanorods~\cite{ren_near-field_2020,PhysRevLett.122.213901,KamandarDezfouli2017}. A Drude model is used to describe the dielectric function of the gold nanorods,
\begin{equation}\label{eq: Drude}
    \epsilon_{\rm Au}(\omega)=1-\frac{\omega_{\rm p}^{2}}{\omega^{2}+i\omega\gamma_{\rm p}},
\end{equation}
with $\hbar\omega_{\rm p}=8.2934$ eV 
and $\hbar\gamma_{\rm p}=0.0928$ eV. 

A dominant single QNM for this system is found when a potential $z-$polarized dipole is placed at the gap center, and we report its complex eigenfrequency and quality factor $Q_c = \omega_c/(2\gamma_c)$ in Table~\ref{tab:tab} (as well as these parameters for all subsequent examples in this section). The mode distribution ($|\tilde{f}_{z}|^2$) is shown in Fig.~\ref{fig: metal_dimer_elli_W10L30d10} (b). We assume a dipole to be placed at the center of the gap, which we set to be the origin $\mathbf{r}=\mathbf{0}$ (and in all subsequent examples we again assume the dipole location is set to be the origin). The QNM phase there is defined here by $\tilde{f}_{z}(\mathbf{r})=|\tilde{f}_{z}(\mathbf{r})|e^{i\phi_0(\mathbf{r})}$ (dominant $z$-component), and we assume a dipole polarized in the $z$-direction such that the projected phase in Eq.~\eqref{eq:zetadef} is equal to $\phi_0$. The distribution of  $2Q_{\rm c}\tan{(2\phi_0)}$ [which quantifies the deviation from the zero-phase spectral density as given by $\zeta_c(\phi_0,\omega)$] is shown Fig.~\ref{fig: metal_dimer_elli_W10L30d10} (c). The value of the QNM at the dipole location as well as its phase is shown in Table~\ref{tab:tab}.

Figure~\ref{fig: metal_dimer_elli_W10L30d10}(e,f) show the dipole decay rate under weak coupling with the reservoir (Eq.~\eqref{eq:gamma_GF}) calculated with full Maxwell simulations (numerical Green's function), as well as using the QNM expansion of Eq.~\eqref{eq:gamma_qnm}. The agreement is quite good for a single mode and shows small improvements when a second mode with higher resonant frequency is included.
We also plot in (f) and in all other similar figures in this section the line $\omega_0/\omega_c$, which is the QNM prediction with zero phase $\phi_0=0$, and in all cases fails to capture the correct trends~\cite{Gustin2025}. 
The value $\tilde{\Omega}_{\rm BB} = 0.1$ indicates that the broadband dissipative regime coincides with USC. The bandwidth of the system dynamics spans a number of bare cavity linewidths $\kappa_c$ equal to $\sim \tilde{\Omega}_{\rm BB}Q_c \approx 2.6$, and comparing with Fig.~\ref{fig: metal_dimer_elli_W10L30d10} (f), the agreement with full Maxwell calculations remains accurate across this bandwidth, suggesting our \emph{ab initio} approach can accurately be used in the single-mode dissipative USC regime to identify corrections beyond phenomenological models. Finally, we can use the spatially unspecified representation of our QNM quantization scheme to estimate the fraction of radiative vs. nonradiative dissipation for the bare (uncoupled) cavity mode. Using Eq.'s~\eqref{eq:s_nrad} and~\eqref{eq:s_rad} (see also Ref.~\cite{ren_near-field_2020}), we find $S_c^{\rm rad}/S_c \approx 9.9$\%, with $S_c = S_c^{\rm rad} + S_c^{\rm nrad} \approx 1$.  

In Appendix~\ref{app:additional}, we also consider a cylindrical gold dimer resonator, similar to the one studied in Ref.~\cite{Gustin2025}, which gives similar qualitative results, though with a much higher radiative photon loss fraction. Additionally, we discuss briefly some gold dimer examples with smaller gap sizes (which can enable USC more easily through stronger field confinement) in Sec.~\ref{sec:exp}.

\begin{figure*}[htb]
    \centering
    \includegraphics[width = 1.68\columnwidth]{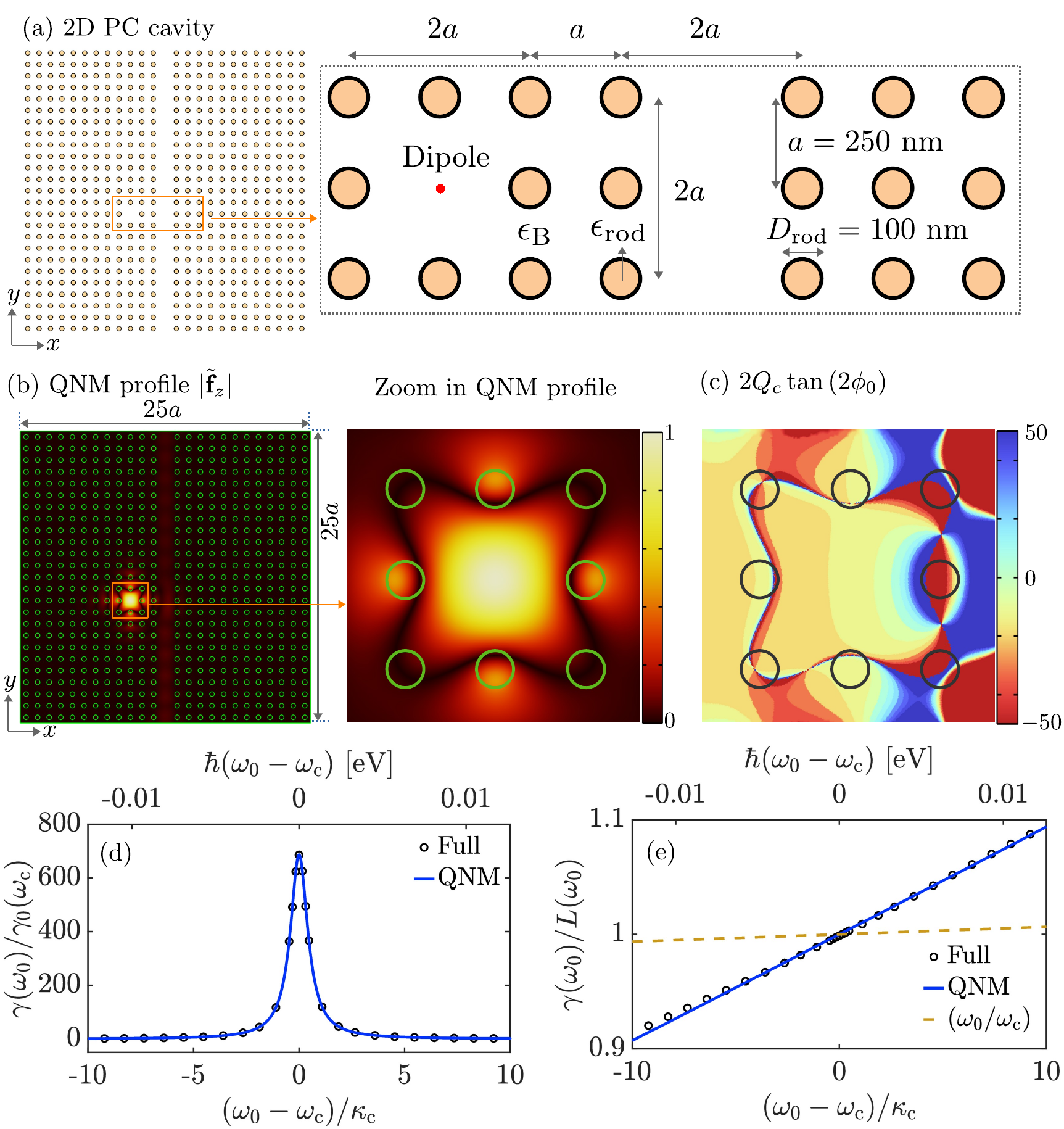}   
    \caption{(a) Schematic of 2D PC cavity structure. $\epsilon_{\rm B}=1.0$. The dielectric function of the nanorods 
    is governed by a Lorentz model; see Eq.~\eqref{eq: Lorentz}. (b) Mode profile for QNM $|\tilde{\mathbf{f}}_z|$ (arb. units; TM mode, the only component). (c) Phase distribution of $2Q_{\rm c}\tan{(2\phi_0)}$ for dominant QNM, where the phase is defined as $\tilde{f}_{z}(\mathbf{r})=|\tilde{f}_{z}(\mathbf{r})|e^{i\phi_{0}(\mathbf{r})}$. (d) Purcell factors for an emitter (line current) at the cavity center, normalized by the 2D free-space rate $\gamma_0^{\rm 2D}(\omega)  =|\mathbf{d}^{\rm 2D}|^2\omega^2/(2 \epsilon_0 \hbar c^2)$.
    (e) Decay rate normalized by Lorentzian function $L(\omega_0)=\gamma_{c}^{\rm QNM}(\omega_c)\times\frac{\kappa_c^2/4}{\kappa_c^2/4+(\omega_0-\omega_c)^2}$. Here, we plot the full dipole decay rate with an approximate background $0.05\gamma_0^{\rm 2D}(\omega_0)$ subtracted out. The dominant QNM parameters are found in Table~\ref{tab:tab}.
    }
    \label{fig: 2DPC_20250508}
\end{figure*}

\subsection{Dielectric resonators}

We now consider dielectric resonators. We first consider a 3D dielectric bowtie cavity design, as shown in Fig.~\ref{fig: bowtie_20250507}, which references the design shown in Refs.~\cite{Kountouris:22} and \cite{Kountouris_thesis}. The thickness is $t=240~$nm, and the width/length is $W=1927.4~$nm and $L=1780.75~$nm. At the cavity center, there are two air holes with diameters of $220~$nm. The surface-to-surface gap distance  (bridge width) between two holes is $60~$nm. The potential $z-$polarized dipole is placed $5~$nm above the bridge. The constant permittivity of the cavity material (Indium phosphide: InP) is $\epsilon_{\rm InP}=3.1649^2\approx10.02$. 
Here, the agreement between full dipole simulations and the QNM expansion in the dipole decay rate predictions is excellent within a few linewidths of the resonance, but deviates beyond that, potentially due to the influence of additional modes.  
Interestingly, the value of $\tilde{\Omega}_{\rm BB}=0.028$ suggests that the broadband dissipative regime, where corrections to phenomenological dissipation models become significant, can occur \emph{well below} the usual threshold of USC. Here, $ \tilde{\Omega}_{\rm BB}Q_c \approx 4.5$, and some deviations at the edge of the band of dynamics with scaled dynamical rate on the threshold of the dissipative broadband regime ($\tilde{\Omega}_{\rm BB}$) can be observed in Fig.~\ref{fig: bowtie_20250507} (e), suggesting some limits to the efficacy of single mode models, though note we have made no effort to optimize the cavity design to mitigate this effect. For USC, however, the bandwidth of dynamics would become larger, and the prediction of the single-mode QNM theory would not be more accurate in this regime than phenomenological models.

Next, we consider a 2D photonic crystal (PC) design with very low material loss, which we show in Fig.~\ref{fig: 2DPC_20250508}. The side length of the square nanorod array is $25a$, where the lattice constant is $a=250~$nm and the diameter of the nanorod is $D_{\rm rod}=0.4a=100~$nm. There is a point defect (cavity) coupled to a line defect (waveguide), where the center-to-center distance between the cavity and the waveguide is $3a$. The potential dipole (line current) is placed at the cavity center. The dielectric function $\epsilon_{\rm rod}(\omega)$ of the nanorods is described by a single Lorentz oscillator model,  
\begin{equation}\label{eq: Lorentz}
\epsilon_{\rm rod}(\omega)=\epsilon_{\infty}-\frac{(\epsilon_{\rm s}-\epsilon_{\infty})\omega_{\rm t}^{2}}{\omega^{2}-\omega_{\rm t}^{2}+i\omega\gamma_{\rm L}},
\end{equation}
with $\epsilon_{\infty}= n_{\rm B}^2=1.0$, $\epsilon_{\rm s}=8.9$, $\hbar\omega_{\rm t}=24.12$ eV, and $\hbar\gamma_{\rm L}=0.131$ eV. The resonance of the Lorentz model is very far away from the frequency regime of interest. As a result, the rods have an average permittivity of $\epsilon_{\rm rod}\approx 8.95+0.0035i$, within $[\omega_c-10\kappa_c,\omega_c+10\kappa_c]$, namely, $10$ linewidths away from the resonance of the dominant single QNM.
Here, the projected QNM phase plot in Fig.~\ref{fig: 2DPC_20250508} (c) shows rich and asymmetric features, highlighting the possibility of engineering broadband dissipative behavior by spatial positioning of the dipole. The full calculation of the dipole decay rates again shows good agreement with a single QNM expansion, and the value $\tilde{\Omega}_{\rm BB} = 0.0036$ reveals that the broadband regime can be reached far below the threshold for USC. However, here, $|\eta^{(1)}|$ takes the value $0.074$, which, being less than the threshold for USC indicates that single-mode models of USC will fail to capture accurate corrections to dynamics beyond phenomenological models of dissipation.  Here we obtain a radiative decay fraction of $S_c^{\rm rad}/S_c \approx 79$\%, where we calculate $S_c^{\rm nrad}$ and approximate $S_c^{\rm rad} = S_c -S_c^{\rm nrad}$ and $S_c \approx 1$.

In Appendix~\ref{app:additional}, we show an additional example for a 3D PC cavity mode, similar to the one studied in Ref.~\cite{Gustin2025}.

\begin{figure*}[htb]
    \centering
    \includegraphics[width = 1.6\columnwidth]{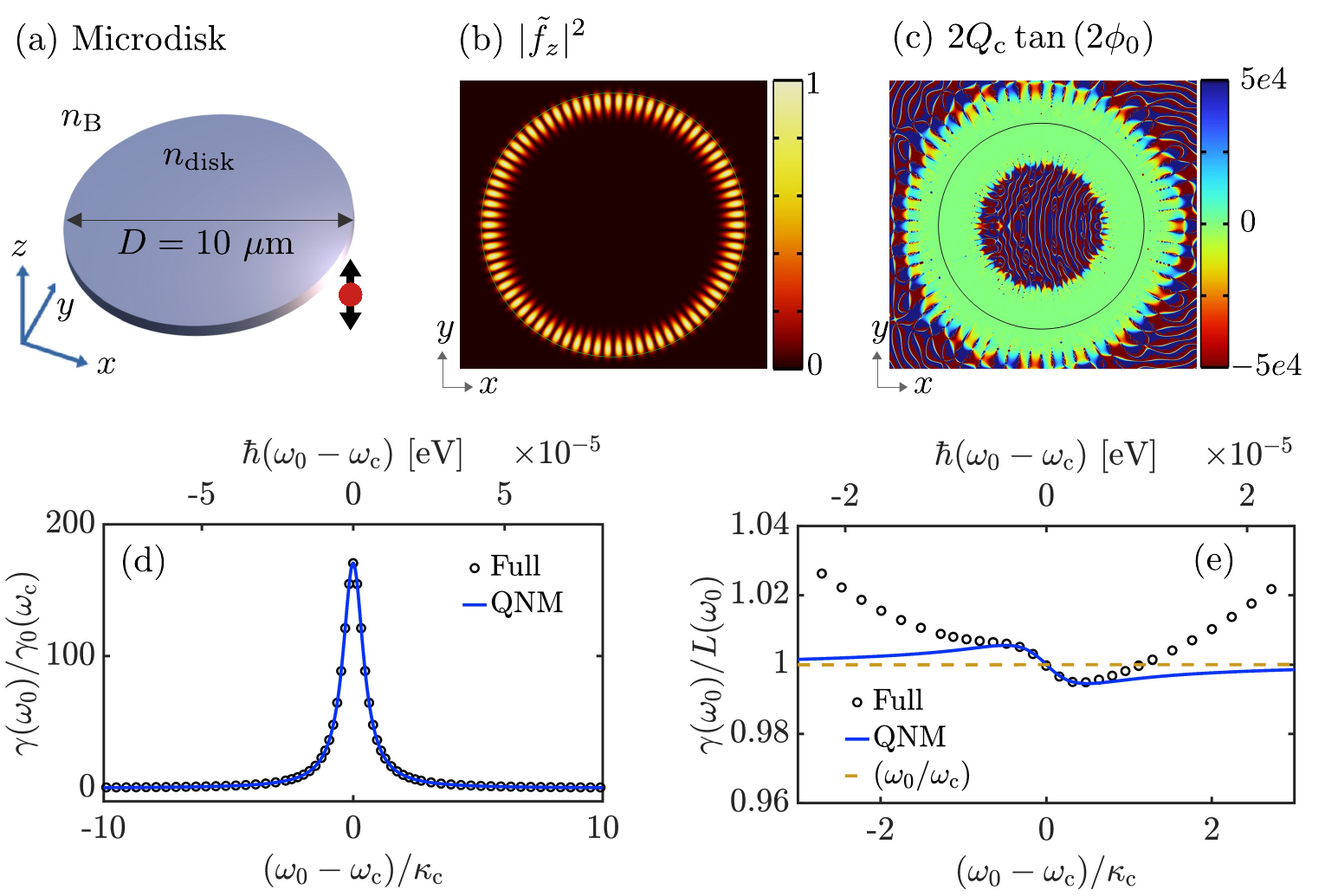}   
    \caption{(a) 3D schematic of microdisk (however, note our simulation is in 2D for this structure). $n_{\rm disk}=2.0+10^{-5}i$. $n_{\rm B}=1$. (b) Mode profile for QNM $|\tilde{{f}}_z|^2$ (arb. units; TM mode, this is the only component). (c) Phase distribution of $2Q_{\rm c}\tan{(2\phi_0)}$ for dominant QNM, where the phase is defined as $\tilde{f}_{z}(\mathbf{r})=|\tilde{f}_{z}(\mathbf{r})|e^{i\phi_{0}(\mathbf{r})}$. (d) Purcell factors for an emitter (line current) $10$ nm away from the disk, normalized by the  2D free-space rate $\gamma_0^{\rm 2D}(\omega)  =|\mathbf{d}^{\rm 2D}|^2\omega^2/(2 \epsilon_0 \hbar c^2)$.
    (e) Decay rate normalized by Lorentzian function $L(\omega_0)=\gamma_{c}^{\rm QNM}(\omega_c)\times\frac{\kappa_c^2/4}{\kappa_c^2/4+(\omega_0-\omega_c)^2}$. Here, we plot the full dipole decay rate with an approximate background $0.5\gamma_0^{\rm 2D}(\omega_0)$ subtracted out. The dominant QNM parameters are found in Table~\ref{tab:tab}.
    }
    \label{fig: WGM}
\end{figure*}

As a final example, we also consider a 2D lossy microdisk with a diameter of $10~\mu$m, where the constant refractive index of the disk is $n_{\rm disk}= 2.0+10^{-5}i$.
We find a dominant whispering-gallery mode (WGM), and we plot the mode and phase distribution of this dominant QNM in Fig.~\ref{fig: WGM} (a-c). We consider a dipole (line current) location 10 nm away from the disk. Here a non-monotonic frequency dependence is observed [when normalized to $L_c(\omega_0)$], and values of $|\eta^{(1)}|=0.088$ and $\tilde{\Omega}_{\rm BB}=0.0042$ again suggest limitations to single-mode dissipative models of USC and broadband dynamics for this mode. Here, in fact, these limitations are \emph{extremely} significant; with a characteristic scaled dynamical rate $\tilde{\Omega}_{\rm BB}$, the bandwidth of the system dynamics spans a number of bare cavity linewidths $\kappa_c$ equal to $\sim \tilde{\Omega}_{\rm BB}Q_c \approx 425$, but from Fig.~\ref{fig: WGM} (e), our \emph{ab initio} QNM correction to the dissipative dynamics remains accurate only around a single linewidth of the cavity mode. Such a result can be likely be attributed to the quasi-1D nature of such ring resonators, leading to a quasi-harmonic distribution of longitudinal modes, and a strong reduction of the potential for broadband dynamics to be constrained to a single-mode subspace. We expect this feature to be a general property of 1D and quasi-1D cavity designs, which also highlights the importance of our \emph{ab initio} approach which is appropriate for full 3D cavity geometries. The emission here is almost entirely nonradiative, with $S_c^{\rm nrad}/S_c \approx 1$;
however, this is really just a consequence of modeling the disk with a complex refractive index and having a 2D WGM. In more practical examples, the disk would be 3D and would thus yield vertical decay emission, and also such resonators are often coupled to an output waveguide mode, such as through an evanescently coupled fiber.

In summary, these results in their entirety (and those of Appendix~\ref{app:additional}) suggest that {\it low-Q plasmonic resonators generally have a broadband dissipation regime that coincides with the USC regime (in terms of resonant light-matter coupling strength)}, and dynamical corrections to the predictions of phenomenological theories of their photon dissipation 
can be captured by our corrected and \emph{ab initio} approach well into the USC regime. In contrast, the dielectric systems we study have larger quality factors, and will likely require multimode models for accurate description in the USC regime (even when other modes are very spectrally isolated from the dominant one). However, these systems also enter the broadband dissipative regime with (resonant) coupling strengths far below the threshold of USC, where our single-mode \emph{ab initio} approach can accurately predict broadband dissipative corrections for certain cavity designs. For 1D and quasi-1D cavity designs (like the microdisk WGM), we expect the quasi-harmonic distribution of longitudinal modes to preclude the possibility of accurate calculations of broadband dissipative corrections to the dynamics using just a single-mode model.

Based on these results, we shall pick the plasmonic dimer systems as our main example for our study of USC using the \emph{ab initio} single-mode master equation, which we shall use going forward as rough guides for our parameter sets to study quantum observables in the next section.

\section{Quantum master equation simulations in the USC regime}\label{sec:quantum_sim}
In this section we numerically solve the Coulomb gauge master equation of Eq.~\eqref{eq:QNM_ME} in the single-mode limit, using the QNM frequency-dependent dissipation rates from Eq.~\eqref{eq:QNM_rate} (in the spatially-specified representation). The system level Hamiltonian is given by Eq.~\eqref{eq:HC_sys}, with the longitudinal term neglected. Explicitly, the master equation is
\begin{align}
\dot{\rho} &= -\frac{i}{\hbar}[\hat{\mathcal{H}}_{\rm C}^{\rm S},\rho] \nonumber \\ 
& + 2\pi\sum_{\alpha,\alpha'}\left( c^\Pi_{\alpha} c^{\Pi *}_{\alpha'}\Lambda^2(\omega_{\alpha})\left[\hat{\sigma}_{\alpha}\rho\hat{\sigma}^{\dagger}_{\alpha'}- \hat{\sigma}^{\dagger}_{\alpha '}\hat{\sigma}_{\alpha}\rho\right] + \text{H.c.}\right),
\end{align}
and here we have not made the secular approximation from Sec.~\ref{subsec:BM} for the purposes of greater accuracy of simulations, though for strong coupling ($|\eta_c| Q_c > 1$) the results are very similar with it. The system Hamiltonian $\hat{\mathcal{H}}_{\rm C}^{\rm S}$ is explicitly in the single mode limit, 
\begin{equation}\label{eq:HC_sys1mode}
    \hat{\mathcal{H}}_{\rm C}^{\rm S} = \hbar\omega_c\hat{a}_c^{\dagger}\hat{a}_c + \frac{\hbar \omega_0}{2}\Big[ \cos{(\hat{\Phi}_{c})}\hat{\sigma}_z+\sin{(\hat{\Phi}_{\rm c})}\hat{\sigma}_y\Big],
\end{equation}
where we have approximated $\chi_{cc} \approx \omega_c$, and $\hat{\Phi}_c = 2\eta_c \hat{a}_c + \text{H.c.}$
We have also made two generalizations to be able to compare with more phenomenological theories of cavity-bath coupling: (i) we have written the spectral density in general terms as $\Lambda^2(\omega_{\alpha})$, where the correct \emph{ab initio} result is given by Eq.~\eqref{eq:spectral_density}, and (ii) we have expressed the matrix element $c^{\Pi}_{\alpha} = \bra{j}\hat{\Pi}\ket{k}$ in terms of a general cavity operator $\hat{\Pi}$ which couples to the reservoir. Specifically, in these terms we are able to compare with a phenomenological model of cavity-bath coupling which can be expressed in terms of an effective system-reservoir Hamiltonian of the form
\begin{equation}
    \hat{H}_{\rm phen} = \hbar\int d\omega \Lambda(\omega)\left[\hat{\Pi}\hat{d}_{\omega} + \text{H.c.}\right],
\end{equation}
where $[\hat{d}_{\omega},\hat{d}^{\dagger}_{\omega'}]=\delta(\omega-\omega')$.

To model excitation, we consider an incoherent excitation drive which can be modeled by adding to the master equation in Eq.~\eqref{eq:QNM_ME} the term
\begin{align}\label{eq:inc}
\dot{\rho} &\rightarrow  \dot{\rho} \ + \nonumber \\ &  2\pi\sum_{\alpha,\alpha'}\left( c_{\alpha} c^*_{\alpha'}\Lambda_{\rm inc}^2(\omega_{\alpha})\left[\hat{\sigma}^{\dagger}_{\alpha}\rho\hat{\sigma}_{\alpha'}- \hat{\sigma}_{\alpha '}\hat{\sigma}^{\dagger}_{\alpha}\rho\right] + \text{H.c.}\right),
\end{align}
where we have chosen to model cavity driving with matrix elements $c_{\alpha}$ calculated from $\hat{a}_c$, and we choose the incoherent excitation spectral density to take the form $\Lambda_{\rm inc}^2(\omega_{\alpha}) = \kappa_c \omega_c/(2\pi\omega_{\alpha})$. Other forms of incoherent excitation are possible, and will generally depend on the physical model of excitation; we choose this simple form as it recovers results from a simple classical heuristic model of incoherent excitation when we compare with the bosonic Hopfield model in the following subsection, which has been previously shown to have a quantum correspondence in a phenomenological dissipation model ~\cite{Hughes2024Jun}. Note one could also consider coherent excitation~\cite{Salmon2022Mar}, which would render the master equation explicitly time-dependent, and provide for a physical excitation model without such heuristic assumptions. Additionally, other sources of incoherent dynamics (i.e., noise sources) may also be present in realistic systems, especially at room temperature, as arising from (e.g.) pure dephasing~\cite{PhysRevLett.130.123601}, phonon scattering in solid-state systems~\cite{reiter_distinctive_2019}, or intramolecular vibrations. For the sake of analyzing the underlying \emph{fundamental} quantum optical characteristics of photon loss, we neglect these for the purpose of this work.

From the master equation solution, 
we can calculate the steady-state intracavity (specifically, near-field) spectrum,
which, following the discussion in Sec.~\ref{sec:photodet}, should be calculated in terms of matrix elements of the the transverse electric field operator $\hat{\bm{\mathcal{E}}}_{\perp}$.
Defining $\hat{x}_{\rm det} = \sum_{\alpha} c^{\rm det}_{\alpha} \hat{\sigma}_{\alpha}$, where 
\begin{equation}\label{eq:cefield}
c^{\rm det}_{\alpha} = \bra{j} i\eta_c \hat{a}_c - i\eta_c^* \hat{a}^{\dagger}_c \ket{k},
\end{equation}
where again $\alpha$ indexes a transition between states $\ket{k}$ and $\ket{j}$ with $\omega_k > \omega_j$, the intracavity spectrum is
\begin{equation}
    S_{c}(\omega) = \lim_{t\rightarrow \infty}\text{Re}\Bigg\{\int_0^{\infty} d\tau e^{i\omega \tau} \langle \hat{x}_{\rm det}^{\dagger}(t)\hat{x}_{\rm det}(t+\tau) \rangle \Bigg\}.
\end{equation}

Here we have assumed the projected QNM phase at the detector location to be approximately the same as at the emitter location (and the detected polarization the same), such that the matrix elements can be calculated from Eq.~\eqref{eq:cefield}. 
Other observables, such as the second order degree of coherence $g^{(2)}(\tau)$ could also be calculated if desired~\cite{Salmon2022Mar}.

\begin{figure}[htb]
    \centering
    \includegraphics[width = 0.99\columnwidth]{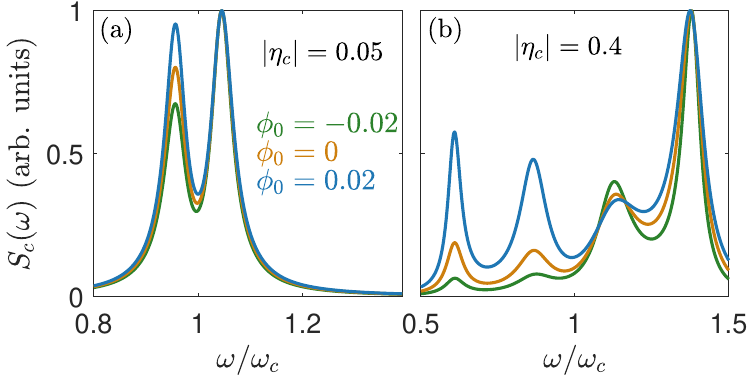}   
    \caption{Normalized intracavity spectrum for a single cavity mode with $Q_c=13$  for a TLS coupled with  (a) $|\eta_c| = 0.05$ and (b) $|\eta_c|=0.4$. Spectra are shown for values of the projected QNM phase at the dipole location of $\phi_0 = -0.02$ (green), $\phi_0 =0$ (orange), and $\phi_0 = 0.02$ (blue; most similar to the dimer value of $\sim 0.014$ in Ref.~\cite{Gustin2025}). Here we use $\gamma_{\rm inc} = \kappa_c/100$, which can probe beyond the weak-excitation regime.}
    \label{fig:Phi_effect}
\end{figure}

\begin{figure}[htb]
    \centering
    \includegraphics[width = 0.99\columnwidth]{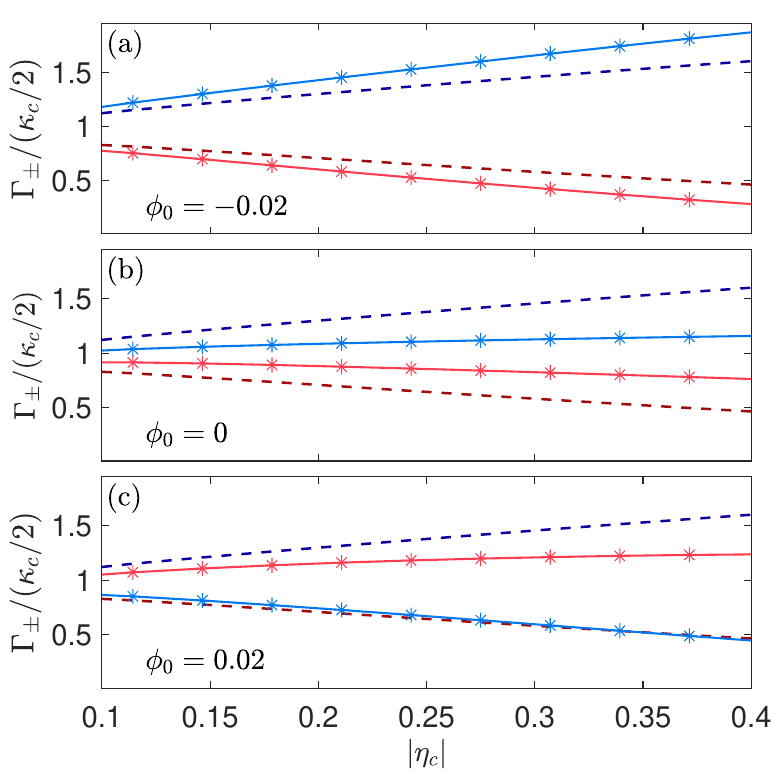}   
    \caption{Linewidths of dominant two peaks in intracavity spectrum (red-shifted peak in red, blue-shifted peak in blue) for a single cavity mode with $Q_c=20$  for a TLS coupled to the mode with a projected QNM phase at the TLS location of (a) $\phi_0=-0.02$, (b) $\phi_0=0$, and (c) $\phi_0 = 0.02$. Here we let the incoherent drive be sufficiently weak such that the trends converge ($\gamma_{\rm inc} = \kappa_c/10^4$ is sufficient). The dark red and blue dashed lines give the result for a naive flat spectral density $\Lambda_{\rm flat}(\omega) = \sqrt{\kappa_c/(2\pi)}$.}
    \label{fig:linewidths}
\end{figure}

    We first simulate intracavity spectra using our \emph{ab initio} quantum master equation model in the spatially specified representation using a value of $Q_c = 13$, similar to the cylindrical gold dimer mode from Ref.~\cite{Gustin2025} and also of the same order of magnitude as the plasmonic dimer examples given in this work (see cylindrical dimer example in Appendix~\ref{app:additional}). In Fig.~\ref{fig:Phi_effect} we show the highly significant effect (even outside of the formal USC regime) of varying the QNM phase at the dipole location on the emission spectra. Here we use the correct single-mode spectral density $\Lambda_c(\omega)$ from Eq.~\eqref{eq:spectral_density}. We stress again that the broadband dissipative regime criteria $|\eta_c| \geq \tilde{\Omega}_{\rm BB}$ (for the $\omega_0=\omega_c$ case studied here) is only a rough metric; indeed, here for $|\eta_c|=0.05$ significant deviations from symmetric spectra (i.e., broadband dissipative effects) can be observed even for the $\phi_0\geq 0$ case, where $\tilde{\Omega}_{\rm BB} \approx 0.1$,

    Next, we investigate the effect of using the correct spectral density and QNM phase on the linewidths (full width at half-maximum) of the emission spectrum in the weak-excitation limit. In Fig.~\ref{fig:linewidths} we fix $Q_c=20$ (somewhat similar to the ellipsoid gold dimer dominant QNM) and vary the coupling strength $|\eta_c|$ for a variety of different projected QNM phases at the dipole location, and fit the resulting peaks to a two-Lorentzian function, plotting the linewidths of the resulting dominant two spectral peaks.   Note that the quantum Rabi Model and its Coulomb gauge equivalent, Eq.~\eqref{eq:HC_sys1mode}, do not have a classical correspondence in the weak-excitation regime due to virtual excitations in the ground state giving rise to fundamental anharmonicity, which manifests as a multi-peak structure in the emission spectra for sufficiently large $|\eta_c|$~\cite{Hughes2024Jun}; here we show the linewidths of the most dominant peaks which correspond to the JC peaks for sufficiently small cavity-TLS coupling.
    
    These trends agree well with the perturbative analytical solution (which we derive in Appendix~\ref{app:BS} valid to first order in $|\eta_c|$), given by Eq.~\eqref{eq:lws}.
Interestingly, this equation predicts the existence of a certain $\phi_0^* \approx 1/(8Q_c)$ such that the linewidths remain symmetrical to leading order in $|\eta_c|$, even in the USC regime.
 We have verified numerically (not shown) for the range of $|\eta_c|$ considered here the linewidths remain indeed approximately constant at $\kappa_c/2$, with only higher-order nonlinear deviations. Note the spectrum itself (i.e., the spectral weights) is not necessarily symmetric as well in this case, as this depends on the specific incoherent excitation model. The persistent asymmetries in spectral linewidths (observable even below the USC regime) has also been predicted using purely classical \emph{ab initio} models~\cite{canales_polaritonic_2023}.

\begin{figure}[htb]
    \centering
    \includegraphics[width = 0.99\columnwidth]{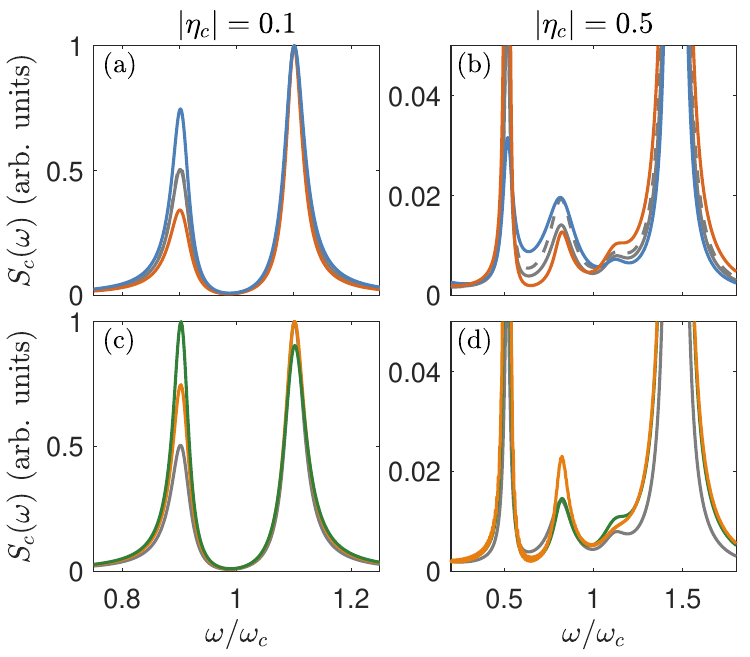}   
    \caption{ Normalized cavity spectrum for gold dimer-like cavity with $Q_c=13$, $\phi_0 = 0.0137$, and (a,b) variable system-bath coupling operator $\Pi = \hat{a}_c$ (grey, solid), $\Pi = (\hat{Q} + \hat{P})/\sqrt{2}$ (grey, dashed), $\hat{\Pi} = \hat{Q}$ (red, solid), and $\hat{\Pi} = \hat{P}$ (blue, solid). (c,d) cavity spectra with fixed (correct) $\hat{\Pi} = \hat{a}_c$ and $\Lambda^2_{n}(\omega) = \frac{\kappa}{2\pi}\left(\omega/\omega_c\right)^n\zeta_c(\phi_0,\omega)$ with $n = -1$ (grey) $n = 0$ (orange), and $n= 1$ (green). 
    }
    \label{fig:Pi_SD}
\end{figure}

Next, we study the effect of the cavity-bath coupling operator $\hat{\Pi}$ (with $\hat{\Pi}=\hat{a}_c$ being the correct choice in our \emph{ab initio} quantized QNM model). Often in the literature a phenomenological choice of $\hat{Q} = \hat{a}_c + \hat{a}_c^{\dagger}$ or $\hat{P} = i(\hat{a}^{\dagger}_c - \hat{a}_c)$ is assumed~\cite{Bamba2014Feb}. In Fig.~\ref{fig:Pi_SD}, we show the cavity spectrum for different choices of $\hat{\Pi}$, showing the significant effect of the choice of bath-coupling operator. Interestingly, the choice of $\hat{\Pi} = (\hat{Q} + \hat{P})/\sqrt{2}$ (and $(\hat{Q} - \hat{P})/\sqrt{2}$ not shown, but gives equivalent results) gives very similar results to the \emph{ab initio} $\hat{\Pi}=\hat{a}_c$.
Recently, both of these forms were shown to give a classical-quantum correspondence in models of ultrastrong coupling to bosonic matter with a phenomenological dissipative normal mode approach~\cite{Hughes2024Jun}
and indeed we show in the following subsection that such a correspondence also exists using an \emph{ab initio} QNM theory. Also in Fig.~\ref{fig:Pi_SD} we show the effect of using different spectral densities on the cavity spectra.
\subsection{Bosonic systems and classical comparison}
In addition to our study of the quantum Rabi Model, it is also interesting to probe the dynamics of ultrastrong light-matter interactions in the many-body regime where, in the thermodynamic limit and assuming isotropic couplings, the matter degrees of freedom become an
{\it effective} single bosonic oscillator~\cite{Garziano2020Aug}, and the system Hamiltonian becomes the quantum Hopfield model. In addition to being the main regime in optics where USC has been experimentally realized to date~\cite{li_vacuum_2018,rajabali_polaritonic_2021,baranov_ultrastrong_2020,mueller_deep_2020}, this regime is also known to have a classical-quantum correspondence, both in terms of oscillator resonances and certain phenomenological dissipation models~\cite{Hughes2024Jun}.

To study such systems, we consider $N$ TLS systems with identical dipole moments $\mathbf{d}'$ and transition energy $\hbar \omega_0$, located in a region of space where the dominant QNM varies sufficiently slowly in phase and magnitude that we can approximate their positions identically as $\mathbf{r}_i \approx \mathbf{0}$ for $i=1,2,...,N$. The thermodynamic limit in these conditions corresponding to taking $N \rightarrow \infty$ and $|\mathbf{d}'| \rightarrow 0$, with $\sqrt{N} \mathbf{d}'\rightarrow \mathbf{d}_{\rm eff}$ remaining finite (in practice $N$ is of course finite, and associated correction terms can be computed~\cite{Garziano2020Aug}). Under these assumptions, the Coulomb gauge system Hamiltonian becomes 
[cf.~Eq.~\eqref{eq:HC_sys1mode}]
\begin{align}\label{eq:HDicke}
\hat{\mathcal{H}}_{\rm C,Dicke}^{\rm S} = \hat{H}^{\rm em}_{\rm QNM} + \frac{\hbar \omega_0}{2}\sum_{i=1}^{N}\left[\cos{\left(\hat{\Phi}'\right)}\hat{\sigma}_z^i+\sin{\left(\hat{\Phi}'\right)}\hat{\sigma}_y^i\right],
\end{align}
which is a form of (gauge-corrected) Dicke-like Hamiltonian. Here, $\hat{\Phi}' = 2\hat{a}_c \eta'_c + \text{H.c.}$, where $\eta'_c = \mathbf{d}' \cdot \tilde{\mathbf{f}}^{\rm s}_c(\mathbf{0})/\sqrt{2\epsilon_0 \hbar \chi_{cc}}$.

Next, we apply the Holstein-Primakoff transformation~\cite{PhysRev.58.1098,PhysRevA.11.981,PhysRevLett.92.073602,Garziano2020Aug}; in anticipation of taking the thermodynamic limit, it is sufficient to here take $\sum_{i=1}^{N}\hat{\sigma}_z^i = 2\hat{b}^{\dagger}\hat{b} - N$, and $\sum_{i=1}^{N}\hat{\sigma}_{y}^i = i\sqrt{N}(\hat{b}-\hat{b}^{\dagger})$, where $[\hat{b},\hat{b}^{\dagger}]=1$. Expanding Eq.~\eqref{eq:HDicke} to order $\hat{\Phi}'^2$ (higher-order terms vanish in the thermodynamic limit), and taking the thermodynamic limit, we obtain (dropping a term proportional to the identity)
\begin{align}\label{eq:H_hop}
\hat{\mathcal{H}}_{\rm C, Hop}^{\rm S} = &\hbar \chi_{cc}\hat{a}^{\dagger}\hat{a}_c + \hbar \omega_0\hat{b}^{\dagger}\hat{b} + \hbar\omega_0\left(\lambda_c \hat{a}_c + \lambda_c^*\hat{a}_c^{\dagger}\right)^2 \nonumber \\ &+ i\hbar\omega_0(\hat{b}-\hat{b}^{\dagger})(\lambda_c \hat{a}_c + \lambda_c^*\hat{a}_c^{\dagger}),
\end{align}
which is a Coulomb-gauge Hopfield Hamiltonian, and we have defined $\lim_{N\rightarrow \infty} \sqrt{N} \eta_c' = \lambda_c$. The system-reservoir Hamiltonian remains
$\hat{\mathcal{H}}_{\rm C,Hop}^{\rm SR} = \hat{H}^{\rm em}_{\rm QNM-R} $.

In addition to this quantum model, we can also compare with a fully classical approach~\cite{Hughes2024Jun}.
In such a model, we can calculate the classical spectrum
from $S_{\rm cl}(\omega) = |\mathbf{E}(\mathbf{r}_{\rm det},\omega)|^2$, where $\mathbf{E}$ is the classical scattered electric field at the detector,
\begin{equation}
\mathbf{E}(\mathbf{r}_{\rm det},\omega) = \mathbf{G}^{\perp}(\mathbf{r}_{\rm det},\mathbf{0},\omega) \cdot {\bm \alpha}(\omega) \cdot \mathbf{E}_0,
\end{equation}
where $\mathbf{E}_0$ is a constant representing a frequency-independent dipole excitation, $\mathbf{G}^{\perp}$
is the medium transverse Green function (we take the transverse part, as we want to compare with the quantum spectrum defined in terms of the transverse QNM cavity operators), and
\begin{equation}
{\bm \alpha}(\omega) = \left[\mathbf{I} - {\bm \alpha}_0(\omega) \cdot \mathbf{G}(\mathbf{0},\mathbf{0},\omega)\right]^{-1} \cdot {\bm \alpha}_0(\omega),
\end{equation}
is the polarizability of the classical dipole (expressed in terms of a tensor inverse), and $ {\bm \alpha}_0(\omega) = 2\mathbf{d}'\mathbf{d}'\omega_0/(\hbar\epsilon_0(\omega_0^2-\omega^2))$ is the bare polarizability (i.e., before radiative coupling). Here we neglect any singular contributions to $\mathbf{G}(\mathbf{0},\mathbf{0},\omega$) (i.e., we take the transverse part and neglect the background terms as we have done throughout).

In this formalism, we can substitute $\mathbf{G}_c(\mathbf{r},\mathbf{0},\omega) = A_c(\omega) \tilde{\mathbf{f}}_c(\mathbf{r})\tilde{\mathbf{f}}_c(\mathbf{0})$ for a single QNM expansion.
We assume that the spectrum is detected at a location $\mathbf{r}_{\rm det}$ within the system region, such that the QNM transverse Green function expansion holds.
 For a single QNM, we find, analytically,
\begin{equation}
{\bm \alpha}_{\rm QNM}(\omega)  ={\bm \alpha}_0(\omega)\left[1 - \frac{4 \omega_0 \omega_{c} A_c(\omega) \eta_c^2}{S_c(\omega_0^2-\omega^2)}\right]^{-1},
\end{equation}
where we approximate $\chi_{cc} \approx \omega_c$. As we shall see, we in fact need to use a modified QNM expansion in the classical theory to obtain agreement with the quantum theory.
In particular, there exists for each positive-eigenfrequency QNM $\tilde{\mathbf{f}}_c$ with $\tilde{\omega}_c$, a corresponding negative-frequency QNM with $\tilde{\mathbf{f}}_c^*$and $-\tilde{\omega}_c^*$. we can add these negative-frequency QNMs to the Green function to obtain the modified expansion
\begin{equation}
\mathbf{G}'_c(\mathbf{r},\mathbf{0},\omega) = A_c(\omega)\tilde{\mathbf{f}}_c(\mathbf{r})\tilde{\mathbf{f}}_c(\mathbf{0}) + A^*_c(-\omega)\tilde{\mathbf{f}}^*_c(\mathbf{r})\tilde{\mathbf{f}}^*_c(\mathbf{0}),
\end{equation}
which leads to 
\begin{align}
{\bm \alpha}'&_{\rm QNM}(\omega)  = \nonumber \\ &{\bm \alpha}_0(\omega)\left[1 - \frac{4 \omega_0 \omega_{c} \left[A_c(\omega) \eta_c^2 + A^*_c(-\omega) \eta_c^{* 2}\right]}{S_c(\omega_0^2-\omega^2)}\right]^{-1}.
\end{align}

\begin{figure}[htb]
    \centering
    \includegraphics[width = 0.99\columnwidth]{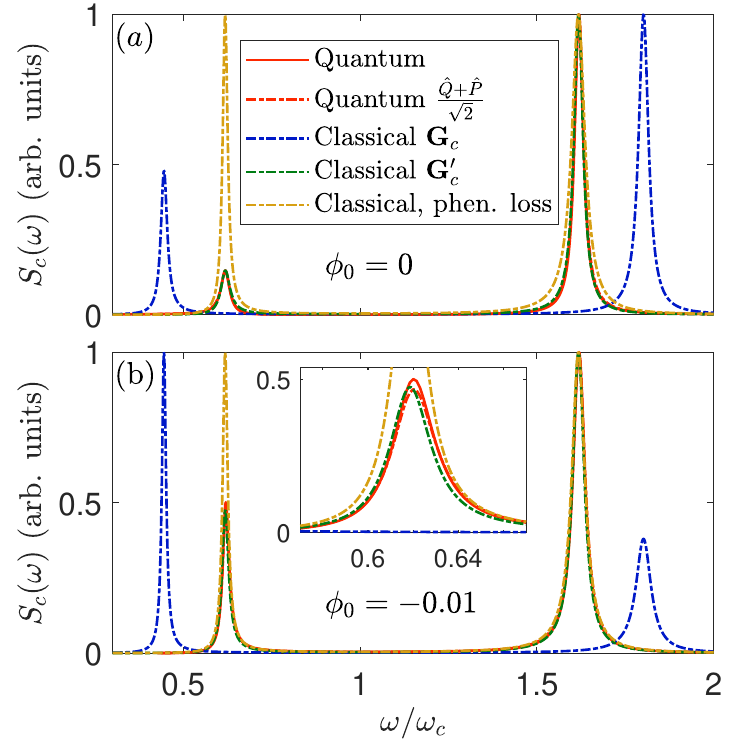}   
    \caption{Normalized intracavity spectrum for the bosonic Hopfield model Hamiltonian. We use $Q_c=16$, (a) $\phi_0 = 0$, and (b) $\phi_0 = -0.01$. We let $|\eta_c| = 0.5$. The inset on (b) shows a zoom-in on the lower-frequency peak. The red dash-dotted line is the quantum theory with phenomenological system-bath coupling operator $\hat{\Pi} = (\hat{Q}+\hat{P})/\sqrt{2}$.}
    \label{fig:classical}
\end{figure}

In Fig.~\ref{fig:classical}, we compare the quantum cavity spectrum $S_c(\omega)$ under weak excitation with the classically calculated spectrum $S_{cl}$ using both $\mathbf{G}_c$ and $\mathbf{G}'_c$. We also plot the spectrum with phenomenological dissipation (i.e., using a normal mode expansion with phenomenological decay rate $\kappa_c$), using the model from Ref.~\cite{Hughes2024Jun}. This model recovers the quantum cavity spectrum under an (incorrect) assumption of a \emph{flat} spectral density. While using the single-mode expansion without the negative frequency QNM in the classical theory does not recover the quantum result (and in fact also predicts different resonances, as well as a reversed asymmetry in the case of negative QNM phase), adding in the negative QNM contribution leads to a spectrum almost entirely in agreement with the QNM theory. Interestingly, {\it the negative frequency QNM does not have to be accounted for in the quantum theory to see agreement with the classical result}, indicating important differences in how the single-mode approximation is properly implemented in the different formalisms.

Very slight deviations between classical and quantum results  
are likely mostly a result of the fact that the classical theory does not use any Markov approximation. 
In the quantum theory, the use of a Born-Markov approximation leads to a time-convolutionless master equation which has strictly Lorentzian spectral peaks in the limit of strong coupling (well-separated peaks). In contrast, the classical theory is able to predict very small modifications of the these lineshapes as it does not rely on such an approximation. 

It is important to note that the  spectral weights of each peak in the spectra shown here are highly sensitive to the form of the incoherent excitation model used in both classical and quantum formalisms. For example, in the quantum theory, we use an incoherent drive spectral density of 
the form: $\Lambda_{\rm inc}^2(\omega_{\alpha}) = \kappa_c \omega_c/(2\pi\omega_{\alpha})$, which is independent of $\phi_0$, and gives a close correspondence with the \emph{phenomenological} classical model of a frequency- and QNM phase-independent excitation constant $\mathbf{E}_0$. Different excitation models would in general lead to different predictions for the relative peak amplitudes---e.g., thermal excitation in the quantum theory, which would have the same $\phi$-dependent spectral density for the incoherent drive as the \emph{ab initio} $\Lambda_c(\omega)$ in the dissipative part. 

Additionally, we note that the quantum-classical correspondence can also be achieved using a quantum model of incoherent excitation of the material oscillator component (instead of the cavity drive). In particular, we find that replacing the $c_{\alpha}$ in Eq.~\eqref{eq:inc},
with $\bra{j} i(\hat{b}-\hat{b}^{\dagger})\ket{k}$, and the spectral density with $\Lambda_{\rm inc}(\omega) = \kappa_c \omega^2/(2\pi \omega_c^2)$ gives identical results to our quantum cavity drive model. This can be understood loosely as follows: in the Coulomb gauge, the transverse light-matter interaction term can be written as $i\omega_0(\hat{b}-\hat{b}^\dagger)\mathbf{d} \cdot \hat{\mathbf{A}}_{\perp}$ (with $\mathbf{d} = \sqrt{N} \mathbf{d}'$). Substituting in the quantized QNM form for $\hat{\mathbf{A}}_{\perp}$ gives the result from Eq.~\eqref{eq:H_hop}, but we can assume an additional coupling to an effective incoherent excitation reservoir 
to model incoherent excitation. In the classical model, we have an effective frequency-independent electric field drive $\mathbf{E}_0$. But the transverse vector potential contains an additional factor of $1/\omega$ (i.e., as $\hat{\mathbf{E}}_{\perp} = -\frac{d}{d t}\hat{\mathbf{A}}_{\perp}$). Thus, we recover the classical model of excitation by assuming an effective incoherent excitation spectral density proportional to $\omega^2$, and with matrix elements computed from the form of the interaction Hamiltonian, in terms of the operator $i(\hat{b}-\hat{b}^{\dagger})$.

Overall, 
in this bosonic system example, the agreement between our classical QNM and quantum QNM results is excellent (as is the agreement with full dipole simulations without any approximations), and highlight important cavity mode effects that are not captured by the typical phenomenological light-matter theories (both classical and quantum). Such effects are not restricted to USC, but more generally apply in broadband regimes where the ratio of a characteristic dynamical rate of the system to the cavity frequency exceeds $\tilde{\Omega}_{\rm BB}$; i.e., when it approaches $\left|1 - 4Q_c\tan{(2\phi_0)}\right|^{-1}$. As we have shown with our dielectric cavity examples, this can occur in realistic systems with (e.g.) TLS-cavity coupling constants orders of magnitude lower than those required to enter the USC regime.

\section{Experimental Prospects}\label{sec:exp}

One of the key advantages of our \emph{ab initio} quantized quasinormal mode approach to the broadband regime of dissipative cavity-QED is its applicability to emerging optical experiments in USC and near-USC. Advances in both plasmonic and subdiffraction limit dielectric resonator design and fabrication~\cite{molesky_inverse_2018,albrechtsen2022,babar_self-assembled_2023,xiong_experimental_2024} are rapidly approaching the USC and near-USC regimes, and indeed, our finding that the broadband dissipative regime can be reached for dielectric systems with coupling strengths orders of magnitude below the usual threshold for USC indicates that the corrections we report in this work may be observable in a variety of currently accessible platforms.

In this section, we give a brief estimate of achievable normalized single-mode coupling strengths $|\eta_c|$ for reasonable dipole moments for the cavity designs we study in this work, followed by a comparison with some experimentally demonstrated subdiffraction dielectric and plasmonic cavities. 

Using the definition of $\eta_c$ from Eq.~\eqref{eq:eta_def}, and letting $\mathbf{\tilde{f}}^{\rm s}_c \approx \mathbf{\tilde{f}}_c$ and $\chi_{c c} \approx \omega_c$, we can express the normalized single-mode coupling strength for a given resonator structure as 
\begin{equation}
    |\eta_c| \approx 9.5 \times 10^{-14} \text{eV}^{-\frac{1}{2}}\text{m}^{\frac{3}{2}}\left[\frac{d}{d_0}\right]\left[\frac{|\tilde{\mathbf{f}}_c|}{\sqrt{\hbar\omega_c}}\right],
\end{equation}
where we assume the field amplitude to be dominated by a single direction, which the dipole moment is oriented parallel to,
and $d_0 = 1$~e$\cdot$nm is a reference dipole moment. With $d=d_0$, the predicted $|\eta_c|$ value is given in Table~\ref{tab:tab_eta}. For the 2D PC cavity and the microdisk modes, we estimate an effective 3D QNM amplitude by scaling the 2D QNM mode profiles by $l_{\rm eff}^{-1/2}$; here we use a value of $l_{\rm eff} = 100$ nm, which is half of the actual thickness of the 3D PC example. Although the predicted $|\eta_c|$ for dipole moment $d_0$ is below the broadband dissipative threshold $\tilde{\Omega}_{\rm BB}$ for all examples, it is important to note that we have made no particular effort to optimize for ultrasmall mode volume in our cavity designs.
Inverse design techniques in nanophotonics could likely easily increase all of these coupling rates significantly, which is another advantage of our arbitrary cavity mode approach.
This has been shown, for example, 
 with waveguide mode Purcell factors, including the design of chiral emitter coupling rates~\cite{PhysRevA.106.033514}.
Similar methods can be used to optimize QNM properties, which will be a topic of future work.

\begin{table}
    \centering
    \begin{tabular}{|l||c|} \hline 
         & $|\eta_c|$ \\ \hline 
         Ellipsoid Dimer & 0.02 \\ \hline
         Cyl. Dimer & 0.006 \\ \hline
         Bowtie & 0.0002 \\ \hline
         2D PC & 0.0007\\ \hline
         3D PC Beam & 0.0001 \\ \hline
         Microdisk & 0.00003 \\ \hline
    \end{tabular}
\caption{$|\eta_c|$ for example cavity structures studied in Sec.~\ref{sec:classical_results} and Appendix~\ref{app:additional} with dipole moment $d = 1$ e$\cdot$nm.
Note for the first example, this coupling rate can easily be increased by decreasing the gap, which is discussed in more detail in the main text. 
}

\label{tab:tab_eta}
\end{table}

Using a recently demonstrated topology-optimized dielectric cavity~\cite{albrechtsen2022,Kountouris:22}  which far surpasses the (bulk-like cavity) diffraction limit, and $d= d_0$, we find $|\eta_c| = 0.003$. Assuming this structure, similar to our bowtie cavity design, has a similar QNM phase above the bridge center ($\tan{(2\phi_0)} \approx 0.007$), using the quality factor $Q_c \approx 1100$ from Ref.~\cite{albrechtsen2022}, we predict that broadband dissipative effects will become observable for $\tilde{\Omega}_{\rm BB} \approx 0.003$. The coinciding of this broadband dissipative region threshold parameter with our estimated $|\eta_c|$ for a dipole moment $d_0$ (of the same order of magnitude as common dipole couplings including organic dye molecules~\cite{chikkaraddy2016single} and InAs/GaAs quantum dots; higher dipole moments can be achived in, e.g., CdSe/CdS quantum dots~\cite{hu_robust_2024}) indicates that observation of these effects may be possible in cavity designs fabricated using existing modern techniques.

Similarly, the PC beam cavity from Ref.~\cite{PhysRevLett.118.223605} gives $|\eta_c| \approx 0.007$ for $d=d_0$, assuming the dipole is placed in air with index of refraction $n=1$. Using the same QNM phase as our PC beam cavity, and assuming a quality-factor $Q \approx 10^6$, this gives $\tilde{\Omega}_{\rm BB} = 2 \times 10^{-5}$.

Moving onto plasmonic systems, near-USC has already been demonstrated using single quantum dots coupled to plasmonic nanoparticles. For example, a USC threshold coupling of $|\eta_c| =0.1$ was recently observed at room temperature~\cite{hu_robust_2024}, and similarly near-USC $|\eta_c|\approx 0.06$ was also achieved recently with gold nanorods~\cite{li_room-temperature_2022}. Beyond quantum dots, organic dye molecules (with dipole moments on the order of $\sim d_0$) were previously observed with $|\eta_c| \approx 0.04$ using a few-molecule system~\cite{chikkaraddy2016single}. Theoretical calculations using time-dependent density functional theory have also predicted the achievability of single-dipole USC coupling with dye molecules~\cite{kuisma_ultrastrong_2022}. Beyond single (or few) dipole coupling, collective USC couplings between nanoparticle plasmons and cavity modes have already been reached~\cite{baranov_ultrastrong_2020,mueller_deep_2020}.

Finally, we remark that by decreasing the gap size in our presented dimer simulations, significant decreases in mode volume (and thus enhanced light-matter coupling strength) can be predicted. For example, using an ellipsoidal gold dimer cavity with dimensions $W_{\rm elli}=8~$nm and $L_{\rm elli}=40~$nm (similar to the one shown in Fig.~\ref{fig: bowtie_20250507}) but decreasing the gap size to $d_{\rm elli} = 0.93$ nm, we find a quality factor $Q_c\approx 18.5$ and corresponding $|\eta_c| \approx 0.18$ for dipole moment $d_0/2=0.5$ e$\cdot$nm.
It should be noted however, that with such small gaps the contribution of quasistatic couplings beyond our single transverse mode model presented in this work likely become significant~\cite{PhysRevLett.131.013602}. Possible breakdown of the electric  approximation in such regimes can also be treated within our formalism~\cite{Gustin2023Jan} with slight model modifications.

\section{Conclusions}\label{sec:conclusions}
In this paper, we have presented an \emph{ab initio} approach to single-mode master equations for coupled cavity-dipole systems that remain valid in the USC regime, which is, to the best of our knowledge, the first time this has been achieved for realistic 3D cavity geometries. We introduced and defined the broadband dissipative regime of cavity-QED, where corrections to quantum observables begin to deviate substantially from predictions made with standard master equation models, of which the USC regime is contained as a subset, but in fact can occur for system dynamical parameters (e.g., light-matter coupling strengths) orders of magnitude below the usual threshold for USC. We assessed the validity of our quantized QNM approach for a variety of plasmonic and dielectric cavity designs, and performed quantum simulations in the USC regime to compare significant differences between the predictions of the \emph{ab initio} quantized QNM theory, when compared with standard phenomenological models. We also posited a set of potential heuristic rules regarding the ability and limitations of single-mode models to accurately model the broadband dissipative regime for dielectric and plasmonic structures in Sec.~\ref{sec:classical_results}. We finally showed how the effects we identify associated with broadband dissipative and USC regimes should be observable in a variety of optical platforms in the near-future.

There are numerous ways our approach could be generalized and built upon, and we expect this work to spur further research in these directions. For one, we have restricted ourselves to a study of near-field observables, leaving the more subtle problem of connecting cavity dynamics to input-output relations of scattered quantum fields to future work. Additionally, we have neglected the influence of background dipole couplings of the TLS to non-cavity reservoir modes, which with some generalization could be self-consistently included in our formalism. Finally, our presentation has been limited to a single QNM and a single dipole (or many dipoles at locations with identical QNM phases) for now; while the system-reservoir Hamiltonian we derived in Sec.~\ref{Sec: Light-Matter} is valid for an arbitrary number of QNMs and dipoles, the generalization of how to transform to the spatially specified representation in a general system with multiple modes and/or multiple dipoles placed at locations with differing QNM phases will require further work. We anticipate that a more rigorous and systematic treatment of the different representations of quantized QNMs (spatially specified vs unspecified) will shed fruitful light on this important question. Generalization of our approach to a multi-QNM system is especially desirable, as our results indicate that many dielectric systems can not be accurately modeled using a single-mode model in the dissipative broadband regime of cavity-QED

\acknowledgements
This work was supported by the
Natural Sciences and Engineering Research Council of
Canada (NSERC), the Canadian Foundation for Innovation
(CFI), Queen’s University, Canada, NSF awards PHY2011363 and CCF-1918549, and the
Alexander von Humboldt Foundation through a Humboldt Research Award. We also thank CMC Microsystems for
the provision of COMSOL Multiphysics and Hideo Mabuchi for useful discussions.

\appendix

\section{Vanishing of background transverse vector potential operator}\label{app:background_A}
Here we show that when using the quantized QNM projection functions $\mathbf{L}_{\mu}(\mathbf{r},\omega_{\rm m})$, the background contribution to the transverse vector potential (the part expressed in terms of the residual reservoir operators $\hat{\mathbf{c}}$, $\hat{\mathbf{c}}^{\dagger}$), $\hat{\mathbf{A}}^{\perp}_{\rm B}(\mathbf{r})$, vanishes.

To do so, we first make use of the following relation~\cite{Franke2020thesis}, valid for all $\mu$:
\begin{align}\label{eq:tbd}
    \int d^3r \int d\omega_{\rm m} \mathbf{L}_{\mu}(\mathbf{r},\omega_{\rm m}) \cdot \hat{\mathbf{c}}(\mathbf{r},\omega_{\rm m}) &=0 \nonumber \\ 
    \sum_{\nu}\left[S^{-\frac{1}{2}}\right]_{\mu\nu}\hat{x}_{\nu}&=0,
\end{align}
where $\hat{x}_{\nu} = \sqrt{2\omega_{\nu}/\pi}\,\hat{y}_{\nu}$, and
\begin{equation}
    \hat{y}_{\nu} \!= \!\int \!\!d^3r' \!\!\!\! \int \! \frac{d\omega_{\rm m}}{\omega_{\rm m}} \sqrt{\epsilon_I(\mathbf{r}',\omega_{\rm m})}A_{\nu}(\omega_{\rm m}) \tilde{\mathbf{F}}'(\mathbf{r}',\omega_{\rm m}) \cdot \hat{\mathbf{c}}(\mathbf{r}',\omega_{\rm m}).
\end{equation}
Now, $S$, and thus $S^{-\frac{1}{2}}$, are positive definite matrices~\cite{Franke2020thesis}, which implies $\sum_{\mu \nu} x_{\mu}^*\left[S^{-\frac{1}{2}}\right]_{\mu \nu}x_{\nu} > 0$ for any complex non-zero vector $x_{\mu}$. We can thus consider \emph{any} given matrix element $x^j_{\nu}$ (indexed by $j$) of $\hat{x}_{\nu}$, and find as a consequence of Eq.~\eqref{eq:tbd} that
\begin{equation}
    \sum_{\mu \nu} x^{j*}_{\mu} \left[S^{-\frac{1}{2}}\right]_{\mu \nu} x^j_{\nu} =0,
\end{equation}
which, as it is true for any arbitrary matrix element, implies $\hat{x}_{\mu}=0$, and thus $\hat{y}_{\mu}=0$.
Next, we note that $\hat{\mathbf{A}}_{\rm B}^{\perp}$  can be written as $\sqrt{\frac{\hbar}{\pi \epsilon_0}}\sum_{\mu} \tilde{\mathbf{f}}_{\mu}(\mathbf{r}) \hat{y}_{\mu}$, which thus vanishes.

\begin{widetext}
\section{Derivation of Hamiltonian with quantized QNM operators defined with respect to an electric field expansion}\label{app:E_field_QNM}
 
In the main text, we worked with a projection function for the quantum QNMs defined such that the vector potential can be expanded directly in a basis of quantized QNMs. 
Alternatively, and as has been done in previous work on quantized QNMs, one could define them such that the transverse electric field (more precisely, $\hat{\mathbf{E}}^{\perp}_{\rm F}$, the component of the transverse field which can be expressed in terms of bosonic field + medium polariton operators) can be expanded directly. In this case, the projection operator is 
\begin{equation}
\mathbf{L}^{(\mathbf{E})}_{\mu}(\mathbf{r},\omega_{\rm m}) = \sum_{\nu}\left[S^{(\mathbf{E})}\right]^{-\frac{1}{2}}_{\mu \nu} \sqrt{\frac{2\epsilon_I(\mathbf{r},\omega_{\rm m})}{\pi \omega_{\nu}}}A_{\nu}(\omega_{\rm m}) \tilde{\mathbf{F}}_{\nu}'(\mathbf{r},\omega_{\rm m}),
\end{equation}
which is equivalent to that in the expansion for the transverse vector potential (Eq.~\eqref{eq:projection_fn})  aside from an additional factor of $\omega/\omega_{\nu}$. Additionally, the $S$ quantization matrix is modified:
\begin{equation}
    S_{\mu \nu}^{(\mathbf{E})} = \frac{2}{\pi\sqrt{\omega_{\mu}\omega_{\nu}}} \int d^3r \int d\omega_{\rm m} \epsilon_I(\mathbf{r},\omega_{\rm m})A_{\mu}(\omega_{\rm m})A^*_{\nu}(\omega_{\rm m})\tilde{\mathbf{F}}'_{\mu}(\mathbf{r},\omega_{\rm m}) \cdot \tilde{\mathbf{F}}'^*_{\nu}(\mathbf{r},\omega_{\rm m}),
\end{equation}
which is the same as $S_{\mu \nu}$ in Eq.~\eqref{eq:S} but with an additional factor of $\omega^2/(\omega_{\mu}\omega_{\nu})$. This leads to an expansion of the transverse field $\hat{\mathbf{E}}^{\perp}_{\rm F}$ of (for $\mathbf{r}$ in the {\it system} region)
\begin{equation}
\hat{\mathbf{E}}^{\perp}_{\rm F}(\mathbf{r}) = i\sum_{\mu}\sqrt{\frac{\hbar \chi_{\mu\mu}}{2\epsilon_0}}\tilde{\mathbf{f}}^{s(\mathbf{E})}_{\mu}(\mathbf{r})\hat{a}_{\mu} + \text{H.c.},
\end{equation}
where
\begin{equation}
    \tilde{\mathbf{f}}^{s(\mathbf{E})}(\mathbf{r}) = \sum_{\nu} \left[S^{(\mathbf{E})}\right]^{\frac{1}{2}}_{\nu \mu} \tilde{\mathbf{f}}_{\nu}(\mathbf{r}) \sqrt{\frac{\omega_{\nu}}{\chi_{\mu \mu}}}.
\end{equation}

Under this definition, the dipole gauge Hamiltonian becomes
\begin{equation}
    \hat{\mathcal{H}}_{\rm d}^{\rm S} = \hat{\mathcal{H}}^{\rm em}_{\rm QNM,(\mathbf{E})} + \frac{\hbar\omega_0}{2}\hat{\sigma}_z+\hat{\mathcal{H}}_{\parallel}^{\rm S},
\end{equation}
\begin{equation}
    \hat{\mathcal{H}}_{\rm d}^{\rm SR}= \hat{\mathcal{H}}^{\rm em}_{\rm QNM-R,(\mathbf{E})} +\hat{B}_{\parallel}\hat{\sigma}_x,
\end{equation}
where in $\hat{\mathcal{H}}^{\rm em}_{\rm QNM,(\mathbf{E})} = \sum_{\mu \nu} \hbar \chi_{\mu \nu}\hat{a}^{\dagger}_{\mu}\hat{a}_{\nu}$ the $\chi$ matrix is now computed with $\mathbf{L}^{(\mathbf{E})}$, and $\hat{\mathcal{H}}^{\rm em}_{\rm QNM-R,(\mathbf{E})} = \sum_{\mu}\hat{C}^{(\mathbf{E})}_{\mu}\hat{a}^{\dagger}_{\mu} + \text{H.c.}$, with
\begin{equation}
 \hat{C}^{(\mathbf{E})}_{\mu} = \hbar \int d^3r \int d\omega_{\rm m} \mathbf{g}^{(\mathbf{E})}_{\mu}(\mathbf{r},\omega_{\rm m}) \cdot \hat{\mathbf{c}}(\mathbf{r},\omega),
\end{equation}
where
\begin{align}\label{eq:projection_fn_ge}
\mathbf{g}^{(\mathbf{E})}_{\mu}(\mathbf{r},\omega_{\rm m}) = \sum_{\nu} \left[S^{(\mathbf{E})}\right]^{-\frac{1}{2}}_{\mu \nu} \sqrt{\frac{2\epsilon_I(\mathbf{r},\omega_{\rm m})}{\pi \omega_{\nu}}}B_{\nu}({\omega_{\rm m}})\tilde{\mathbf{F}}'_{\nu}(\mathbf{r},\omega_{\rm m}),
\end{align}
which is the same as Eq.~\eqref{eq:projection_fn_g}, but with an additional factor of $\omega_{\rm m}/\omega_{\nu}$. Following this change through to the calculation of the QNM decay rate, we find
\begin{equation}
\Gamma^{ (\mathbf{E})}_{\alpha} = \kappa_c|c_{\alpha}^c|^2\frac{\omega_{\alpha}}{\omega_{c}}\zeta_c(\phi_0,\omega_{\alpha}),
\end{equation}
which is the same as $\Gamma_{\alpha}$ in Eq.~\eqref{eq:QNM_rate}, but with an additional factor of $\omega_{\alpha}^2/\omega_c^2$. This gives a spectral density of $\Lambda_{(\mathbf{E})}^2(\omega_{\alpha}) = \kappa_c \omega_{\alpha}/(2\pi\omega_c)$, in agreement with the finding in Ref.~\cite{Gustin2025} that when the bosonic cavity operators $\hat{a}_c$, $\hat{a}^{\dagger}_c$ couple directly to the reservoir field in the system-reservoir Hamiltonian, the resulting spectral density should scale with $\sim \omega$. 

Using these results, one could transform to the Coulomb gauge using the $\hat{\mathcal{W}}$ operators, with the vector potential expressed in terms of the $\mathbf{L}^{(\mathbf{E})}$ projector functions. 
This is however, much less convenient than using the vector potential definition of the quantized QNMs from the main text, as in that case the transformation $\hat{\mathcal{W}}$ is expressed in terms of the transverse vector potential. Using $\mathbf{L}^{(\mathbf{E})}$, the transverse vector potential must be expressed in terms of both quantized QNM operators and residual bath operators, the latter of which have a non-local commutation relation which must be carefully accounted for, and additional parameters similar to the $\chi$ matrix would have to be defined.

\section{Justification of the local bosonic commutator assumption}\label{app:boson}
In this part, we give justification for the local bosonic commutator approximation of the reservoir operators $\hat{\mathbf{c}}(\mathbf{r},\omega_{\rm m})$, which have the exact commutator, given by Eq.~\eqref{eq:commutator}.
The time derivative of $\hat{c}_i(\mathbf{r},\omega_{\rm m})$ with respect to the bare reservoir Hamiltonian (interaction picture evolution) is given by 
\begin{equation}
    -\frac{i}{\hbar}[\hat{c}_i(\mathbf{r},\omega_{\rm m}), H_{\rm R}]=-i\omega_{\rm m}\hat{c}_i(\mathbf{r},\omega_{\rm m})+i\sum_\nu L_{\nu,i}^*(\mathbf{r},\omega_{\rm m})\sum_j\int_{0}^\infty{d}\omega_{\rm m}'\int{d}^3r'~\omega_{\rm m}' L_{\nu,j}(\mathbf{r}',\omega_{\rm m}')\hat{c}_j(\mathbf{r}',\omega_{\rm m}').
\end{equation}
The first term is the usual bosonic contribution (free evolution), while the second term comes from the non-local commutator nature of $\hat{\mathbf{c}}(\mathbf{r},\omega_{\rm m})$.
Using the projection rule $\int_{0}^\infty{d}\omega_{\rm m}\int{d}^3r \mathbf{L}_{\nu}(\mathbf{r},\omega_{\rm m})\cdot \hat{\mathbf{c}}(\mathbf{r},\omega_{\rm m})=0$ (valid for all $\nu$), the time evolution of the (exact) reservoir operators in the interaction picture can be calculated as
\begin{equation}\label{eq:c_exact}
\hat{c}_i(\mathbf{r},\omega_{\rm m},t) = e^{-i \omega_{\rm m}(t-t_0)}\hat{c}_i(\mathbf{r},\omega_{\rm m}) + i\sum_\nu L_{\nu,i}^*(\mathbf{r},\omega_{\rm m})\sum_j\int d^3r' \int_0^{\infty} d\omega_{\rm m}' \int_{t_0}^{t}dt' g_{\nu,j}(\mathbf{r}',\omega_{\rm m}') \hat{c}_{j}(\mathbf{r}',\omega_{\rm m}',t')e^{-i\omega_{\rm m}(t-t')},
\end{equation}
and so
\begin{align}
    \hat{C}_\mu(t) =& \hbar\int d^3r \int_0^{\infty} d\omega_{\rm m} \mathbf{g}_\mu(\mathbf{r},\omega_{\rm m}) \cdot \hat{\mathbf{c}}(\mathbf{r},\omega_{\rm m}) e^{-i\omega_{\rm m}(t-t_0)} \nonumber\\
    &+ i \sum_{\nu \nu' \mu'} [S^{-1/2}]_{\mu\mu'}[S^{-1/2}]_{\nu'\nu} \int_{t_0}^{t}dt' \hat{C}_\nu(t') \int_0^{\infty} { d}\omega_{\rm m}\frac{\tilde{g}_{\mu'\nu'}(\omega_{\rm m})}{\omega_{\rm m}-\tilde{\omega}_{\nu'}^*} e^{-i\omega_{\rm m}(t-t')},
\end{align}
where $\hat{C}_\mu(t)=\hbar\int d^3r \int_0^{\infty} d\omega \mathbf{g}_{\mu}(\mathbf{r},\omega_{\rm m})\cdot\hat{\mathbf{c}}(\mathbf{r},\omega_{\rm m},t)$, and $g_{\mu\eta}(\omega_{\rm m})=\sum_{\mu'\nu'}\left[S^{-1/2}\right]_{\mu\mu'}\left[S^{-1/2}\right]_{\nu'\nu}\tilde{g}_{\mu'\nu'}(\omega_{\rm m})$.

Next, if we can make a Markov approximation $\tilde{g}_{\mu'\nu'}(\omega_{\rm m}) \approx\tilde{g}_{\mu'\nu'}$ (assuming $\tilde{g}_{\mu'\nu'}(\omega_{\rm m})$ to vary slowly over the scale of $\gamma_\mu,\gamma_\nu$, equivalent to a relatively high $Q$ approximation), and extend the frequency integral to $-\infty$, we can use the residue theorem to see
\begin{equation}
    \int_{-\infty}^{\infty} d\omega_{\rm m}\frac{e^{-i\omega_{\rm m}(t-t')}}{\omega_{\rm m}-\tilde{\omega}_{\nu'}^*} =0,
\end{equation}
since $t - t' > 0$, and $\tilde{\omega}_\nu^*$ is always located in the upper complex half plane. Thus, within the bounds of the Markov approximation,
\begin{equation}
\hat{C}_\mu(t)=\hbar\int d^3r \int_0^{\infty} d\omega \mathbf{g}_{\mu}(\mathbf{r},\omega_{\rm m})\cdot\hat{\mathbf{c}}(\mathbf{r},\omega_{\rm m})e^{-i\omega_{\rm m}(t-t_0)}.
\end{equation} 

Subsequently, we show that the trace over the reservoir subsystem in the master equation can also be evaluated as if the reservoir operators had local bosonic commutation relations.
We assume that the state of the reservoir is the vacuum state [which is implicitly defined from the modes of the universe operators $\hat{\mathbf{b}}(\mathbf{r},\omega_{\rm m})$], and evaluate:
\begin{align}
    \frac{1}{\hbar^2}\text{Tr}_{R}\left[\hat{C}_{\mu}(t)\hat{C}_{\nu}^{\dagger}(t-\tau)\rho_R\right] =& \int_0^{\infty} d\omega_{\rm m} g_{\mu\eta}(\omega_{\rm m})e^{-i \omega_{\rm m} \tau} \nonumber\\
    -\sum_{\mu' \nu' \mu'' \nu''} &\left[[S^{-1/2}]_{\mu\mu'}\int_0^{\infty} d\omega_{\rm m} \frac{\tilde{g}_{\mu'\nu'}(\omega_{\rm m})e^{-i\omega_{\rm m}(t-t_0)}}{\omega_{\rm m}-\tilde{\omega}_{\nu'}^*}\right]\nonumber\\
    &\times\left[[S^{-1/2}]_{\mu''\nu}[S^{-1}]_{\nu'\nu''}\int_0^{\infty} d\omega_{\rm m} \frac{\tilde{g}_{\nu''\mu''}(\omega_{\rm m})e^{i\omega_{\rm m}(t-\tau-t_0)}}{\omega_{\rm m}-\tilde{\omega}_{\nu''}}\right].
\end{align}
Making the Markov approximation again and extending the frequency integral bounds, the second term vanishes (within the same argumentation as further above), and we are left with the same result as if we assumed bosonic commutation relations.

The arguments developed in this section should also be applicable to the case where longitudinal interactions are included via $\hat{B}_{\parallel}$, although in this case some more assumptions on the specific form of $\mathbf{G}^{\parallel}(\mathbf{0},\mathbf{r},\omega_{\rm m})$ may be needed, in principle.

\end{widetext}

\begin{figure*}[htb]
    \centering
    \includegraphics[width = 1.69\columnwidth]{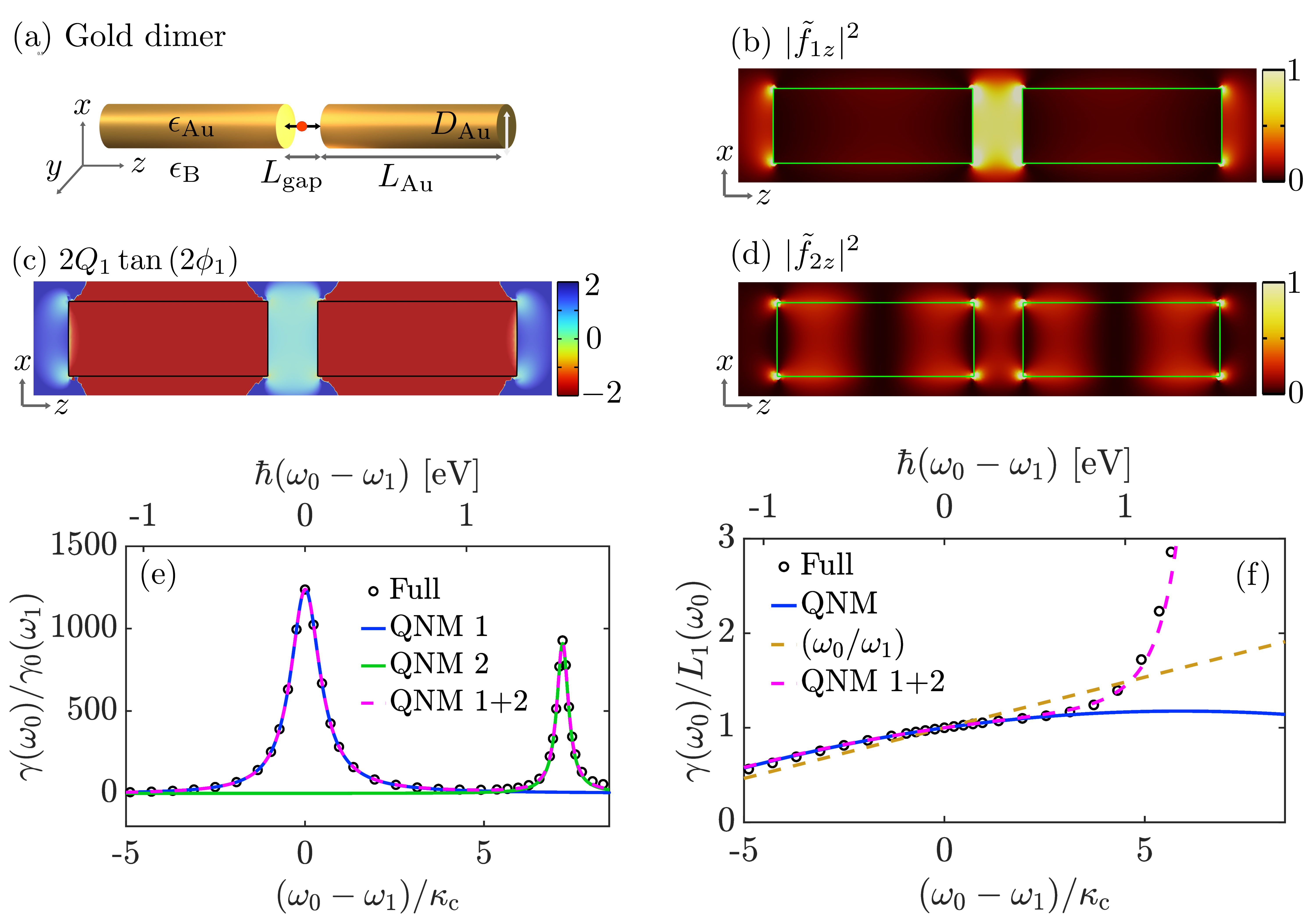}   
    \caption{
    (a) Schematic of gold cylinder dimer in free space ($\epsilon_{\rm B}=1.0$) with dimensions $D_{\rm Au}=30~$nm, $L_{\rm Au}=80~$nm, and $L_{\rm gap}=20~$nm. The dielectric function of the gold nanorods is governed by the Drude model Eq.~\eqref{eq: Drude}. QNM profile (b) dominant QNM (1) and (d) second QNM (2) (arb. units). (c) Phase distribution of $2Q_{\rm 1}\tan{(2\phi_1)}$ for dominant QNM, where the phase is defined as $\tilde{f}_{1z}(\mathbf{r})=|\tilde{f}_{1z}(\mathbf{r})|e^{i\phi_{1}(\mathbf{r})}$. (e) Purcell factors for a $z-$polarized emitter at the gap center, normalized by the free-space rate $\gamma_0(\omega)  =|\mathbf{d}|^2\omega^3/(3\pi \epsilon_0 \hbar c^3)$. (f) The decay rate normalized by  Lorentzian function $L_c(\omega_0)$. Here, we plot the full dipole decay rate with the background $\gamma_0(\omega_0)$ subtracted out. The dominant QNM (1) parameters are found in Table~\ref{tab:tab_appendix}, and QNM 2 has (complex) eigenfrequency $\hbar\tilde{\omega}_2 = (3.665 - 0.0438i)$ eV and projected phase at the dipole location $\phi_2(\mathbf{r}_0) = 0.000612$.
    }
    \label{fig: metal_dimer_r15h80g20}
\end{figure*}

\begin{figure*}[htb]
    \centering
    \includegraphics[width = 1.49\columnwidth]{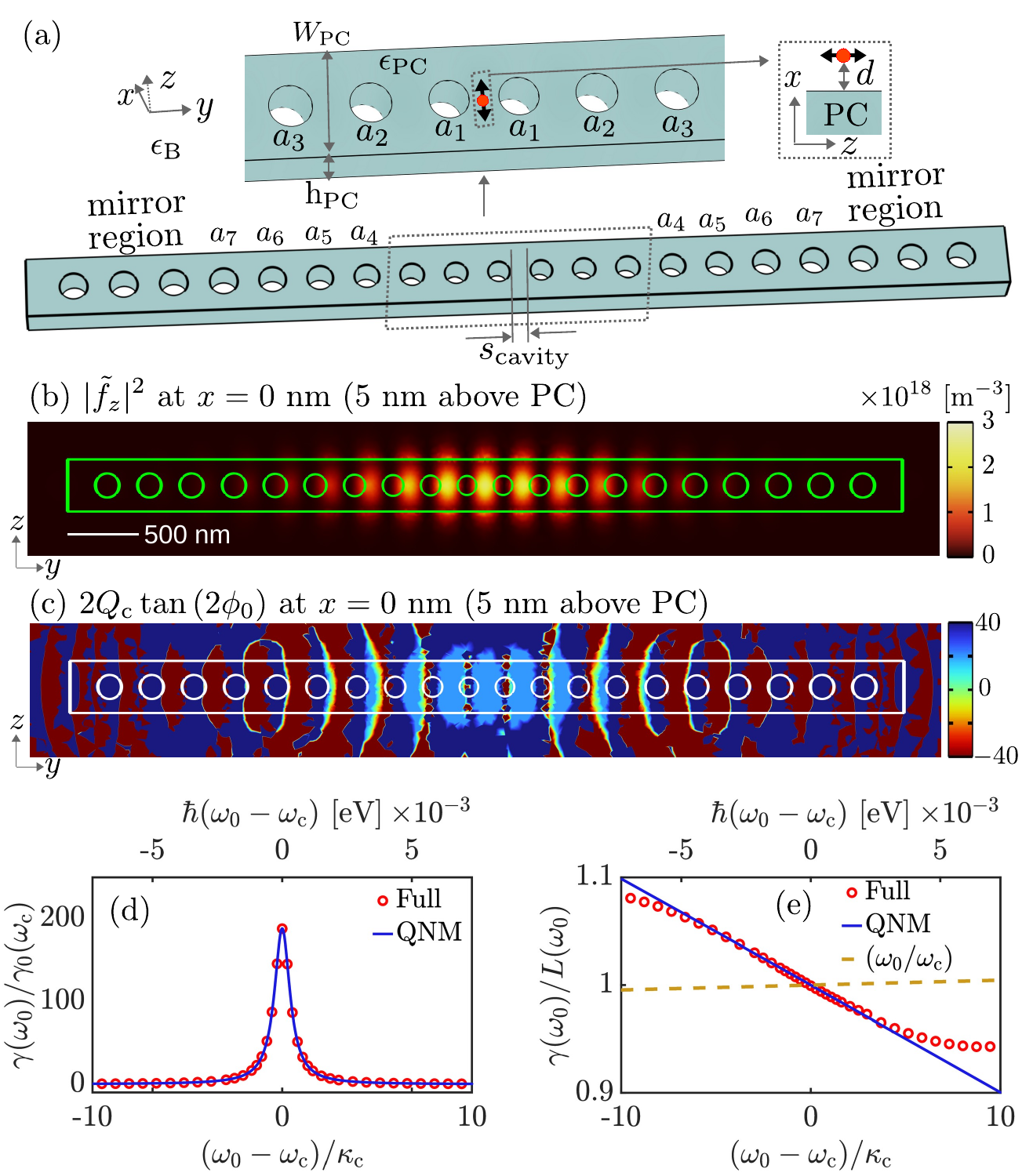}   
    \caption{(a) Schematic of PC beam cavity in free space ($\epsilon_{\rm B}=1$). The permittivity $\epsilon_{\rm PC}$ is governed by a single Lorentz oscillator model. (b) Mode profile of dominant QNM. (c) Phase distribution of $2Q_c\tan{(2\phi_0)}$. (d) Purcell factors for a $z-$polarized emitter $5$ nm above the cavity center, normalized by the free-space rate $\gamma_0(\omega)  =\mathbf{d}^2\omega^3/(3\pi \epsilon_0 \hbar c^3)$. (e) Decay rate normalized by Lorentzian function $L_c(\omega_0)$. Here, we plot the full dipole decay rate with an approximate background $1.78\gamma_0(\omega_0)$ subtracted out.
    }
    \label{fig: PC_t7m3}
\end{figure*}

\section{Perturbative analytic expression for linewidths under weak-excitation}\label{app:BS}

In this Appendix, we derive a perturbative solution for the linewidths of the intracavity spectrum under weak excitation. Our starting point is the Coulomb gauge system Hamiltonian of Eq.~\eqref{eq:HC_sys1mode}, expanded to order $|\eta_c|^2$ (here we let $\omega_0=\omega_c$ for simplicity, though this is not necessary),
\begin{align}
\hat{\mathcal{H}}_{\rm C}^{\rm S} = &\hbar\omega_0\hat{a}^{\dagger}_c\hat{a}_c + \frac{\hbar\omega_0}{2}\left[1 - |\eta_c|^2\left(\hat{a}_c+\hat{a}_c^{\dagger}\right)^2\right]\hat{\sigma}_z \nonumber \\ &+ \hbar \omega_0 |\eta_c|(\hat{a}_c+\hat{a}_c^{\dagger})\hat{\sigma}_x + \mathcal{O}(|\eta_c|^3),
\end{align}
and here we have performed a unitary transformation on the cavity and TLS operators to eliminate the phase of $\eta_c$, which does not change any of our results due to the phase-insensitivity of the cavity-reservoir coupling. We can perturbatively diagonalize this Hamiltonian by means of the Bloch-Siegert (BS) transformation~\cite{Salmon2022Mar}
\begin{equation}
\hat{U}_{\rm BS} = \text{exp}\left[\frac{|\eta_c|}{2}\left(\hat{a}_c\hat{\sigma}^- - \hat{a}^{\dagger}_c \hat{\sigma}^+\right)\right],
\end{equation}
and we find
\begin{align}\label{eq:HBS}
\hat{\mathcal{H}}_{\rm BS} &= \hat{U}^{\dagger}_{\rm BS} \hat{\mathcal{H}}_{\rm C}^{\rm S}\hat{U}_{\rm BS} \nonumber \\ & =  \hbar\omega_0\left[1 + \frac{3}{2}|\eta_c|^2\right]\hat{a}^{\dagger}_c\hat{a}_c + \hbar\omega_0\left[1 - \frac{3}{2}|\eta_c|^2\right]\hat{\sigma}^+\hat{\sigma}^- \nonumber \\ &+ \hbar\omega_0|\eta_c|\left(\hat{a}_c\hat{\sigma}^+ + \hat{a}^{\dagger}_c\hat{\sigma}^-\right),
\end{align}
and we have dropped both terms proportional to $\hat{a}_c^2$ and $\hat{a}_c^{\dagger 2}$, which can be removed by another transformation without any change in the Hamiltonian to order $|\eta_c|^2$, as well as terms diagonal in higher-order states beyond the lowest 2 excited energy levels~\cite{Salmon2022Mar,Hughes2024Jun}. We have also added a constant energy term to the Hamiltonian to set the ground state $\ket{G}$ energy at zero. By diagonalizing Eq.~\eqref{eq:HBS}, we obtain, for the two lowest-energy excited states,
\begin{subequations}
\begin{equation}
    E_{\pm} = \hbar\omega_0 \pm \hbar |\eta_c|\sqrt{1+\frac{9}{4}|\eta_c|^2},
\end{equation}
\begin{equation}
\ket{\pm}=\frac{1}{\sqrt{2}}\left(1\pm \frac{3}{4}|\eta_c|\right)\ket{1,g} + \frac{1}{\sqrt{2}}\left(\pm 1 - \frac{3}{4}|\eta_c|\right)
\ket{0,g},
\end{equation}
\end{subequations}
and the state vectors have been expanded to order $|\eta_c|$. In the limit of well-separated spectral peaks (strong coupling), the full width at half maxima are~\cite{Salmon2022Mar}
\begin{equation}
\Gamma_{\pm} = \frac{\Lambda^2(E_{\pm}/\hbar)}{2\pi}\bra{G}\hat{U}_{\rm BS}^{\dagger}\hat{a}_c \hat{U}_{\rm BS}\ket{\pm},
\end{equation}
which, upon expanding to leading order in $|\eta_c|$, gives the result from Eq.~\eqref{eq:lws}.
\section{Additional Resonator Examples}\label{app:additional}
In this Appendix, we show two more examples of resonator structures from which we can identify dominant QNMs and compare the weak-coupling dipole decay rate calculated under a QNM expansion with full Maxwell simulations. We first consider a cylindrical gold dimer setup similar to the one studied in Ref.~\cite{Gustin2025}, but with slightly different dimensions, where the dielectric function of the gold nanoparticle is also governed by the Drude model shown in Eq.~\eqref{eq: Drude}; this is shown in Fig.~\ref{fig: metal_dimer_r15h80g20}. The bare cavity mode has a radiative photon loss fraction of (using the spatially unspecified representation of the quantized QNMs) $S_c^{\rm rad}/S_c \approx 68$\%, with $S_c \approx 1$.

We then show a 3D PC beam cavity example in Fig.~\ref{fig: PC_t7m3}, also similar to a cavity shown in Ref.~\cite{Gustin2025,ren_near-field_2020,PhysRevA.95.013846,franke2020fluctuation}. The finite length of the PC beam cavity is $6.052~\mu$m, and the height/width of the beam are $h_{\rm PC}=200~$nm and $W_{\rm PC}=376~$nm. In the taper region, there are $7$ air holes ($a_1$, $a_2$, $a_3$, $a_4$, $a_5$, $a_6$, $a_7$), where the radius/center-to-center distance are increasing linearly from $68$/$264$~nm to $86$/$299$ nm.
In addition, $3$ identical air holes with a radius of $86$ nm and center-to-center spacing of $306$ nm are added in the mirror region. The surface-to-surface distance between two middle holes (two $a_1$) is $s_{\rm cavity}=126$ nm. The potential $z-$polarized dipole is placed at $d=5~$nm above the cavity center. The dielectric function $\epsilon_{\rm PC}(\omega)$ of the PC beam is also modeled by a single Lorentz oscillator model, similar to Eq.~\eqref{eq: Lorentz}, but with $\epsilon_{\rm s}=2.04^2=4.1616$, $\omega_{\rm t}=12$ eV and $\gamma_{\rm L}=0.0131$ eV.

\begin{table}
    \centering
    \begin{tabular}{|l||c|c|} \hline 
         & Cyl. Dimer & PC Beam \\ \hline 
          $\hbar\tilde{\omega}_c$ (eV) & $2.070 - 0.1108i$ & $1.598 - 0.0003662i$  \\ \hline 
          $Q_c$ & 9.34 & 2182  \\ \hline 
          \multirow{2}{*}{$\tilde{f}_z$ (m$^{-\frac{3}{2}}$)} & $9.009 \times 10^{10}$ & $1.557 \times 10^9$   \\
           & $+ 1.066 \times 10^9i$ & $+4.039\times 10^6i$  \\ \hline 
          $\tan{(2\phi_0)} \approx  2\phi_0$ & 0.0237 & 0.00519  \\ \hline
      $|\eta_c^{(1)}|$ & 2.3 & 0.044  \\ \hline $\tilde{\Omega}_{\rm BB}$ & 0.1 & 0.023 \\ \hline
    \end{tabular}
\caption{Parameters of dominant QNM for additional cavity structures.
}
\label{tab:tab_appendix}
\end{table}

For both examples, the parameters associated with the dominant QNMs are shown in Table~\ref{tab:tab_appendix}.
In both cases, our observations are similar to the ones made in both the main text and Ref.~\cite{Gustin2025}. The plasmonic dimer example, with its large $|\eta_c^{(1)}|$ should be compatible with our \emph{ab initio} quantized single-mode QNM theory in the USC regime, and its broadband regime coincides with USC for resonant coupling. The PC cavity, in contrast, we predict requires a multimode model to accurately predict spectral quantities in the USC regime. Moreover, as $\tilde{\Omega}_{\rm BB}Q_c \approx 50.2$, but deviations from the single-QNM theory occur beyond $\sim$ 10 linewidths from the resonance, single mode models are also insufficient to calculate accurate observables beyond phenomenological approaches for this dominant QNM.

\vspace{1cm}

\clearpage

\bibliography{references}

%apsrev4-2.bst 2019-01-14 (MD) hand-edited version of apsrev4-1.bst
%Control: key (0)
%Control: author (8) initials jnrlst
%Control: editor formatted (1) identically to author
%Control: production of article title (0) allowed
%Control: page (0) single
%Control: year (1) truncated
%Control: production of eprint (0) enabled
\begin{thebibliography}{127}%
\makeatletter
\providecommand \@ifxundefined [1]{%
 \@ifx{#1\undefined}
}%
\providecommand \@ifnum [1]{%
 \ifnum #1\expandafter \@firstoftwo
 \else \expandafter \@secondoftwo
 \fi
}%
\providecommand \@ifx [1]{%
 \ifx #1\expandafter \@firstoftwo
 \else \expandafter \@secondoftwo
 \fi
}%
\providecommand \natexlab [1]{#1}%
\providecommand \enquote  [1]{``#1''}%
\providecommand \bibnamefont  [1]{#1}%
\providecommand \bibfnamefont [1]{#1}%
\providecommand \citenamefont [1]{#1}%
\providecommand \href@noop [0]{\@secondoftwo}%
\providecommand \href [0]{\begingroup \@sanitize@url \@href}%
\providecommand \@href[1]{\@@startlink{#1}\@@href}%
\providecommand \@@href[1]{\endgroup#1\@@endlink}%
\providecommand \@sanitize@url [0]{\catcode `\\12\catcode `\$12\catcode `\&12\catcode `\#12\catcode `\^12\catcode `\_12\catcode `\%12\relax}%
\providecommand \@@startlink[1]{}%
\providecommand \@@endlink[0]{}%
\providecommand \url  [0]{\begingroup\@sanitize@url \@url }%
\providecommand \@url [1]{\endgroup\@href {#1}{\urlprefix }}%
\providecommand \urlprefix  [0]{URL }%
\providecommand \Eprint [0]{\href }%
\providecommand \doibase [0]{https://doi.org/}%
\providecommand \selectlanguage [0]{\@gobble}%
\providecommand \bibinfo  [0]{\@secondoftwo}%
\providecommand \bibfield  [0]{\@secondoftwo}%
\providecommand \translation [1]{[#1]}%
\providecommand \BibitemOpen [0]{}%
\providecommand \bibitemStop [0]{}%
\providecommand \bibitemNoStop [0]{.\EOS\space}%
\providecommand \EOS [0]{\spacefactor3000\relax}%
\providecommand \BibitemShut  [1]{\csname bibitem#1\endcsname}%
\let\auto@bib@innerbib\@empty
%</preamble>
\bibitem [{\citenamefont {Mabuchi}\ and\ \citenamefont {Doherty}(2002)}]{Mabuchi2002Nov}%
  \BibitemOpen
  \bibfield  {author} {\bibinfo {author} {\bibfnamefont {H.}~\bibnamefont {Mabuchi}}\ and\ \bibinfo {author} {\bibfnamefont {A.~C.}\ \bibnamefont {Doherty}},\ }\bibfield  {title} {\bibinfo {title} {{Cavity Quantum Electrodynamics: Coherence in Context}},\ }\href {https://doi.org/10.1126/science.1078446} {\bibfield  {journal} {\bibinfo  {journal} {Science}\ }\textbf {\bibinfo {volume} {298}},\ \bibinfo {pages} {1372} (\bibinfo {year} {2002})}\BibitemShut {NoStop}%
\bibitem [{\citenamefont {Browne}\ \emph {et~al.}(2017)\citenamefont {Browne}, \citenamefont {Bose}, \citenamefont {Mintert},\ and\ \citenamefont {Kim}}]{Browne2017}%
  \BibitemOpen
  \bibfield  {author} {\bibinfo {author} {\bibfnamefont {D.}~\bibnamefont {Browne}}, \bibinfo {author} {\bibfnamefont {S.}~\bibnamefont {Bose}}, \bibinfo {author} {\bibfnamefont {F.}~\bibnamefont {Mintert}},\ and\ \bibinfo {author} {\bibfnamefont {M.}~\bibnamefont {Kim}},\ }\bibfield  {title} {\bibinfo {title} {From quantum optics to quantum technologies},\ }\href {https://doi.org/10.1016/j.pquantelec.2017.06.002} {\bibfield  {journal} {\bibinfo  {journal} {Prog. Quantum Electron.}\ }\textbf {\bibinfo {volume} {54}},\ \bibinfo {pages} {2} (\bibinfo {year} {2017})}\BibitemShut {NoStop}%
\bibitem [{\citenamefont {Mandal}\ \emph {et~al.}(2023)\citenamefont {Mandal}, \citenamefont {Taylor}, \citenamefont {Weight}, \citenamefont {Koessler}, \citenamefont {Li},\ and\ \citenamefont {Huo}}]{doi:10.1021/acs.chemrev.2c00855}%
  \BibitemOpen
  \bibfield  {author} {\bibinfo {author} {\bibfnamefont {A.}~\bibnamefont {Mandal}}, \bibinfo {author} {\bibfnamefont {M.~A.}\ \bibnamefont {Taylor}}, \bibinfo {author} {\bibfnamefont {B.~M.}\ \bibnamefont {Weight}}, \bibinfo {author} {\bibfnamefont {E.~R.}\ \bibnamefont {Koessler}}, \bibinfo {author} {\bibfnamefont {X.}~\bibnamefont {Li}},\ and\ \bibinfo {author} {\bibfnamefont {P.}~\bibnamefont {Huo}},\ }\bibfield  {title} {\bibinfo {title} {{Theoretical Advances in Polariton Chemistry and Molecular Cavity Quantum Electrodynamics}},\ }\href {https://doi.org/10.1021/acs.chemrev.2c00855} {\bibfield  {journal} {\bibinfo  {journal} {Chem. Rev.}\ }\textbf {\bibinfo {volume} {123}},\ \bibinfo {pages} {9786} (\bibinfo {year} {2023})}\BibitemShut {NoStop}%
\bibitem [{\citenamefont {van Enk}\ \emph {et~al.}(2004)\citenamefont {van Enk}, \citenamefont {Kimble},\ and\ \citenamefont {Mabuchi}}]{van_enk_quantum_2004}%
  \BibitemOpen
  \bibfield  {author} {\bibinfo {author} {\bibfnamefont {S.~J.}\ \bibnamefont {van Enk}}, \bibinfo {author} {\bibfnamefont {H.~J.}\ \bibnamefont {Kimble}},\ and\ \bibinfo {author} {\bibfnamefont {H.}~\bibnamefont {Mabuchi}},\ }\bibfield  {title} {\bibinfo {title} {Quantum {Information} {Processing} in {Cavity}-{QED}},\ }\href {https://doi.org/10.1007/s11128-004-3104-2} {\bibfield  {journal} {\bibinfo  {journal} {Quantum Inf. Process.}\ }\textbf {\bibinfo {volume} {3}},\ \bibinfo {pages} {75} (\bibinfo {year} {2004})}\BibitemShut {NoStop}%
\bibitem [{\citenamefont {Metelmann}\ and\ \citenamefont {Clerk}(2015)}]{PhysRevX.5.021025}%
  \BibitemOpen
  \bibfield  {author} {\bibinfo {author} {\bibfnamefont {A.}~\bibnamefont {Metelmann}}\ and\ \bibinfo {author} {\bibfnamefont {A.~A.}\ \bibnamefont {Clerk}},\ }\bibfield  {title} {\bibinfo {title} {{Nonreciprocal Photon Transmission and Amplification via Reservoir Engineering}},\ }\href {https://doi.org/10.1103/PhysRevX.5.021025} {\bibfield  {journal} {\bibinfo  {journal} {Phys. Rev. X}\ }\textbf {\bibinfo {volume} {5}},\ \bibinfo {pages} {021025} (\bibinfo {year} {2015})}\BibitemShut {NoStop}%
\bibitem [{\citenamefont {Verstraete}\ \emph {et~al.}(2009)\citenamefont {Verstraete}, \citenamefont {Wolf},\ and\ \citenamefont {Ignacio~Cirac}}]{verstraete_quantum_2009}%
  \BibitemOpen
  \bibfield  {author} {\bibinfo {author} {\bibfnamefont {F.}~\bibnamefont {Verstraete}}, \bibinfo {author} {\bibfnamefont {M.~M.}\ \bibnamefont {Wolf}},\ and\ \bibinfo {author} {\bibfnamefont {J.}~\bibnamefont {Ignacio~Cirac}},\ }\bibfield  {title} {\bibinfo {title} {Quantum computation and quantum-state engineering driven by dissipation},\ }\href {https://doi.org/10.1038/nphys1342} {\bibfield  {journal} {\bibinfo  {journal} {Nat. Phys.}\ }\textbf {\bibinfo {volume} {5}},\ \bibinfo {pages} {633} (\bibinfo {year} {2009})}\BibitemShut {NoStop}%
\bibitem [{\citenamefont {González-Tudela}\ \emph {et~al.}(2024)\citenamefont {González-Tudela}, \citenamefont {Reiserer}, \citenamefont {García-Ripoll},\ and\ \citenamefont {García-Vidal}}]{gonzalez-tudela_lightmatter_2024}%
  \BibitemOpen
  \bibfield  {author} {\bibinfo {author} {\bibfnamefont {A.}~\bibnamefont {González-Tudela}}, \bibinfo {author} {\bibfnamefont {A.}~\bibnamefont {Reiserer}}, \bibinfo {author} {\bibfnamefont {J.~J.}\ \bibnamefont {García-Ripoll}},\ and\ \bibinfo {author} {\bibfnamefont {F.~J.}\ \bibnamefont {García-Vidal}},\ }\bibfield  {title} {\bibinfo {title} {Light–matter interactions in quantum nanophotonic devices},\ }\href {https://doi.org/10.1038/s42254-023-00681-1} {\bibfield  {journal} {\bibinfo  {journal} {Nat. Rev. Phys.}\ }\textbf {\bibinfo {volume} {6}},\ \bibinfo {pages} {166} (\bibinfo {year} {2024})}\BibitemShut {NoStop}%
\bibitem [{\citenamefont {Gardiner}\ and\ \citenamefont {Zoller}(2004)}]{gardiner_quantum_2004}%
  \BibitemOpen
  \bibfield  {author} {\bibinfo {author} {\bibfnamefont {C.}~\bibnamefont {Gardiner}}\ and\ \bibinfo {author} {\bibfnamefont {P.}~\bibnamefont {Zoller}},\ }\href@noop {} {\emph {\bibinfo {title} {Quantum {Noise}: {A} {Handbook} of {Markovian} and {Non}-{Markovian} {Quantum} {Stochastic} {Methods} with {Applications} to {Quantum} {Optics}}}}\ (\bibinfo  {publisher} {Springer Science \& Business Media},\ \bibinfo {year} {2004})\BibitemShut {NoStop}%
\bibitem [{\citenamefont {Carmichael}(013)}]{carmichael2013statistical}%
  \BibitemOpen
  \bibfield  {author} {\bibinfo {author} {\bibfnamefont {H.~J.}\ \bibnamefont {Carmichael}},\ }\href@noop {} {\emph {\bibinfo {title} {{Statistical Methods in Quantum Optics 1: Master Equations and Fokker-Planck Equations}}}}\ (\bibinfo  {publisher} {Springer Science \& Business Media},\ \bibinfo {year} {013})\BibitemShut {NoStop}%
\bibitem [{\citenamefont {Carmichael}(2009)}]{carmichael2009statistical}%
  \BibitemOpen
  \bibfield  {author} {\bibinfo {author} {\bibfnamefont {H.~J.}\ \bibnamefont {Carmichael}},\ }\href@noop {} {\emph {\bibinfo {title} {{Statistical Methods in Quantum Optics 2: Non-Classical Fields}}}}\ (\bibinfo  {publisher} {Springer Science \& Business Media},\ \bibinfo {year} {2009})\BibitemShut {NoStop}%
\bibitem [{\citenamefont {Breuer}\ and\ \citenamefont {Petruccione}(2002)}]{breuer2002theory}%
  \BibitemOpen
  \bibfield  {author} {\bibinfo {author} {\bibfnamefont {H.-P.}\ \bibnamefont {Breuer}}\ and\ \bibinfo {author} {\bibfnamefont {F.}~\bibnamefont {Petruccione}},\ }\href@noop {} {\emph {\bibinfo {title} {{The Theory of Open Quantum Systems}}}}\ (\bibinfo  {publisher} {Oxford University Press on Demand},\ \bibinfo {year} {2002})\BibitemShut {NoStop}%
\bibitem [{\citenamefont {Larson}\ and\ \citenamefont {Mavrogordatos}(2024)}]{larson_jaynescummings_2024}%
  \BibitemOpen
  \bibfield  {author} {\bibinfo {author} {\bibfnamefont {J.}~\bibnamefont {Larson}}\ and\ \bibinfo {author} {\bibfnamefont {T.}~\bibnamefont {Mavrogordatos}},\ }\href {https://iopscience-iop-org.stanford.idm.oclc.org/book/mono/978-0-7503-6452-2} {\emph {\bibinfo {title} {The {Jaynes}–{Cummings} {Model} and its {Descendants} ({Second} {Edition}): {Modern} research directions}}}\ (\bibinfo  {publisher} {IOP Publishing},\ \bibinfo {year} {2024})\BibitemShut {NoStop}%
\bibitem [{\citenamefont {Dicke}(1954)}]{PhysRev.93.99}%
  \BibitemOpen
  \bibfield  {author} {\bibinfo {author} {\bibfnamefont {R.~H.}\ \bibnamefont {Dicke}},\ }\bibfield  {title} {\bibinfo {title} {{Coherence in Spontaneous Radiation Processes}},\ }\href {https://doi.org/10.1103/PhysRev.93.99} {\bibfield  {journal} {\bibinfo  {journal} {Phys. Rev.}\ }\textbf {\bibinfo {volume} {93}},\ \bibinfo {pages} {99} (\bibinfo {year} {1954})}\BibitemShut {NoStop}%
\bibitem [{\citenamefont {Kirton}\ \emph {et~al.}(2019)\citenamefont {Kirton}, \citenamefont {Roses}, \citenamefont {Keeling},\ and\ \citenamefont {Dalla~Torre}}]{https://doi.org/10.1002/qute.201800043}%
  \BibitemOpen
  \bibfield  {author} {\bibinfo {author} {\bibfnamefont {P.}~\bibnamefont {Kirton}}, \bibinfo {author} {\bibfnamefont {M.~M.}\ \bibnamefont {Roses}}, \bibinfo {author} {\bibfnamefont {J.}~\bibnamefont {Keeling}},\ and\ \bibinfo {author} {\bibfnamefont {E.~G.}\ \bibnamefont {Dalla~Torre}},\ }\bibfield  {title} {\bibinfo {title} {{Introduction to the Dicke Model: From Equilibrium to Nonequilibrium, and Vice Versa}},\ }\href {https://advanced.onlinelibrary.wiley.com/doi/abs/10.1002/qute.201800043} {\bibfield  {journal} {\bibinfo  {journal} {Adv. Quantum Technol.}\ }\textbf {\bibinfo {volume} {2}},\ \bibinfo {pages} {1800043} (\bibinfo {year} {2019})}\BibitemShut {NoStop}%
\bibitem [{\citenamefont {Gardiner}\ and\ \citenamefont {Collett}(1985)}]{Gardiner1}%
  \BibitemOpen
  \bibfield  {author} {\bibinfo {author} {\bibfnamefont {C.~W.}\ \bibnamefont {Gardiner}}\ and\ \bibinfo {author} {\bibfnamefont {M.~J.}\ \bibnamefont {Collett}},\ }\bibfield  {title} {\bibinfo {title} {Input and output in damped quantum systems: Quantum stochastic differential equations and the master equation},\ }\href {https://doi.org/10.1103/PhysRevA.31.3761} {\bibfield  {journal} {\bibinfo  {journal} {Phys. Rev. A}\ }\textbf {\bibinfo {volume} {31}},\ \bibinfo {pages} {3761} (\bibinfo {year} {1985})}\BibitemShut {NoStop}%
\bibitem [{\citenamefont {Combes}\ \emph {et~al.}(2017)\citenamefont {Combes}, \citenamefont {Joseph},\ and\ \citenamefont {Sarovar}}]{combes_slh_2017}%
  \BibitemOpen
  \bibfield  {author} {\bibinfo {author} {\bibfnamefont {J.}~\bibnamefont {Combes}}, \bibinfo {author} {\bibfnamefont {K.}~\bibnamefont {Joseph}},\ and\ \bibinfo {author} {\bibfnamefont {M.}~\bibnamefont {Sarovar}},\ }\bibfield  {title} {\bibinfo {title} {The {SLH} framework for modeling quantum input-output networks},\ }\href {https://doi.org/10.1080/23746149.2017.1343097} {\bibfield  {journal} {\bibinfo  {journal} {Adv. Phys.: X}\ }\textbf {\bibinfo {volume} {2}},\ \bibinfo {pages} {784} (\bibinfo {year} {2017})}\BibitemShut {NoStop}%
\bibitem [{\citenamefont {Lindblad}(1976)}]{lindblad_generators_1976}%
  \BibitemOpen
  \bibfield  {author} {\bibinfo {author} {\bibfnamefont {G.}~\bibnamefont {Lindblad}},\ }\bibfield  {title} {\bibinfo {title} {On the generators of quantum dynamical semigroups},\ }\href {https://projecteuclid.org/journals/communications-in-mathematical-physics/volume-48/issue-2/On-the-generators-of-quantum-dynamical-semigroups/cmp/1103899849.full} {\bibfield  {journal} {\bibinfo  {journal} {Communications in Mathematical Physics}\ }\textbf {\bibinfo {volume} {48}},\ \bibinfo {pages} {119} (\bibinfo {year} {1976})}\BibitemShut {NoStop}%
\bibitem [{\citenamefont {Gustin}\ \emph {et~al.}(2025)\citenamefont {Gustin}, \citenamefont {Ren},\ and\ \citenamefont {Hughes}}]{Gustin2025}%
  \BibitemOpen
  \bibfield  {author} {\bibinfo {author} {\bibfnamefont {C.}~\bibnamefont {Gustin}}, \bibinfo {author} {\bibfnamefont {J.}~\bibnamefont {Ren}},\ and\ \bibinfo {author} {\bibfnamefont {S.}~\bibnamefont {Hughes}},\ }\bibfield  {title} {\bibinfo {title} {{What Is the Spectral Density of the Reservoir for a Lossy Quantized Cavity?}},\ }\href {https://doi.org/10.1103/PhysRevLett.134.123601} {\bibfield  {journal} {\bibinfo  {journal} {Phys. Rev. Lett.}\ }\textbf {\bibinfo {volume} {134}},\ \bibinfo {pages} {123601} (\bibinfo {year} {2025})}\BibitemShut {NoStop}%
\bibitem [{\citenamefont {Forn-D{\'i}az}\ \emph {et~al.}(2019)\citenamefont {Forn-D{\'i}az}, \citenamefont {Lamata}, \citenamefont {Rico}, \citenamefont {Kono},\ and\ \citenamefont {Solano}}]{forn-diaz_ultrastrong_2019}%
  \BibitemOpen
  \bibfield  {author} {\bibinfo {author} {\bibfnamefont {P.}~\bibnamefont {Forn-D{\'i}az}}, \bibinfo {author} {\bibfnamefont {L.}~\bibnamefont {Lamata}}, \bibinfo {author} {\bibfnamefont {E.}~\bibnamefont {Rico}}, \bibinfo {author} {\bibfnamefont {J.}~\bibnamefont {Kono}},\ and\ \bibinfo {author} {\bibfnamefont {E.}~\bibnamefont {Solano}},\ }\bibfield  {title} {\bibinfo {title} {Ultrastrong coupling regimes of light-matter interaction},\ }\href {https://doi.org/10.1103/RevModPhys.91.025005} {\bibfield  {journal} {\bibinfo  {journal} {Rev. Mod. Phys.}\ }\textbf {\bibinfo {volume} {91}},\ \bibinfo {pages} {025005} (\bibinfo {year} {2019})}\BibitemShut {NoStop}%
\bibitem [{\citenamefont {Forn-D\'{\i}az}\ \emph {et~al.}(2019)\citenamefont {Forn-D\'{\i}az}, \citenamefont {Lamata}, \citenamefont {Rico}, \citenamefont {Kono},\ and\ \citenamefont {Solano}}]{RevModPhys.91.025005}%
  \BibitemOpen
  \bibfield  {author} {\bibinfo {author} {\bibfnamefont {P.}~\bibnamefont {Forn-D\'{\i}az}}, \bibinfo {author} {\bibfnamefont {L.}~\bibnamefont {Lamata}}, \bibinfo {author} {\bibfnamefont {E.}~\bibnamefont {Rico}}, \bibinfo {author} {\bibfnamefont {J.}~\bibnamefont {Kono}},\ and\ \bibinfo {author} {\bibfnamefont {E.}~\bibnamefont {Solano}},\ }\bibfield  {title} {\bibinfo {title} {Ultrastrong coupling regimes of light-matter interaction},\ }\href {https://doi.org/10.1103/RevModPhys.91.025005} {\bibfield  {journal} {\bibinfo  {journal} {Rev. Mod. Phys.}\ }\textbf {\bibinfo {volume} {91}},\ \bibinfo {pages} {025005} (\bibinfo {year} {2019})}\BibitemShut {NoStop}%
\bibitem [{\citenamefont {Frisk~Kockum}\ \emph {et~al.}(2019)\citenamefont {Frisk~Kockum}, \citenamefont {Miranowicz}, \citenamefont {De~Liberato}, \citenamefont {Savasta},\ and\ \citenamefont {Nori}}]{frisk_kockum_ultrastrong_2019}%
  \BibitemOpen
  \bibfield  {author} {\bibinfo {author} {\bibfnamefont {A.}~\bibnamefont {Frisk~Kockum}}, \bibinfo {author} {\bibfnamefont {A.}~\bibnamefont {Miranowicz}}, \bibinfo {author} {\bibfnamefont {S.}~\bibnamefont {De~Liberato}}, \bibinfo {author} {\bibfnamefont {S.}~\bibnamefont {Savasta}},\ and\ \bibinfo {author} {\bibfnamefont {F.}~\bibnamefont {Nori}},\ }\bibfield  {title} {\bibinfo {title} {Ultrastrong coupling between light and matter},\ }\href {https://doi.org/10.1038/s42254-018-0006-2} {\bibfield  {journal} {\bibinfo  {journal} {Nat. Rev. Phys.}\ }\textbf {\bibinfo {volume} {1}},\ \bibinfo {pages} {19} (\bibinfo {year} {2019})}\BibitemShut {NoStop}%
\bibitem [{\citenamefont {Le~Boit{\ifmmode\acute{e}\else\'{e}\fi}}(2020)}]{LeBoite2020Jul}%
  \BibitemOpen
  \bibfield  {author} {\bibinfo {author} {\bibfnamefont {A.}~\bibnamefont {Le~Boit{\ifmmode\acute{e}\else\'{e}\fi}}},\ }\bibfield  {title} {\bibinfo {title} {{Theoretical Methods for Ultrastrong Light{\textendash}Matter Interactions}},\ }\href {https://doi.org/10.1002/qute.201900140} {\bibfield  {journal} {\bibinfo  {journal} {Adv. Quantum Technol.}\ }\textbf {\bibinfo {volume} {3}},\ \bibinfo {pages} {1900140} (\bibinfo {year} {2020})}\BibitemShut {NoStop}%
\bibitem [{\citenamefont {Purcell}\ \emph {et~al.}(1946)\citenamefont {Purcell}, \citenamefont {Torrey},\ and\ \citenamefont {Pound}}]{purcell1946resonance}%
  \BibitemOpen
  \bibfield  {author} {\bibinfo {author} {\bibfnamefont {E.~M.}\ \bibnamefont {Purcell}}, \bibinfo {author} {\bibfnamefont {H.~C.}\ \bibnamefont {Torrey}},\ and\ \bibinfo {author} {\bibfnamefont {R.~V.}\ \bibnamefont {Pound}},\ }\bibfield  {title} {\bibinfo {title} {{Resonance Absorption by Nuclear Magnetic Moments in a Solid}},\ }\href {https://doi.org/10.1103/PhysRev.69.37} {\bibfield  {journal} {\bibinfo  {journal} {Phys. Rev.}\ }\textbf {\bibinfo {volume} {69}},\ \bibinfo {pages} {37} (\bibinfo {year} {1946})}\BibitemShut {NoStop}%
\bibitem [{\citenamefont {Ciuti}\ \emph {et~al.}(2005)\citenamefont {Ciuti}, \citenamefont {Bastard},\ and\ \citenamefont {Carusotto}}]{ciuti_quantum_2005}%
  \BibitemOpen
  \bibfield  {author} {\bibinfo {author} {\bibfnamefont {C.}~\bibnamefont {Ciuti}}, \bibinfo {author} {\bibfnamefont {G.}~\bibnamefont {Bastard}},\ and\ \bibinfo {author} {\bibfnamefont {I.}~\bibnamefont {Carusotto}},\ }\bibfield  {title} {\bibinfo {title} {Quantum vacuum properties of the intersubband cavity polariton field},\ }\href {https://doi.org/10.1103/PhysRevB.72.115303} {\bibfield  {journal} {\bibinfo  {journal} {Phys. Rev. B}\ }\textbf {\bibinfo {volume} {72}},\ \bibinfo {pages} {115303} (\bibinfo {year} {2005})}\BibitemShut {NoStop}%
\bibitem [{\citenamefont {Anappara}\ \emph {et~al.}(2009)\citenamefont {Anappara}, \citenamefont {De~Liberato}, \citenamefont {Tredicucci}, \citenamefont {Ciuti}, \citenamefont {Biasiol}, \citenamefont {Sorba},\ and\ \citenamefont {Beltram}}]{anappara_signatures_2009}%
  \BibitemOpen
  \bibfield  {author} {\bibinfo {author} {\bibfnamefont {A.~A.}\ \bibnamefont {Anappara}}, \bibinfo {author} {\bibfnamefont {S.}~\bibnamefont {De~Liberato}}, \bibinfo {author} {\bibfnamefont {A.}~\bibnamefont {Tredicucci}}, \bibinfo {author} {\bibfnamefont {C.}~\bibnamefont {Ciuti}}, \bibinfo {author} {\bibfnamefont {G.}~\bibnamefont {Biasiol}}, \bibinfo {author} {\bibfnamefont {L.}~\bibnamefont {Sorba}},\ and\ \bibinfo {author} {\bibfnamefont {F.}~\bibnamefont {Beltram}},\ }\bibfield  {title} {\bibinfo {title} {Signatures of the ultrastrong light-matter coupling regime},\ }\href {https://doi.org/10.1103/PhysRevB.79.201303} {\bibfield  {journal} {\bibinfo  {journal} {Phys. Rev. B}\ }\textbf {\bibinfo {volume} {79}},\ \bibinfo {pages} {201303(R)} (\bibinfo {year} {2009})}\BibitemShut {NoStop}%
\bibitem [{\citenamefont {Zaks}\ \emph {et~al.}(2011)\citenamefont {Zaks}, \citenamefont {Stehr}, \citenamefont {Truong}, \citenamefont {Petroff}, \citenamefont {Hughes},\ and\ \citenamefont {Sherwin}}]{zaks_thz-driven_2011}%
  \BibitemOpen
  \bibfield  {author} {\bibinfo {author} {\bibfnamefont {B.}~\bibnamefont {Zaks}}, \bibinfo {author} {\bibfnamefont {D.}~\bibnamefont {Stehr}}, \bibinfo {author} {\bibfnamefont {T.-A.}\ \bibnamefont {Truong}}, \bibinfo {author} {\bibfnamefont {P.~M.}\ \bibnamefont {Petroff}}, \bibinfo {author} {\bibfnamefont {S.}~\bibnamefont {Hughes}},\ and\ \bibinfo {author} {\bibfnamefont {M.~S.}\ \bibnamefont {Sherwin}},\ }\bibfield  {title} {\bibinfo {title} {{THz}-driven quantum wells: {Coulomb} interactions and {Stark} shifts in the ultrastrong coupling regime},\ }\href {https://doi.org/10.1088/1367-2630/13/8/083009} {\bibfield  {journal} {\bibinfo  {journal} {New J. Phys.}\ }\textbf {\bibinfo {volume} {13}},\ \bibinfo {pages} {083009} (\bibinfo {year} {2011})}\BibitemShut {NoStop}%
\bibitem [{\citenamefont {Scarlino}\ \emph {et~al.}(2022)\citenamefont {Scarlino}, \citenamefont {Ungerer}, \citenamefont {van Woerkom}, \citenamefont {Mancini}, \citenamefont {Stano}, \citenamefont {M\"uller}, \citenamefont {Landig}, \citenamefont {Koski}, \citenamefont {Reichl}, \citenamefont {Wegscheider}, \citenamefont {Ihn}, \citenamefont {Ensslin},\ and\ \citenamefont {Wallraff}}]{PhysRevX.12.031004}%
  \BibitemOpen
  \bibfield  {author} {\bibinfo {author} {\bibfnamefont {P.}~\bibnamefont {Scarlino}}, \bibinfo {author} {\bibfnamefont {J.~H.}\ \bibnamefont {Ungerer}}, \bibinfo {author} {\bibfnamefont {D.~J.}\ \bibnamefont {van Woerkom}}, \bibinfo {author} {\bibfnamefont {M.}~\bibnamefont {Mancini}}, \bibinfo {author} {\bibfnamefont {P.}~\bibnamefont {Stano}}, \bibinfo {author} {\bibfnamefont {C.}~\bibnamefont {M\"uller}}, \bibinfo {author} {\bibfnamefont {A.~J.}\ \bibnamefont {Landig}}, \bibinfo {author} {\bibfnamefont {J.~V.}\ \bibnamefont {Koski}}, \bibinfo {author} {\bibfnamefont {C.}~\bibnamefont {Reichl}}, \bibinfo {author} {\bibfnamefont {W.}~\bibnamefont {Wegscheider}}, \bibinfo {author} {\bibfnamefont {T.}~\bibnamefont {Ihn}}, \bibinfo {author} {\bibfnamefont {K.}~\bibnamefont {Ensslin}},\ and\ \bibinfo {author} {\bibfnamefont {A.}~\bibnamefont {Wallraff}},\ }\bibfield  {title} {\bibinfo {title} {{In situ Tuning of the Electric-Dipole Strength of a Double-Dot Charge Qubit: Charge-Noise Protection and Ultrastrong
  Coupling}},\ }\href {https://doi.org/10.1103/PhysRevX.12.031004} {\bibfield  {journal} {\bibinfo  {journal} {Phys. Rev. X}\ }\textbf {\bibinfo {volume} {12}},\ \bibinfo {pages} {031004} (\bibinfo {year} {2022})}\BibitemShut {NoStop}%
\bibitem [{\citenamefont {Scalari}\ \emph {et~al.}(2012)\citenamefont {Scalari}, \citenamefont {Maissen}, \citenamefont {Turčinková}, \citenamefont {Hagenmüller}, \citenamefont {De~Liberato}, \citenamefont {Ciuti}, \citenamefont {Reichl}, \citenamefont {Schuh}, \citenamefont {Wegscheider}, \citenamefont {Beck},\ and\ \citenamefont {Faist}}]{scalari_ultrastrong_2012}%
  \BibitemOpen
  \bibfield  {author} {\bibinfo {author} {\bibfnamefont {G.}~\bibnamefont {Scalari}}, \bibinfo {author} {\bibfnamefont {C.}~\bibnamefont {Maissen}}, \bibinfo {author} {\bibfnamefont {D.}~\bibnamefont {Turčinková}}, \bibinfo {author} {\bibfnamefont {D.}~\bibnamefont {Hagenmüller}}, \bibinfo {author} {\bibfnamefont {S.}~\bibnamefont {De~Liberato}}, \bibinfo {author} {\bibfnamefont {C.}~\bibnamefont {Ciuti}}, \bibinfo {author} {\bibfnamefont {C.}~\bibnamefont {Reichl}}, \bibinfo {author} {\bibfnamefont {D.}~\bibnamefont {Schuh}}, \bibinfo {author} {\bibfnamefont {W.}~\bibnamefont {Wegscheider}}, \bibinfo {author} {\bibfnamefont {M.}~\bibnamefont {Beck}},\ and\ \bibinfo {author} {\bibfnamefont {J.}~\bibnamefont {Faist}},\ }\bibfield  {title} {\bibinfo {title} {Ultrastrong {Coupling} of the {Cyclotron} {Transition} of a {2D} {Electron} {Gas} to a {THz} {Metamaterial}},\ }\href {https://doi.org/10.1126/science.1216022} {\bibfield  {journal} {\bibinfo  {journal} {Science}\ }\textbf {\bibinfo {volume} {335}},\
  \bibinfo {pages} {1323} (\bibinfo {year} {2012})}\BibitemShut {NoStop}%
\bibitem [{\citenamefont {Rajabali}\ \emph {et~al.}(2022)\citenamefont {Rajabali}, \citenamefont {Markmann}, \citenamefont {Jöchl}, \citenamefont {Beck}, \citenamefont {Lehner}, \citenamefont {Wegscheider}, \citenamefont {Faist},\ and\ \citenamefont {Scalari}}]{rajabali_ultrastrongly_2022}%
  \BibitemOpen
  \bibfield  {author} {\bibinfo {author} {\bibfnamefont {S.}~\bibnamefont {Rajabali}}, \bibinfo {author} {\bibfnamefont {S.}~\bibnamefont {Markmann}}, \bibinfo {author} {\bibfnamefont {E.}~\bibnamefont {Jöchl}}, \bibinfo {author} {\bibfnamefont {M.}~\bibnamefont {Beck}}, \bibinfo {author} {\bibfnamefont {C.~A.}\ \bibnamefont {Lehner}}, \bibinfo {author} {\bibfnamefont {W.}~\bibnamefont {Wegscheider}}, \bibinfo {author} {\bibfnamefont {J.}~\bibnamefont {Faist}},\ and\ \bibinfo {author} {\bibfnamefont {G.}~\bibnamefont {Scalari}},\ }\bibfield  {title} {\bibinfo {title} {An ultrastrongly coupled single terahertz meta-atom},\ }\href {https://doi.org/10.1038/s41467-022-29974-2} {\bibfield  {journal} {\bibinfo  {journal} {Nat. Commun.}\ }\textbf {\bibinfo {volume} {13}},\ \bibinfo {pages} {2528} (\bibinfo {year} {2022})}\BibitemShut {NoStop}%
\bibitem [{\citenamefont {Tay}\ \emph {et~al.}(2025)\citenamefont {Tay}, \citenamefont {Mojibpour}, \citenamefont {Sanders}, \citenamefont {Liang}, \citenamefont {Xu}, \citenamefont {Gardner}, \citenamefont {Baydin}, \citenamefont {Manfra}, \citenamefont {Alabastri}, \citenamefont {Hagenmüller},\ and\ \citenamefont {Kono}}]{tay_multimode_2025}%
  \BibitemOpen
  \bibfield  {author} {\bibinfo {author} {\bibfnamefont {F.}~\bibnamefont {Tay}}, \bibinfo {author} {\bibfnamefont {A.}~\bibnamefont {Mojibpour}}, \bibinfo {author} {\bibfnamefont {S.}~\bibnamefont {Sanders}}, \bibinfo {author} {\bibfnamefont {S.}~\bibnamefont {Liang}}, \bibinfo {author} {\bibfnamefont {H.}~\bibnamefont {Xu}}, \bibinfo {author} {\bibfnamefont {G.~C.}\ \bibnamefont {Gardner}}, \bibinfo {author} {\bibfnamefont {A.}~\bibnamefont {Baydin}}, \bibinfo {author} {\bibfnamefont {M.~J.}\ \bibnamefont {Manfra}}, \bibinfo {author} {\bibfnamefont {A.}~\bibnamefont {Alabastri}}, \bibinfo {author} {\bibfnamefont {D.}~\bibnamefont {Hagenmüller}},\ and\ \bibinfo {author} {\bibfnamefont {J.}~\bibnamefont {Kono}},\ }\bibfield  {title} {\bibinfo {title} {Multimode ultrastrong coupling in three-dimensional photonic-crystal cavities},\ }\href {https://doi.org/10.1038/s41467-025-58835-x} {\bibfield  {journal} {\bibinfo  {journal} {Nat. Commun.}\ }\textbf {\bibinfo {volume} {16}},\ \bibinfo {pages} {3603} (\bibinfo
  {year} {2025})}\BibitemShut {NoStop}%
\bibitem [{\citenamefont {Li}\ \emph {et~al.}(2018)\citenamefont {Li}, \citenamefont {Bamba}, \citenamefont {Zhang}, \citenamefont {Fallahi}, \citenamefont {Gardner}, \citenamefont {Gao}, \citenamefont {Lou}, \citenamefont {Yoshioka}, \citenamefont {Manfra},\ and\ \citenamefont {Kono}}]{li_vacuum_2018}%
  \BibitemOpen
  \bibfield  {author} {\bibinfo {author} {\bibfnamefont {X.}~\bibnamefont {Li}}, \bibinfo {author} {\bibfnamefont {M.}~\bibnamefont {Bamba}}, \bibinfo {author} {\bibfnamefont {Q.}~\bibnamefont {Zhang}}, \bibinfo {author} {\bibfnamefont {S.}~\bibnamefont {Fallahi}}, \bibinfo {author} {\bibfnamefont {G.~C.}\ \bibnamefont {Gardner}}, \bibinfo {author} {\bibfnamefont {W.}~\bibnamefont {Gao}}, \bibinfo {author} {\bibfnamefont {M.}~\bibnamefont {Lou}}, \bibinfo {author} {\bibfnamefont {K.}~\bibnamefont {Yoshioka}}, \bibinfo {author} {\bibfnamefont {M.~J.}\ \bibnamefont {Manfra}},\ and\ \bibinfo {author} {\bibfnamefont {J.}~\bibnamefont {Kono}},\ }\bibfield  {title} {\bibinfo {title} {Vacuum {Bloch}–{Siegert} shift in {Landau} polaritons with ultra-high cooperativity},\ }\href {https://doi.org/10.1038/s41566-018-0153-0} {\bibfield  {journal} {\bibinfo  {journal} {Nat. Photon.}\ }\textbf {\bibinfo {volume} {12}},\ \bibinfo {pages} {324} (\bibinfo {year} {2018})}\BibitemShut {NoStop}%
\bibitem [{\citenamefont {Mueller}\ \emph {et~al.}(2020)\citenamefont {Mueller}, \citenamefont {Okamura}, \citenamefont {Vieira}, \citenamefont {Juergensen}, \citenamefont {Lange}, \citenamefont {Barros}, \citenamefont {Schulz},\ and\ \citenamefont {Reich}}]{mueller_deep_2020}%
  \BibitemOpen
  \bibfield  {author} {\bibinfo {author} {\bibfnamefont {N.~S.}\ \bibnamefont {Mueller}}, \bibinfo {author} {\bibfnamefont {Y.}~\bibnamefont {Okamura}}, \bibinfo {author} {\bibfnamefont {B.~G.~M.}\ \bibnamefont {Vieira}}, \bibinfo {author} {\bibfnamefont {S.}~\bibnamefont {Juergensen}}, \bibinfo {author} {\bibfnamefont {H.}~\bibnamefont {Lange}}, \bibinfo {author} {\bibfnamefont {E.~B.}\ \bibnamefont {Barros}}, \bibinfo {author} {\bibfnamefont {F.}~\bibnamefont {Schulz}},\ and\ \bibinfo {author} {\bibfnamefont {S.}~\bibnamefont {Reich}},\ }\bibfield  {title} {\bibinfo {title} {Deep strong light–matter coupling in plasmonic nanoparticle crystals},\ }\href {https://doi.org/10.1038/s41586-020-2508-1} {\bibfield  {journal} {\bibinfo  {journal} {Nature}\ }\textbf {\bibinfo {volume} {583}},\ \bibinfo {pages} {780} (\bibinfo {year} {2020})}\BibitemShut {NoStop}%
\bibitem [{\citenamefont {Hu}\ \emph {et~al.}(2024)\citenamefont {Hu}, \citenamefont {Huang}, \citenamefont {Arul}, \citenamefont {Sánchez-Iglesias}, \citenamefont {Xiong}, \citenamefont {Liz-Marzán},\ and\ \citenamefont {Baumberg}}]{hu_robust_2024}%
  \BibitemOpen
  \bibfield  {author} {\bibinfo {author} {\bibfnamefont {S.}~\bibnamefont {Hu}}, \bibinfo {author} {\bibfnamefont {J.}~\bibnamefont {Huang}}, \bibinfo {author} {\bibfnamefont {R.}~\bibnamefont {Arul}}, \bibinfo {author} {\bibfnamefont {A.}~\bibnamefont {Sánchez-Iglesias}}, \bibinfo {author} {\bibfnamefont {Y.}~\bibnamefont {Xiong}}, \bibinfo {author} {\bibfnamefont {L.~M.}\ \bibnamefont {Liz-Marzán}},\ and\ \bibinfo {author} {\bibfnamefont {J.~J.}\ \bibnamefont {Baumberg}},\ }\bibfield  {title} {\bibinfo {title} {Robust consistent single quantum dot strong coupling in plasmonic nanocavities},\ }\href {https://doi.org/10.1038/s41467-024-51170-7} {\bibfield  {journal} {\bibinfo  {journal} {Nat. Commun.}\ }\textbf {\bibinfo {volume} {15}},\ \bibinfo {pages} {6835} (\bibinfo {year} {2024})}\BibitemShut {NoStop}%
\bibitem [{\citenamefont {Chang}\ \emph {et~al.}(2025)\citenamefont {Chang}, \citenamefont {Roman}, \citenamefont {Paul}, \citenamefont {Sakotic}, \citenamefont {Vora}, \citenamefont {Kim}, \citenamefont {Hurst}, \citenamefont {Wasserman}, \citenamefont {Truskett},\ and\ \citenamefont {Milliron}}]{chang_ultrastrong_2025}%
  \BibitemOpen
  \bibfield  {author} {\bibinfo {author} {\bibfnamefont {W.~J.}\ \bibnamefont {Chang}}, \bibinfo {author} {\bibfnamefont {B.~J.}\ \bibnamefont {Roman}}, \bibinfo {author} {\bibfnamefont {T.}~\bibnamefont {Paul}}, \bibinfo {author} {\bibfnamefont {Z.}~\bibnamefont {Sakotic}}, \bibinfo {author} {\bibfnamefont {P.}~\bibnamefont {Vora}}, \bibinfo {author} {\bibfnamefont {K.}~\bibnamefont {Kim}}, \bibinfo {author} {\bibfnamefont {L.~E.}\ \bibnamefont {Hurst}}, \bibinfo {author} {\bibfnamefont {D.}~\bibnamefont {Wasserman}}, \bibinfo {author} {\bibfnamefont {T.~M.}\ \bibnamefont {Truskett}},\ and\ \bibinfo {author} {\bibfnamefont {D.~J.}\ \bibnamefont {Milliron}},\ }\bibfield  {title} {\bibinfo {title} {Ultrastrong {Coupling} by {Assembling} {Plasmonic} {Metal} {Oxide} {Nanocrystals} in {Open} {Cavities}},\ }\href {https://doi.org/10.1021/acsnano.5c01913} {\bibfield  {journal} {\bibinfo  {journal} {ACS Nano}\ }\textbf {\bibinfo {volume} {19}},\ \bibinfo {pages} {12332} (\bibinfo {year} {2025})}\BibitemShut
  {NoStop}%
\bibitem [{\citenamefont {Yoo}\ \emph {et~al.}(2021)\citenamefont {Yoo}, \citenamefont {de~León-Pérez}, \citenamefont {Pelton}, \citenamefont {Lee}, \citenamefont {Mohr}, \citenamefont {Raschke}, \citenamefont {Caldwell}, \citenamefont {Martín-Moreno},\ and\ \citenamefont {Oh}}]{yoo_ultrastrong_2021}%
  \BibitemOpen
  \bibfield  {author} {\bibinfo {author} {\bibfnamefont {D.}~\bibnamefont {Yoo}}, \bibinfo {author} {\bibfnamefont {F.}~\bibnamefont {de~León-Pérez}}, \bibinfo {author} {\bibfnamefont {M.}~\bibnamefont {Pelton}}, \bibinfo {author} {\bibfnamefont {I.-H.}\ \bibnamefont {Lee}}, \bibinfo {author} {\bibfnamefont {D.~A.}\ \bibnamefont {Mohr}}, \bibinfo {author} {\bibfnamefont {M.~B.}\ \bibnamefont {Raschke}}, \bibinfo {author} {\bibfnamefont {J.~D.}\ \bibnamefont {Caldwell}}, \bibinfo {author} {\bibfnamefont {L.}~\bibnamefont {Martín-Moreno}},\ and\ \bibinfo {author} {\bibfnamefont {S.-H.}\ \bibnamefont {Oh}},\ }\bibfield  {title} {\bibinfo {title} {Ultrastrong plasmon–phonon coupling via epsilon-near-zero nanocavities},\ }\href {https://doi.org/10.1038/s41566-020-00731-5} {\bibfield  {journal} {\bibinfo  {journal} {Nat. Photon.}\ }\textbf {\bibinfo {volume} {15}},\ \bibinfo {pages} {125} (\bibinfo {year} {2021})}\BibitemShut {NoStop}%
\bibitem [{\citenamefont {Di~Stefano}\ \emph {et~al.}(2019)\citenamefont {Di~Stefano}, \citenamefont {Settineri}, \citenamefont {Macr{\ifmmode\grave{\imath}\else\`{\i}\fi}}, \citenamefont {Garziano}, \citenamefont {Stassi}, \citenamefont {Savasta},\ and\ \citenamefont {Nori}}]{DiStefano2019Aug}%
  \BibitemOpen
  \bibfield  {author} {\bibinfo {author} {\bibfnamefont {O.}~\bibnamefont {Di~Stefano}}, \bibinfo {author} {\bibfnamefont {A.}~\bibnamefont {Settineri}}, \bibinfo {author} {\bibfnamefont {V.}~\bibnamefont {Macr{\ifmmode\grave{\imath}\else\`{\i}\fi}}}, \bibinfo {author} {\bibfnamefont {L.}~\bibnamefont {Garziano}}, \bibinfo {author} {\bibfnamefont {R.}~\bibnamefont {Stassi}}, \bibinfo {author} {\bibfnamefont {S.}~\bibnamefont {Savasta}},\ and\ \bibinfo {author} {\bibfnamefont {F.}~\bibnamefont {Nori}},\ }\bibfield  {title} {\bibinfo {title} {{Resolution of gauge ambiguities in ultrastrong-coupling cavity quantum electrodynamics}},\ }\href {https://doi.org/10.1038/s41567-019-0534-4} {\bibfield  {journal} {\bibinfo  {journal} {Nat. Phys.}\ }\textbf {\bibinfo {volume} {15}},\ \bibinfo {pages} {803} (\bibinfo {year} {2019})}\BibitemShut {NoStop}%
\bibitem [{\citenamefont {Salmon}\ \emph {et~al.}(2022)\citenamefont {Salmon}, \citenamefont {Gustin}, \citenamefont {Settineri}, \citenamefont {Di~Stefano}, \citenamefont {Zueco}, \citenamefont {Savasta}, \citenamefont {Nori},\ and\ \citenamefont {Hughes}}]{Salmon2022Mar}%
  \BibitemOpen
  \bibfield  {author} {\bibinfo {author} {\bibfnamefont {W.}~\bibnamefont {Salmon}}, \bibinfo {author} {\bibfnamefont {C.}~\bibnamefont {Gustin}}, \bibinfo {author} {\bibfnamefont {A.}~\bibnamefont {Settineri}}, \bibinfo {author} {\bibfnamefont {O.}~\bibnamefont {Di~Stefano}}, \bibinfo {author} {\bibfnamefont {D.}~\bibnamefont {Zueco}}, \bibinfo {author} {\bibfnamefont {S.}~\bibnamefont {Savasta}}, \bibinfo {author} {\bibfnamefont {F.}~\bibnamefont {Nori}},\ and\ \bibinfo {author} {\bibfnamefont {S.}~\bibnamefont {Hughes}},\ }\bibfield  {title} {\bibinfo {title} {{Gauge-independent emission spectra and quantum correlations in the ultrastrong coupling regime of open system cavity-QED}},\ }\href {https://doi.org/10.1515/nanoph-2021-0718} {\bibfield  {journal} {\bibinfo  {journal} {Nanophotonics}\ }\textbf {\bibinfo {volume} {11}},\ \bibinfo {pages} {1573} (\bibinfo {year} {2022})}\BibitemShut {NoStop}%
\bibitem [{\citenamefont {Dutra}\ and\ \citenamefont {Nienhuis}(2000)}]{Dutra2000Oct}%
  \BibitemOpen
  \bibfield  {author} {\bibinfo {author} {\bibfnamefont {S.~M.}\ \bibnamefont {Dutra}}\ and\ \bibinfo {author} {\bibfnamefont {G.}~\bibnamefont {Nienhuis}},\ }\bibfield  {title} {\bibinfo {title} {{Derivation of a Hamiltonian for photon decay in a}},\ }\href {https://doi.org/10.1088/1464-4266/2/5/305} {\bibfield  {journal} {\bibinfo  {journal} {J. Opt. B: Quantum Semiclassical Opt.}\ }\textbf {\bibinfo {volume} {2}},\ \bibinfo {pages} {584} (\bibinfo {year} {2000})}\BibitemShut {NoStop}%
\bibitem [{\citenamefont {Raymer}\ and\ \citenamefont {McKinstrie}(2013)}]{PhysRevA.88.043819}%
  \BibitemOpen
  \bibfield  {author} {\bibinfo {author} {\bibfnamefont {M.~G.}\ \bibnamefont {Raymer}}\ and\ \bibinfo {author} {\bibfnamefont {C.~J.}\ \bibnamefont {McKinstrie}},\ }\bibfield  {title} {\bibinfo {title} {Quantum input-output theory for optical cavities with arbitrary coupling strength: Application to two-photon wave-packet shaping},\ }\href {https://doi.org/10.1103/PhysRevA.88.043819} {\bibfield  {journal} {\bibinfo  {journal} {Phys. Rev. A}\ }\textbf {\bibinfo {volume} {88}},\ \bibinfo {pages} {043819} (\bibinfo {year} {2013})}\BibitemShut {NoStop}%
\bibitem [{\citenamefont {Khanbekyan}\ \emph {et~al.}(2005)\citenamefont {Khanbekyan}, \citenamefont {Kn\"oll}, \citenamefont {Welsch}, \citenamefont {Semenov},\ and\ \citenamefont {Vogel}}]{khanbekyan}%
  \BibitemOpen
  \bibfield  {author} {\bibinfo {author} {\bibfnamefont {M.}~\bibnamefont {Khanbekyan}}, \bibinfo {author} {\bibfnamefont {L.}~\bibnamefont {Kn\"oll}}, \bibinfo {author} {\bibfnamefont {D.-G.}\ \bibnamefont {Welsch}}, \bibinfo {author} {\bibfnamefont {A.~A.}\ \bibnamefont {Semenov}},\ and\ \bibinfo {author} {\bibfnamefont {W.}~\bibnamefont {Vogel}},\ }\bibfield  {title} {\bibinfo {title} {{QED of lossy cavities: Operator and quantum-state input-output relations}},\ }\href {https://doi.org/10.1103/PhysRevA.72.053813} {\bibfield  {journal} {\bibinfo  {journal} {Phys. Rev. A}\ }\textbf {\bibinfo {volume} {72}},\ \bibinfo {pages} {053813} (\bibinfo {year} {2005})}\BibitemShut {NoStop}%
\bibitem [{\citenamefont {Viviescas}\ and\ \citenamefont {Hackenbroich}(2003)}]{Viviescas2003Jan}%
  \BibitemOpen
  \bibfield  {author} {\bibinfo {author} {\bibfnamefont {C.}~\bibnamefont {Viviescas}}\ and\ \bibinfo {author} {\bibfnamefont {G.}~\bibnamefont {Hackenbroich}},\ }\bibfield  {title} {\bibinfo {title} {{Field quantization for open optical cavities}},\ }\href {https://doi.org/10.1103/PhysRevA.67.013805} {\bibfield  {journal} {\bibinfo  {journal} {Phys. Rev. A}\ }\textbf {\bibinfo {volume} {67}},\ \bibinfo {pages} {013805} (\bibinfo {year} {2003})}\BibitemShut {NoStop}%
\bibitem [{\citenamefont {Lentrodt}\ and\ \citenamefont {Evers}(2020)}]{PhysRevX.10.011008}%
  \BibitemOpen
  \bibfield  {author} {\bibinfo {author} {\bibfnamefont {D.}~\bibnamefont {Lentrodt}}\ and\ \bibinfo {author} {\bibfnamefont {J.}~\bibnamefont {Evers}},\ }\bibfield  {title} {\bibinfo {title} {{Ab Initio Few-Mode Theory for Quantum Potential Scattering Problems}},\ }\href {https://doi.org/10.1103/PhysRevX.10.011008} {\bibfield  {journal} {\bibinfo  {journal} {Phys. Rev. X}\ }\textbf {\bibinfo {volume} {10}},\ \bibinfo {pages} {011008} (\bibinfo {year} {2020})}\BibitemShut {NoStop}%
\bibitem [{\citenamefont {Yuen}\ and\ \citenamefont {Demetriadou}(2024)}]{PhysRevLett.133.203604}%
  \BibitemOpen
  \bibfield  {author} {\bibinfo {author} {\bibfnamefont {B.}~\bibnamefont {Yuen}}\ and\ \bibinfo {author} {\bibfnamefont {A.}~\bibnamefont {Demetriadou}},\ }\bibfield  {title} {\bibinfo {title} {{Exact Quantum Electrodynamics of Radiative Photonic Environments}},\ }\href {https://doi.org/10.1103/PhysRevLett.133.203604} {\bibfield  {journal} {\bibinfo  {journal} {Phys. Rev. Lett.}\ }\textbf {\bibinfo {volume} {133}},\ \bibinfo {pages} {203604} (\bibinfo {year} {2024})}\BibitemShut {NoStop}%
\bibitem [{\citenamefont {Hughes}\ \emph {et~al.}(2024)\citenamefont {Hughes}, \citenamefont {Gustin},\ and\ \citenamefont {Nori}}]{Hughes2024Jun}%
  \BibitemOpen
  \bibfield  {author} {\bibinfo {author} {\bibfnamefont {S.}~\bibnamefont {Hughes}}, \bibinfo {author} {\bibfnamefont {C.}~\bibnamefont {Gustin}},\ and\ \bibinfo {author} {\bibfnamefont {F.}~\bibnamefont {Nori}},\ }\bibfield  {title} {\bibinfo {title} {{Reconciling quantum and classical spectral theories of ultrastrong coupling: role of cavity bath coupling and gauge corrections}},\ }\href {https://doi.org/10.1364/OPTICAQ.519395} {\bibfield  {journal} {\bibinfo  {journal} {Opt. Quantum}\ }\textbf {\bibinfo {volume} {2}},\ \bibinfo {pages} {133} (\bibinfo {year} {2024})}\BibitemShut {NoStop}%
\bibitem [{\citenamefont {Lednev}\ \emph {et~al.}(2024)\citenamefont {Lednev}, \citenamefont {Garc\'{\i}a-Vidal},\ and\ \citenamefont {Feist}}]{PhysRevLett.132.106902}%
  \BibitemOpen
  \bibfield  {author} {\bibinfo {author} {\bibfnamefont {M.}~\bibnamefont {Lednev}}, \bibinfo {author} {\bibfnamefont {F.~J.}\ \bibnamefont {Garc\'{\i}a-Vidal}},\ and\ \bibinfo {author} {\bibfnamefont {J.}~\bibnamefont {Feist}},\ }\bibfield  {title} {\bibinfo {title} {{Lindblad Master Equation Capable of Describing Hybrid Quantum Systems in the Ultrastrong Coupling Regime}},\ }\href {https://doi.org/10.1103/PhysRevLett.132.106902} {\bibfield  {journal} {\bibinfo  {journal} {Phys. Rev. Lett.}\ }\textbf {\bibinfo {volume} {132}},\ \bibinfo {pages} {106902} (\bibinfo {year} {2024})}\BibitemShut {NoStop}%
\bibitem [{\citenamefont {Lentrodt}\ \emph {et~al.}(2020)\citenamefont {Lentrodt}, \citenamefont {Heeg}, \citenamefont {Keitel},\ and\ \citenamefont {Evers}}]{PhysRevResearch.2.023396}%
  \BibitemOpen
  \bibfield  {author} {\bibinfo {author} {\bibfnamefont {D.}~\bibnamefont {Lentrodt}}, \bibinfo {author} {\bibfnamefont {K.~P.}\ \bibnamefont {Heeg}}, \bibinfo {author} {\bibfnamefont {C.~H.}\ \bibnamefont {Keitel}},\ and\ \bibinfo {author} {\bibfnamefont {J.}~\bibnamefont {Evers}},\ }\bibfield  {title} {\bibinfo {title} {Ab initio quantum models for thin-film x-ray cavity {QED}},\ }\href {https://doi.org/10.1103/PhysRevResearch.2.023396} {\bibfield  {journal} {\bibinfo  {journal} {Phys. Rev. Res.}\ }\textbf {\bibinfo {volume} {2}},\ \bibinfo {pages} {023396} (\bibinfo {year} {2020})}\BibitemShut {NoStop}%
\bibitem [{\citenamefont {Ching}\ \emph {et~al.}(1998)\citenamefont {Ching}, \citenamefont {Leung}, \citenamefont {Maassen van~den Brink}, \citenamefont {Suen}, \citenamefont {Tong},\ and\ \citenamefont {Young}}]{2ndquant2}%
  \BibitemOpen
  \bibfield  {author} {\bibinfo {author} {\bibfnamefont {E.~S.~C.}\ \bibnamefont {Ching}}, \bibinfo {author} {\bibfnamefont {P.~T.}\ \bibnamefont {Leung}}, \bibinfo {author} {\bibfnamefont {A.}~\bibnamefont {Maassen van~den Brink}}, \bibinfo {author} {\bibfnamefont {W.~M.}\ \bibnamefont {Suen}}, \bibinfo {author} {\bibfnamefont {S.~S.}\ \bibnamefont {Tong}},\ and\ \bibinfo {author} {\bibfnamefont {K.}~\bibnamefont {Young}},\ }\bibfield  {title} {\bibinfo {title} {Quasinormal-mode expansion for waves in open systems},\ }\href {https://doi.org/10.1103/RevModPhys.70.1545} {\bibfield  {journal} {\bibinfo  {journal} {Rev. Mod. Phys.}\ }\textbf {\bibinfo {volume} {70}},\ \bibinfo {pages} {1545} (\bibinfo {year} {1998})}\BibitemShut {NoStop}%
\bibitem [{\citenamefont {Kristensen}\ and\ \citenamefont {Hughes}(2014)}]{NormKristHughes}%
  \BibitemOpen
  \bibfield  {author} {\bibinfo {author} {\bibfnamefont {P.~T.}\ \bibnamefont {Kristensen}}\ and\ \bibinfo {author} {\bibfnamefont {S.}~\bibnamefont {Hughes}},\ }\bibfield  {title} {\bibinfo {title} {{Modes and Mode Volumes of Leaky Optical Cavities and Plasmonic Nanoresonators}},\ }\href {https://pubs.acs.org/doi/10.1021/ph400114e} {\bibfield  {journal} {\bibinfo  {journal} {ACS Photonics}\ }\textbf {\bibinfo {volume} {1}},\ \bibinfo {pages} {2} (\bibinfo {year} {2014})}\BibitemShut {NoStop}%
\bibitem [{\citenamefont {Lalanne}\ \emph {et~al.}(2018)\citenamefont {Lalanne}, \citenamefont {Yan}, \citenamefont {Vynck}, \citenamefont {Sauvan},\ and\ \citenamefont {Hugonin}}]{Lalanne_review}%
  \BibitemOpen
  \bibfield  {author} {\bibinfo {author} {\bibfnamefont {P.}~\bibnamefont {Lalanne}}, \bibinfo {author} {\bibfnamefont {W.}~\bibnamefont {Yan}}, \bibinfo {author} {\bibfnamefont {K.}~\bibnamefont {Vynck}}, \bibinfo {author} {\bibfnamefont {C.}~\bibnamefont {Sauvan}},\ and\ \bibinfo {author} {\bibfnamefont {J.-P.}\ \bibnamefont {Hugonin}},\ }\bibfield  {title} {\bibinfo {title} {{Light Interaction with Photonic and Plasmonic Resonances}},\ }\href {https://onlinelibrary.wiley.com/doi/full/10.1002/lpor.201700113} {\bibfield  {journal} {\bibinfo  {journal} {Laser \& Photonics Reviews}\ }\textbf {\bibinfo {volume} {12}},\ \bibinfo {pages} {1700113} (\bibinfo {year} {2018})}\BibitemShut {NoStop}%
\bibitem [{\citenamefont {Kristensen}\ \emph {et~al.}(2020)\citenamefont {Kristensen}, \citenamefont {Herrmann}, \citenamefont {Intravaia},\ and\ \citenamefont {Busch}}]{Kristensen:20}%
  \BibitemOpen
  \bibfield  {author} {\bibinfo {author} {\bibfnamefont {P.~T.}\ \bibnamefont {Kristensen}}, \bibinfo {author} {\bibfnamefont {K.}~\bibnamefont {Herrmann}}, \bibinfo {author} {\bibfnamefont {F.}~\bibnamefont {Intravaia}},\ and\ \bibinfo {author} {\bibfnamefont {K.}~\bibnamefont {Busch}},\ }\bibfield  {title} {\bibinfo {title} {Modeling electromagnetic resonators using quasinormal modes},\ }\href {https://doi.org/10.1364/AOP.377940} {\bibfield  {journal} {\bibinfo  {journal} {Adv. Opt. Photon.}\ }\textbf {\bibinfo {volume} {12}},\ \bibinfo {pages} {612} (\bibinfo {year} {2020})}\BibitemShut {NoStop}%
\bibitem [{\citenamefont {Ge}\ \emph {et~al.}(2014)\citenamefont {Ge}, \citenamefont {Kristensen}, \citenamefont {Young},\ and\ \citenamefont {Hughes}}]{GeNJP2014}%
  \BibitemOpen
  \bibfield  {author} {\bibinfo {author} {\bibfnamefont {R.-C.}\ \bibnamefont {Ge}}, \bibinfo {author} {\bibfnamefont {P.~T.}\ \bibnamefont {Kristensen}}, \bibinfo {author} {\bibfnamefont {J.~F.}\ \bibnamefont {Young}},\ and\ \bibinfo {author} {\bibfnamefont {S.}~\bibnamefont {Hughes}},\ }\bibfield  {title} {\bibinfo {title} {Quasinormal mode approach to modelling light-emission and propagation in nanoplasmonics},\ }\href {http://stacks.iop.org/1367-2630/16/i=11/a=113048} {\bibfield  {journal} {\bibinfo  {journal} {New J. Phys.}\ }\textbf {\bibinfo {volume} {16}},\ \bibinfo {pages} {113048} (\bibinfo {year} {2014})}\BibitemShut {NoStop}%
\bibitem [{\citenamefont {Franke}\ \emph {et~al.}(2019)\citenamefont {Franke}, \citenamefont {Hughes}, \citenamefont {Dezfouli}, \citenamefont {Kristensen}, \citenamefont {Busch}, \citenamefont {Knorr},\ and\ \citenamefont {Richter}}]{PhysRevLett.122.213901}%
  \BibitemOpen
  \bibfield  {author} {\bibinfo {author} {\bibfnamefont {S.}~\bibnamefont {Franke}}, \bibinfo {author} {\bibfnamefont {S.}~\bibnamefont {Hughes}}, \bibinfo {author} {\bibfnamefont {M.~K.}\ \bibnamefont {Dezfouli}}, \bibinfo {author} {\bibfnamefont {P.~T.}\ \bibnamefont {Kristensen}}, \bibinfo {author} {\bibfnamefont {K.}~\bibnamefont {Busch}}, \bibinfo {author} {\bibfnamefont {A.}~\bibnamefont {Knorr}},\ and\ \bibinfo {author} {\bibfnamefont {M.}~\bibnamefont {Richter}},\ }\bibfield  {title} {\bibinfo {title} {{Quantization of Quasinormal Modes for Open Cavities and Plasmonic Cavity Quantum Electrodynamics}},\ }\href {https://doi.org/10.1103/PhysRevLett.122.213901} {\bibfield  {journal} {\bibinfo  {journal} {Phys. Rev. Lett.}\ }\textbf {\bibinfo {volume} {122}},\ \bibinfo {pages} {213901} (\bibinfo {year} {2019})}\BibitemShut {NoStop}%
\bibitem [{\citenamefont {Ge}\ and\ \citenamefont {Hughes}(2015)}]{PhysRevB.92.205420}%
  \BibitemOpen
  \bibfield  {author} {\bibinfo {author} {\bibfnamefont {R.-C.}\ \bibnamefont {Ge}}\ and\ \bibinfo {author} {\bibfnamefont {S.}~\bibnamefont {Hughes}},\ }\bibfield  {title} {\bibinfo {title} {Quantum dynamics of two quantum dots coupled through localized plasmons: An intuitive and accurate quantum optics approach using quasinormal modes},\ }\href {https://doi.org/10.1103/PhysRevB.92.205420} {\bibfield  {journal} {\bibinfo  {journal} {Phys. Rev. B}\ }\textbf {\bibinfo {volume} {92}},\ \bibinfo {pages} {205420} (\bibinfo {year} {2015})}\BibitemShut {NoStop}%
\bibitem [{\citenamefont {Meschede}\ \emph {et~al.}(2025)\citenamefont {Meschede}, \citenamefont {Clarke},\ and\ \citenamefont {Hess}}]{meschede2025quantumquasinormalmodetheory}%
  \BibitemOpen
  \bibfield  {author} {\bibinfo {author} {\bibfnamefont {L.}~\bibnamefont {Meschede}}, \bibinfo {author} {\bibfnamefont {D.~D.~A.}\ \bibnamefont {Clarke}},\ and\ \bibinfo {author} {\bibfnamefont {O.}~\bibnamefont {Hess}},\ }\href {https://arxiv.org/abs/2507.05233} {\bibinfo {title} {Quantum quasinormal mode theory for dissipative nano-optics and magnetodielectric cavity quantum electrodynamics}} (\bibinfo {year} {2025}),\ \Eprint {https://arxiv.org/abs/2507.05233} {arXiv:2507.05233 [physics.optics]} \BibitemShut {NoStop}%
\bibitem [{\citenamefont {Hughes}\ \emph {et~al.}(2019)\citenamefont {Hughes}, \citenamefont {Franke}, \citenamefont {Gustin}, \citenamefont {Kamandar~Dezfouli}, \citenamefont {Knorr},\ and\ \citenamefont {Richter}}]{Hughes_SPS_2019}%
  \BibitemOpen
  \bibfield  {author} {\bibinfo {author} {\bibfnamefont {S.}~\bibnamefont {Hughes}}, \bibinfo {author} {\bibfnamefont {S.}~\bibnamefont {Franke}}, \bibinfo {author} {\bibfnamefont {C.}~\bibnamefont {Gustin}}, \bibinfo {author} {\bibfnamefont {M.}~\bibnamefont {Kamandar~Dezfouli}}, \bibinfo {author} {\bibfnamefont {A.}~\bibnamefont {Knorr}},\ and\ \bibinfo {author} {\bibfnamefont {M.}~\bibnamefont {Richter}},\ }\bibfield  {title} {\bibinfo {title} {{Theory and Limits of On-Demand Single-Photon Sources Using Plasmonic Resonators: A Quantized Quasinormal Mode Approach}},\ }\href {https://doi.org/10.1021/acsphotonics.9b00849} {\bibfield  {journal} {\bibinfo  {journal} {{ACS} Photonics}\ }\textbf {\bibinfo {volume} {6}},\ \bibinfo {pages} {2168} (\bibinfo {year} {2019})}\BibitemShut {NoStop}%
\bibitem [{\citenamefont {Franke}\ \emph {et~al.}(2020{\natexlab{a}})\citenamefont {Franke}, \citenamefont {Richter}, \citenamefont {Ren}, \citenamefont {Knorr},\ and\ \citenamefont {Hughes}}]{franke2020quantized}%
  \BibitemOpen
  \bibfield  {author} {\bibinfo {author} {\bibfnamefont {S.}~\bibnamefont {Franke}}, \bibinfo {author} {\bibfnamefont {M.}~\bibnamefont {Richter}}, \bibinfo {author} {\bibfnamefont {J.}~\bibnamefont {Ren}}, \bibinfo {author} {\bibfnamefont {A.}~\bibnamefont {Knorr}},\ and\ \bibinfo {author} {\bibfnamefont {S.}~\bibnamefont {Hughes}},\ }\bibfield  {title} {\bibinfo {title} {{Quantized quasinormal-mode description of nonlinear cavity-QED effects from coupled resonators with a Fano-like resonance}},\ }\href {https://doi.org/10.1103/PhysRevResearch.2.033456} {\bibfield  {journal} {\bibinfo  {journal} {Phys. Rev. Res.}\ }\textbf {\bibinfo {volume} {2}},\ \bibinfo {pages} {033456} (\bibinfo {year} {2020}{\natexlab{a}})}\BibitemShut {NoStop}%
\bibitem [{\citenamefont {Ren}\ \emph {et~al.}(2022{\natexlab{a}})\citenamefont {Ren}, \citenamefont {Franke},\ and\ \citenamefont {Hughes}}]{Ren2022}%
  \BibitemOpen
  \bibfield  {author} {\bibinfo {author} {\bibfnamefont {J.}~\bibnamefont {Ren}}, \bibinfo {author} {\bibfnamefont {S.}~\bibnamefont {Franke}},\ and\ \bibinfo {author} {\bibfnamefont {S.}~\bibnamefont {Hughes}},\ }\bibfield  {title} {\bibinfo {title} {{Connecting Classical and Quantum Mode Theories for Coupled Lossy Cavity Resonators Using Quasinormal Modes}},\ }\href {https://doi.org/10.1021/acsphotonics.1c01274} {\bibfield  {journal} {\bibinfo  {journal} {{ACS} Photonics}\ }\textbf {\bibinfo {volume} {9}},\ \bibinfo {pages} {138} (\bibinfo {year} {2022}{\natexlab{a}})}\BibitemShut {NoStop}%
\bibitem [{\citenamefont {Ren}\ \emph {et~al.}(2022{\natexlab{b}})\citenamefont {Ren}, \citenamefont {Franke},\ and\ \citenamefont {Hughes}}]{Ren2022b}%
  \BibitemOpen
  \bibfield  {author} {\bibinfo {author} {\bibfnamefont {J.}~\bibnamefont {Ren}}, \bibinfo {author} {\bibfnamefont {S.}~\bibnamefont {Franke}},\ and\ \bibinfo {author} {\bibfnamefont {S.}~\bibnamefont {Hughes}},\ }\bibfield  {title} {\bibinfo {title} {{Quasinormal Mode Theory of Chiral Power Flow from Linearly Polarized Dipole Emitters Coupled to Index-Modulated Microring Resonators Close to an Exceptional Point}},\ }\href {https://doi.org/10.1021/acsphotonics.1c01848} {\bibfield  {journal} {\bibinfo  {journal} {{ACS} Photonics}\ }\textbf {\bibinfo {volume} {9}},\ \bibinfo {pages} {1315} (\bibinfo {year} {2022}{\natexlab{b}})}\BibitemShut {NoStop}%
\bibitem [{\citenamefont {Franke}\ \emph {et~al.}(2021)\citenamefont {Franke}, \citenamefont {Ren}, \citenamefont {Richter}, \citenamefont {Knorr},\ and\ \citenamefont {Hughes}}]{PhysRevLett.127.013602}%
  \BibitemOpen
  \bibfield  {author} {\bibinfo {author} {\bibfnamefont {S.}~\bibnamefont {Franke}}, \bibinfo {author} {\bibfnamefont {J.}~\bibnamefont {Ren}}, \bibinfo {author} {\bibfnamefont {M.}~\bibnamefont {Richter}}, \bibinfo {author} {\bibfnamefont {A.}~\bibnamefont {Knorr}},\ and\ \bibinfo {author} {\bibfnamefont {S.}~\bibnamefont {Hughes}},\ }\bibfield  {title} {\bibinfo {title} {{Fermi's Golden Rule for Spontaneous Emission in Absorptive and Amplifying Media}},\ }\href {https://doi.org/10.1103/PhysRevLett.127.013602} {\bibfield  {journal} {\bibinfo  {journal} {Phys. Rev. Lett.}\ }\textbf {\bibinfo {volume} {127}},\ \bibinfo {pages} {013602} (\bibinfo {year} {2021})}\BibitemShut {NoStop}%
\bibitem [{\citenamefont {Ren}\ \emph {et~al.}(2021)\citenamefont {Ren}, \citenamefont {Franke},\ and\ \citenamefont {Hughes}}]{PhysRevX.11.041020}%
  \BibitemOpen
  \bibfield  {author} {\bibinfo {author} {\bibfnamefont {J.}~\bibnamefont {Ren}}, \bibinfo {author} {\bibfnamefont {S.}~\bibnamefont {Franke}},\ and\ \bibinfo {author} {\bibfnamefont {S.}~\bibnamefont {Hughes}},\ }\bibfield  {title} {\bibinfo {title} {{Quasinormal Modes, Local Density of States, and Classical Purcell Factors for Coupled Loss-Gain Resonators}},\ }\href {https://doi.org/10.1103/PhysRevX.11.041020} {\bibfield  {journal} {\bibinfo  {journal} {Phys. Rev. X}\ }\textbf {\bibinfo {volume} {11}},\ \bibinfo {pages} {041020} (\bibinfo {year} {2021})}\BibitemShut {NoStop}%
\bibitem [{\citenamefont {Franke}\ \emph {et~al.}(2022)\citenamefont {Franke}, \citenamefont {Ren},\ and\ \citenamefont {Hughes}}]{PhysRevA.105.023702}%
  \BibitemOpen
  \bibfield  {author} {\bibinfo {author} {\bibfnamefont {S.}~\bibnamefont {Franke}}, \bibinfo {author} {\bibfnamefont {J.}~\bibnamefont {Ren}},\ and\ \bibinfo {author} {\bibfnamefont {S.}~\bibnamefont {Hughes}},\ }\bibfield  {title} {\bibinfo {title} {Quantized quasinormal-mode theory of coupled lossy and amplifying resonators},\ }\href {https://doi.org/10.1103/PhysRevA.105.023702} {\bibfield  {journal} {\bibinfo  {journal} {Phys. Rev. A}\ }\textbf {\bibinfo {volume} {105}},\ \bibinfo {pages} {023702} (\bibinfo {year} {2022})}\BibitemShut {NoStop}%
\bibitem [{\citenamefont {VanDrunen}\ \emph {et~al.}(2024)\citenamefont {VanDrunen}, \citenamefont {Ren}, \citenamefont {Franke},\ and\ \citenamefont {Hughes}}]{VanDrunen:24}%
  \BibitemOpen
  \bibfield  {author} {\bibinfo {author} {\bibfnamefont {B.}~\bibnamefont {VanDrunen}}, \bibinfo {author} {\bibfnamefont {J.}~\bibnamefont {Ren}}, \bibinfo {author} {\bibfnamefont {S.}~\bibnamefont {Franke}},\ and\ \bibinfo {author} {\bibfnamefont {S.}~\bibnamefont {Hughes}},\ }\bibfield  {title} {\bibinfo {title} {Gain-compensated metal cavity modes and a million-fold improvement of {Purcell} factors},\ }\href {https://doi.org/10.1364/OPTICAQ.504834} {\bibfield  {journal} {\bibinfo  {journal} {Opt. Quantum}\ }\textbf {\bibinfo {volume} {2}},\ \bibinfo {pages} {85} (\bibinfo {year} {2024})}\BibitemShut {NoStop}%
\bibitem [{\citenamefont {VanDrunen}\ \emph {et~al.}(2025)\citenamefont {VanDrunen}, \citenamefont {Ren}, \citenamefont {Franke},\ and\ \citenamefont {Hughes}}]{rrwd-47xr}%
  \BibitemOpen
  \bibfield  {author} {\bibinfo {author} {\bibfnamefont {B.}~\bibnamefont {VanDrunen}}, \bibinfo {author} {\bibfnamefont {J.}~\bibnamefont {Ren}}, \bibinfo {author} {\bibfnamefont {S.}~\bibnamefont {Franke}},\ and\ \bibinfo {author} {\bibfnamefont {S.}~\bibnamefont {Hughes}},\ }\bibfield  {title} {\bibinfo {title} {Gain-modified emission dynamics between two quantum emitters in a plasmonic gain cavity system},\ }\href {https://doi.org/10.1103/rrwd-47xr} {\bibfield  {journal} {\bibinfo  {journal} {Phys. Rev. A}\ }\textbf {\bibinfo {volume} {112}},\ \bibinfo {pages} {013532} (\bibinfo {year} {2025})}\BibitemShut {NoStop}%
\bibitem [{\citenamefont {Medina}\ \emph {et~al.}(2021)\citenamefont {Medina}, \citenamefont {Garc\'{\i}a-Vidal}, \citenamefont {Fern\'andez-Dom\'{\i}nguez},\ and\ \citenamefont {Feist}}]{PhysRevLett.126.093601}%
  \BibitemOpen
  \bibfield  {author} {\bibinfo {author} {\bibfnamefont {I.}~\bibnamefont {Medina}}, \bibinfo {author} {\bibfnamefont {F.~J.}\ \bibnamefont {Garc\'{\i}a-Vidal}}, \bibinfo {author} {\bibfnamefont {A.~I.}\ \bibnamefont {Fern\'andez-Dom\'{\i}nguez}},\ and\ \bibinfo {author} {\bibfnamefont {J.}~\bibnamefont {Feist}},\ }\bibfield  {title} {\bibinfo {title} {{Few-Mode Field Quantization of Arbitrary Electromagnetic Spectral Densities}},\ }\href {https://doi.org/10.1103/PhysRevLett.126.093601} {\bibfield  {journal} {\bibinfo  {journal} {Phys. Rev. Lett.}\ }\textbf {\bibinfo {volume} {126}},\ \bibinfo {pages} {093601} (\bibinfo {year} {2021})}\BibitemShut {NoStop}%
\bibitem [{\citenamefont {Sánchez-Barquilla}\ \emph {et~al.}(2022)\citenamefont {Sánchez-Barquilla}, \citenamefont {García-Vidal}, \citenamefont {Fernández-Domínguez},\ and\ \citenamefont {Feist}}]{sanchez-barquilla_few-mode_2022}%
  \BibitemOpen
  \bibfield  {author} {\bibinfo {author} {\bibfnamefont {M.}~\bibnamefont {Sánchez-Barquilla}}, \bibinfo {author} {\bibfnamefont {F.~J.}\ \bibnamefont {García-Vidal}}, \bibinfo {author} {\bibfnamefont {A.~I.}\ \bibnamefont {Fernández-Domínguez}},\ and\ \bibinfo {author} {\bibfnamefont {J.}~\bibnamefont {Feist}},\ }\bibfield  {title} {\bibinfo {title} {Few-mode field quantization for multiple emitters},\ }\href {https://doi.org/10.1515/nanoph-2021-0795} {\bibfield  {journal} {\bibinfo  {journal} {Nanophotonics}\ }\textbf {\bibinfo {volume} {11}},\ \bibinfo {pages} {4363} (\bibinfo {year} {2022})}\BibitemShut {NoStop}%
\bibitem [{\citenamefont {Gustin}\ \emph {et~al.}(2023)\citenamefont {Gustin}, \citenamefont {Franke},\ and\ \citenamefont {Hughes}}]{Gustin2023Jan}%
  \BibitemOpen
  \bibfield  {author} {\bibinfo {author} {\bibfnamefont {C.}~\bibnamefont {Gustin}}, \bibinfo {author} {\bibfnamefont {S.}~\bibnamefont {Franke}},\ and\ \bibinfo {author} {\bibfnamefont {S.}~\bibnamefont {Hughes}},\ }\bibfield  {title} {\bibinfo {title} {{Gauge-invariant theory of truncated quantum light-matter interactions in arbitrary media}},\ }\href {https://doi.org/10.1103/PhysRevA.107.013722} {\bibfield  {journal} {\bibinfo  {journal} {Phys. Rev. A}\ }\textbf {\bibinfo {volume} {107}},\ \bibinfo {pages} {013722} (\bibinfo {year} {2023})}\BibitemShut {NoStop}%
\bibitem [{\citenamefont {Dung}\ \emph {et~al.}(1998)\citenamefont {Dung}, \citenamefont {Kn\"oll},\ and\ \citenamefont {Welsch}}]{Dung}%
  \BibitemOpen
  \bibfield  {author} {\bibinfo {author} {\bibfnamefont {H.~T.}\ \bibnamefont {Dung}}, \bibinfo {author} {\bibfnamefont {L.}~\bibnamefont {Kn\"oll}},\ and\ \bibinfo {author} {\bibfnamefont {D.-G.}\ \bibnamefont {Welsch}},\ }\bibfield  {title} {\bibinfo {title} {Three-dimensional quantization of the electromagnetic field in dispersive and absorbing inhomogeneous dielectrics},\ }\href {https://doi.org/10.1103/PhysRevA.57.3931} {\bibfield  {journal} {\bibinfo  {journal} {Phys. Rev. A}\ }\textbf {\bibinfo {volume} {57}},\ \bibinfo {pages} {3931} (\bibinfo {year} {1998})}\BibitemShut {NoStop}%
\bibitem [{\citenamefont {Gruner}\ and\ \citenamefont {Welsch}(1996)}]{grunwel}%
  \BibitemOpen
  \bibfield  {author} {\bibinfo {author} {\bibfnamefont {T.}~\bibnamefont {Gruner}}\ and\ \bibinfo {author} {\bibfnamefont {D.-G.}\ \bibnamefont {Welsch}},\ }\bibfield  {title} {\bibinfo {title} {Green-function approach to the radiation-field quantization for homogeneous and inhomogeneous {Kramers-Kronig} dielectrics},\ }\href {https://doi.org/10.1103/PhysRevA.53.1818} {\bibfield  {journal} {\bibinfo  {journal} {Phys. Rev. A}\ }\textbf {\bibinfo {volume} {53}},\ \bibinfo {pages} {1818} (\bibinfo {year} {1996})}\BibitemShut {NoStop}%
\bibitem [{\citenamefont {Suttorp}\ and\ \citenamefont {van Wonderen}(2004)}]{Suttorp}%
  \BibitemOpen
  \bibfield  {author} {\bibinfo {author} {\bibfnamefont {L.~G.}\ \bibnamefont {Suttorp}}\ and\ \bibinfo {author} {\bibfnamefont {A.~J.}\ \bibnamefont {van Wonderen}},\ }\bibfield  {title} {\bibinfo {title} {Fano diagonalization of a polariton model for an inhomogeneous absorptive dielectric},\ }\href {http://stacks.iop.org/0295-5075/67/i=5/a=766} {\bibfield  {journal} {\bibinfo  {journal} {EPL (Europhysics Letters)}\ }\textbf {\bibinfo {volume} {67}},\ \bibinfo {pages} {766} (\bibinfo {year} {2004})}\BibitemShut {NoStop}%
\bibitem [{\citenamefont {Vogel}\ and\ \citenamefont {Welsch}(2006)}]{vogel2006}%
  \BibitemOpen
  \bibfield  {author} {\bibinfo {author} {\bibfnamefont {W.}~\bibnamefont {Vogel}}\ and\ \bibinfo {author} {\bibfnamefont {D.-G.}\ \bibnamefont {Welsch}},\ }\href@noop {} {\emph {\bibinfo {title} {{Quantum Optics}}}}\ (\bibinfo  {publisher} {John Wiley \& Sons},\ \bibinfo {year} {2006})\BibitemShut {NoStop}%
\bibitem [{\citenamefont {Philbin}(2010)}]{philbin2010canonical}%
  \BibitemOpen
  \bibfield  {author} {\bibinfo {author} {\bibfnamefont {T.~G.}\ \bibnamefont {Philbin}},\ }\bibfield  {title} {\bibinfo {title} {Canonical quantization of macroscopic electromagnetism},\ }\href {https://iopscience.iop.org/article/10.1088/1367-2630/12/12/123008} {\bibfield  {journal} {\bibinfo  {journal} {New J. Phys.}\ }\textbf {\bibinfo {volume} {12}},\ \bibinfo {pages} {123008} (\bibinfo {year} {2010})}\BibitemShut {NoStop}%
\bibitem [{\citenamefont {Franke}\ \emph {et~al.}(2020{\natexlab{b}})\citenamefont {Franke}, \citenamefont {Ren}, \citenamefont {Hughes},\ and\ \citenamefont {Richter}}]{franke2020fluctuation}%
  \BibitemOpen
  \bibfield  {author} {\bibinfo {author} {\bibfnamefont {S.}~\bibnamefont {Franke}}, \bibinfo {author} {\bibfnamefont {J.}~\bibnamefont {Ren}}, \bibinfo {author} {\bibfnamefont {S.}~\bibnamefont {Hughes}},\ and\ \bibinfo {author} {\bibfnamefont {M.}~\bibnamefont {Richter}},\ }\bibfield  {title} {\bibinfo {title} {Fluctuation-dissipation theorem and fundamental photon commutation relations in lossy nanostructures using quasinormal modes},\ }\href {https://doi.org/10.1103/PhysRevResearch.2.033332} {\bibfield  {journal} {\bibinfo  {journal} {Phys. Rev. Res.}\ }\textbf {\bibinfo {volume} {2}},\ \bibinfo {pages} {033332} (\bibinfo {year} {2020}{\natexlab{b}})}\BibitemShut {NoStop}%
\bibitem [{\citenamefont {Dezfouli}\ \emph {et~al.}(2019)\citenamefont {Dezfouli}, \citenamefont {Gordon},\ and\ \citenamefont {Hughes}}]{Dezfouli2019}%
  \BibitemOpen
  \bibfield  {author} {\bibinfo {author} {\bibfnamefont {M.~K.}\ \bibnamefont {Dezfouli}}, \bibinfo {author} {\bibfnamefont {R.}~\bibnamefont {Gordon}},\ and\ \bibinfo {author} {\bibfnamefont {S.}~\bibnamefont {Hughes}},\ }\bibfield  {title} {\bibinfo {title} {{Molecular Optomechanics in the Anharmonic Cavity-{QED} Regime Using Hybrid Metal{\textendash}Dielectric Cavity Modes}},\ }\href {https://doi.org/10.1021/acsphotonics.8b01091} {\bibfield  {journal} {\bibinfo  {journal} {{ACS} Photonics}\ }\textbf {\bibinfo {volume} {6}},\ \bibinfo {pages} {1400} (\bibinfo {year} {2019})}\BibitemShut {NoStop}%
\bibitem [{\citenamefont {Dezfouli}\ \emph {et~al.}(2017)\citenamefont {Dezfouli}, \citenamefont {Tserkezis}, \citenamefont {Mortensen},\ and\ \citenamefont {Hughes}}]{KamandarDezfouli2017b}%
  \BibitemOpen
  \bibfield  {author} {\bibinfo {author} {\bibfnamefont {M.~K.}\ \bibnamefont {Dezfouli}}, \bibinfo {author} {\bibfnamefont {C.}~\bibnamefont {Tserkezis}}, \bibinfo {author} {\bibfnamefont {N.~A.}\ \bibnamefont {Mortensen}},\ and\ \bibinfo {author} {\bibfnamefont {S.}~\bibnamefont {Hughes}},\ }\bibfield  {title} {\bibinfo {title} {Nonlocal quasinormal modes for arbitrarily shaped three-dimensional plasmonic resonators},\ }\href {https://doi.org/10.1364/optica.4.001503} {\bibfield  {journal} {\bibinfo  {journal} {Optica}\ }\textbf {\bibinfo {volume} {4}},\ \bibinfo {pages} {1503} (\bibinfo {year} {2017})}\BibitemShut {NoStop}%
\bibitem [{\citenamefont {Zschiedrich}\ \emph {et~al.}(2018)\citenamefont {Zschiedrich}, \citenamefont {Binkowski}, \citenamefont {Nikolay}, \citenamefont {Benson}, \citenamefont {Kewes},\ and\ \citenamefont {Burger}}]{PhysRevA.98.043806}%
  \BibitemOpen
  \bibfield  {author} {\bibinfo {author} {\bibfnamefont {L.}~\bibnamefont {Zschiedrich}}, \bibinfo {author} {\bibfnamefont {F.}~\bibnamefont {Binkowski}}, \bibinfo {author} {\bibfnamefont {N.}~\bibnamefont {Nikolay}}, \bibinfo {author} {\bibfnamefont {O.}~\bibnamefont {Benson}}, \bibinfo {author} {\bibfnamefont {G.}~\bibnamefont {Kewes}},\ and\ \bibinfo {author} {\bibfnamefont {S.}~\bibnamefont {Burger}},\ }\bibfield  {title} {\bibinfo {title} {Riesz-projection-based theory of light-matter interaction in dispersive nanoresonators},\ }\href {https://doi.org/10.1103/PhysRevA.98.043806} {\bibfield  {journal} {\bibinfo  {journal} {Phys. Rev. A}\ }\textbf {\bibinfo {volume} {98}},\ \bibinfo {pages} {043806} (\bibinfo {year} {2018})}\BibitemShut {NoStop}%
\bibitem [{\citenamefont {Li}\ \emph {et~al.}(2021)\citenamefont {Li}, \citenamefont {Zhou}, \citenamefont {Zhang},\ and\ \citenamefont {Chen}}]{PhysRevLett.126.257401}%
  \BibitemOpen
  \bibfield  {author} {\bibinfo {author} {\bibfnamefont {W.}~\bibnamefont {Li}}, \bibinfo {author} {\bibfnamefont {Q.}~\bibnamefont {Zhou}}, \bibinfo {author} {\bibfnamefont {P.}~\bibnamefont {Zhang}},\ and\ \bibinfo {author} {\bibfnamefont {X.-W.}\ \bibnamefont {Chen}},\ }\bibfield  {title} {\bibinfo {title} {{Bright Optical Eigenmode of $1\text{ }\text{ }{\mathrm{nm}}^{3}$ Mode Volume}},\ }\href {https://doi.org/10.1103/PhysRevLett.126.257401} {\bibfield  {journal} {\bibinfo  {journal} {Phys. Rev. Lett.}\ }\textbf {\bibinfo {volume} {126}},\ \bibinfo {pages} {257401} (\bibinfo {year} {2021})}\BibitemShut {NoStop}%
\bibitem [{\citenamefont {Lee}\ \emph {et~al.}(1999)\citenamefont {Lee}, \citenamefont {Leung},\ and\ \citenamefont {Pang}}]{lee_dyadic_1999}%
  \BibitemOpen
  \bibfield  {author} {\bibinfo {author} {\bibfnamefont {K.~M.}\ \bibnamefont {Lee}}, \bibinfo {author} {\bibfnamefont {P.~T.}\ \bibnamefont {Leung}},\ and\ \bibinfo {author} {\bibfnamefont {K.~M.}\ \bibnamefont {Pang}},\ }\bibfield  {title} {\bibinfo {title} {Dyadic formulation of morphology-dependent resonances. i. completeness relation},\ }\href {https://doi.org/10.1364/JOSAB.16.001409} {\bibfield  {journal} {\bibinfo  {journal} {JOSA B}\ }\textbf {\bibinfo {volume} {16}},\ \bibinfo {pages} {1409} (\bibinfo {year} {1999})}\BibitemShut {NoStop}%
\bibitem [{\citenamefont {Leung}\ \emph {et~al.}(1994)\citenamefont {Leung}, \citenamefont {Liu},\ and\ \citenamefont {Young}}]{LeungSP1D}%
  \BibitemOpen
  \bibfield  {author} {\bibinfo {author} {\bibfnamefont {P.~T.}\ \bibnamefont {Leung}}, \bibinfo {author} {\bibfnamefont {S.~Y.}\ \bibnamefont {Liu}},\ and\ \bibinfo {author} {\bibfnamefont {K.}~\bibnamefont {Young}},\ }\bibfield  {title} {\bibinfo {title} {Completeness and orthogonality of quasinormal modes in leaky optical cavities},\ }\href {https://doi.org/10.1103/PhysRevA.49.3057} {\bibfield  {journal} {\bibinfo  {journal} {Phys. Rev. A}\ }\textbf {\bibinfo {volume} {49}},\ \bibinfo {pages} {3057} (\bibinfo {year} {1994})}\BibitemShut {NoStop}%
\bibitem [{\citenamefont {Ge}\ and\ \citenamefont {Hughes}(2014{\natexlab{a}})}]{Ge2014}%
  \BibitemOpen
  \bibfield  {author} {\bibinfo {author} {\bibfnamefont {R.-C.}\ \bibnamefont {Ge}}\ and\ \bibinfo {author} {\bibfnamefont {S.}~\bibnamefont {Hughes}},\ }\bibfield  {title} {\bibinfo {title} {{Design of an efficient single photon source from a metallic nanorod dimer: a quasi-normal mode finite-difference time-domain approach}},\ }\href {https://doi.org/10.1364/OL.39.004235} {\bibfield  {journal} {\bibinfo  {journal} {Opt. Lett.}\ }\textbf {\bibinfo {volume} {39}},\ \bibinfo {pages} {4235} (\bibinfo {year} {2014}{\natexlab{a}})}\BibitemShut {NoStop}%
\bibitem [{\citenamefont {{Kamandar Dezfouli}}\ \emph {et~al.}(2017)\citenamefont {{Kamandar Dezfouli}}, \citenamefont {Gordon},\ and\ \citenamefont {Hughes}}]{KamandarDezfouli2017}%
  \BibitemOpen
  \bibfield  {author} {\bibinfo {author} {\bibfnamefont {M.}~\bibnamefont {{Kamandar Dezfouli}}}, \bibinfo {author} {\bibfnamefont {R.}~\bibnamefont {Gordon}},\ and\ \bibinfo {author} {\bibfnamefont {S.}~\bibnamefont {Hughes}},\ }\bibfield  {title} {\bibinfo {title} {{Modal theory of modified spontaneous emission for a hybrid plasmonic photonic-crystal cavity system}},\ }\href {https://doi.org/10.1103/PhysRevA.95.013846} {\bibfield  {journal} {\bibinfo  {journal} {Phys. Rev. A}\ }\textbf {\bibinfo {volume} {95}},\ \bibinfo {pages} {013846} (\bibinfo {year} {2017})}\BibitemShut {NoStop}%
\bibitem [{\citenamefont {Carlson}\ and\ \citenamefont {Hughes}(2020)}]{carlson2019dissipative}%
  \BibitemOpen
  \bibfield  {author} {\bibinfo {author} {\bibfnamefont {C.}~\bibnamefont {Carlson}}\ and\ \bibinfo {author} {\bibfnamefont {S.}~\bibnamefont {Hughes}},\ }\bibfield  {title} {\bibinfo {title} {Dissipative modes, purcell factors, and directional beta factors in gold bowtie nanoantenna structures},\ }\href {https://doi.org/10.1103/PhysRevB.102.155301} {\bibfield  {journal} {\bibinfo  {journal} {Phys. Rev. B}\ }\textbf {\bibinfo {volume} {102}},\ \bibinfo {pages} {155301} (\bibinfo {year} {2020})}\BibitemShut {NoStop}%
\bibitem [{\citenamefont {Ren}\ \emph {et~al.}(2020{\natexlab{a}})\citenamefont {Ren}, \citenamefont {Franke}, \citenamefont {Knorr}, \citenamefont {Richter},\ and\ \citenamefont {Hughes}}]{ren_near-field_2020}%
  \BibitemOpen
  \bibfield  {author} {\bibinfo {author} {\bibfnamefont {J.}~\bibnamefont {Ren}}, \bibinfo {author} {\bibfnamefont {S.}~\bibnamefont {Franke}}, \bibinfo {author} {\bibfnamefont {A.}~\bibnamefont {Knorr}}, \bibinfo {author} {\bibfnamefont {M.}~\bibnamefont {Richter}},\ and\ \bibinfo {author} {\bibfnamefont {S.}~\bibnamefont {Hughes}},\ }\bibfield  {title} {\bibinfo {title} {Near-field to far-field transformations of optical quasinormal modes and efficient calculation of quantized quasinormal modes for open cavities and plasmonic resonators},\ }\href {https://doi.org/10.1103/PhysRevB.101.205402} {\bibfield  {journal} {\bibinfo  {journal} {Phys. Rev. B}\ }\textbf {\bibinfo {volume} {101}},\ \bibinfo {pages} {205402} (\bibinfo {year} {2020}{\natexlab{a}})}\BibitemShut {NoStop}%
\bibitem [{\citenamefont {Franke}\ \emph {et~al.}(2023)\citenamefont {Franke}, \citenamefont {Ren},\ and\ \citenamefont {Hughes}}]{PhysRevA.108.043502}%
  \BibitemOpen
  \bibfield  {author} {\bibinfo {author} {\bibfnamefont {S.}~\bibnamefont {Franke}}, \bibinfo {author} {\bibfnamefont {J.}~\bibnamefont {Ren}},\ and\ \bibinfo {author} {\bibfnamefont {S.}~\bibnamefont {Hughes}},\ }\bibfield  {title} {\bibinfo {title} {Impact of mode regularization for quasinormal-mode perturbation theories},\ }\href {https://doi.org/10.1103/PhysRevA.108.043502} {\bibfield  {journal} {\bibinfo  {journal} {Phys. Rev. A}\ }\textbf {\bibinfo {volume} {108}},\ \bibinfo {pages} {043502} (\bibinfo {year} {2023})}\BibitemShut {NoStop}%
\bibitem [{\citenamefont {Ren}\ \emph {et~al.}(2020{\natexlab{b}})\citenamefont {Ren}, \citenamefont {Franke}, \citenamefont {Knorr}, \citenamefont {Richter},\ and\ \citenamefont {Hughes}}]{PhysRevB.101.205402}%
  \BibitemOpen
  \bibfield  {author} {\bibinfo {author} {\bibfnamefont {J.}~\bibnamefont {Ren}}, \bibinfo {author} {\bibfnamefont {S.}~\bibnamefont {Franke}}, \bibinfo {author} {\bibfnamefont {A.}~\bibnamefont {Knorr}}, \bibinfo {author} {\bibfnamefont {M.}~\bibnamefont {Richter}},\ and\ \bibinfo {author} {\bibfnamefont {S.}~\bibnamefont {Hughes}},\ }\bibfield  {title} {\bibinfo {title} {Near-field to far-field transformations of optical quasinormal modes and efficient calculation of quantized quasinormal modes for open cavities and plasmonic resonators},\ }\href {https://doi.org/10.1103/PhysRevB.101.205402} {\bibfield  {journal} {\bibinfo  {journal} {Phys. Rev. B}\ }\textbf {\bibinfo {volume} {101}},\ \bibinfo {pages} {205402} (\bibinfo {year} {2020}{\natexlab{b}})}\BibitemShut {NoStop}%
\bibitem [{\citenamefont {Schelkunoff}(1936)}]{schelkunoff1936some}%
  \BibitemOpen
  \bibfield  {author} {\bibinfo {author} {\bibfnamefont {S.}~\bibnamefont {Schelkunoff}},\ }\bibfield  {title} {\bibinfo {title} {Some equivalence theorems of electromagnetics and their application to radiation problems},\ }\href {https://ieeexplore.ieee.org/document/6768910} {\bibfield  {journal} {\bibinfo  {journal} {Bell Sys. Tech. J}\ }\textbf {\bibinfo {volume} {15}},\ \bibinfo {pages} {92} (\bibinfo {year} {1936})}\BibitemShut {NoStop}%
\bibitem [{\citenamefont {Scheel}\ \emph {et~al.}(1999)\citenamefont {Scheel}, \citenamefont {Kn\"oll},\ and\ \citenamefont {Welsch}}]{Scheel}%
  \BibitemOpen
  \bibfield  {author} {\bibinfo {author} {\bibfnamefont {S.}~\bibnamefont {Scheel}}, \bibinfo {author} {\bibfnamefont {L.}~\bibnamefont {Kn\"oll}},\ and\ \bibinfo {author} {\bibfnamefont {D.-G.}\ \bibnamefont {Welsch}},\ }\bibfield  {title} {\bibinfo {title} {Spontaneous decay of an excited atom in an absorbing dielectric},\ }\href {https://doi.org/10.1103/PhysRevA.60.4094} {\bibfield  {journal} {\bibinfo  {journal} {Phys. Rev. A}\ }\textbf {\bibinfo {volume} {60}},\ \bibinfo {pages} {4094} (\bibinfo {year} {1999})}\BibitemShut {NoStop}%
\bibitem [{\citenamefont {Manga~Rao}\ and\ \citenamefont {Hughes}(2007)}]{PhysRevB.75.205437}%
  \BibitemOpen
  \bibfield  {author} {\bibinfo {author} {\bibfnamefont {V.~S.~C.}\ \bibnamefont {Manga~Rao}}\ and\ \bibinfo {author} {\bibfnamefont {S.}~\bibnamefont {Hughes}},\ }\bibfield  {title} {\bibinfo {title} {{Single quantum-dot Purcell factor and $\ensuremath{\beta}$ factor in a photonic crystal waveguide}},\ }\href {https://doi.org/10.1103/PhysRevB.75.205437} {\bibfield  {journal} {\bibinfo  {journal} {Phys. Rev. B}\ }\textbf {\bibinfo {volume} {75}},\ \bibinfo {pages} {205437} (\bibinfo {year} {2007})}\BibitemShut {NoStop}%
\bibitem [{\citenamefont {Ge}\ and\ \citenamefont {Hughes}(2014{\natexlab{b}})}]{Ge:14Dimer}%
  \BibitemOpen
  \bibfield  {author} {\bibinfo {author} {\bibfnamefont {R.-C.}\ \bibnamefont {Ge}}\ and\ \bibinfo {author} {\bibfnamefont {S.}~\bibnamefont {Hughes}},\ }\bibfield  {title} {\bibinfo {title} {Design of an efficient single photon source from a metallic nanorod dimer: a quasi-normal mode finite-difference time-domain approach},\ }\href {https://doi.org/10.1364/OL.39.004235} {\bibfield  {journal} {\bibinfo  {journal} {Opt. Lett.}\ }\textbf {\bibinfo {volume} {39}},\ \bibinfo {pages} {4235} (\bibinfo {year} {2014}{\natexlab{b}})}\BibitemShut {NoStop}%
\bibitem [{\citenamefont {Savasta}\ \emph {et~al.}(2021)\citenamefont {Savasta}, \citenamefont {Di~Stefano}, \citenamefont {Settineri}, \citenamefont {Zueco}, \citenamefont {Hughes},\ and\ \citenamefont {Nori}}]{Savasta2021May}%
  \BibitemOpen
  \bibfield  {author} {\bibinfo {author} {\bibfnamefont {S.}~\bibnamefont {Savasta}}, \bibinfo {author} {\bibfnamefont {O.}~\bibnamefont {Di~Stefano}}, \bibinfo {author} {\bibfnamefont {A.}~\bibnamefont {Settineri}}, \bibinfo {author} {\bibfnamefont {D.}~\bibnamefont {Zueco}}, \bibinfo {author} {\bibfnamefont {S.}~\bibnamefont {Hughes}},\ and\ \bibinfo {author} {\bibfnamefont {F.}~\bibnamefont {Nori}},\ }\bibfield  {title} {\bibinfo {title} {{Gauge principle and gauge invariance in two-level systems}},\ }\href {https://doi.org/10.1103/PhysRevA.103.053703} {\bibfield  {journal} {\bibinfo  {journal} {Phys. Rev. A}\ }\textbf {\bibinfo {volume} {103}},\ \bibinfo {pages} {053703} (\bibinfo {year} {2021})}\BibitemShut {NoStop}%
\bibitem [{\citenamefont {Settineri}\ \emph {et~al.}(2021)\citenamefont {Settineri}, \citenamefont {Di~Stefano}, \citenamefont {Zueco}, \citenamefont {Hughes}, \citenamefont {Savasta},\ and\ \citenamefont {Nori}}]{PhysRevResearch.3.023079}%
  \BibitemOpen
  \bibfield  {author} {\bibinfo {author} {\bibfnamefont {A.}~\bibnamefont {Settineri}}, \bibinfo {author} {\bibfnamefont {O.}~\bibnamefont {Di~Stefano}}, \bibinfo {author} {\bibfnamefont {D.}~\bibnamefont {Zueco}}, \bibinfo {author} {\bibfnamefont {S.}~\bibnamefont {Hughes}}, \bibinfo {author} {\bibfnamefont {S.}~\bibnamefont {Savasta}},\ and\ \bibinfo {author} {\bibfnamefont {F.}~\bibnamefont {Nori}},\ }\bibfield  {title} {\bibinfo {title} {Gauge freedom, quantum measurements, and time-dependent interactions in cavity qed},\ }\href {https://doi.org/10.1103/PhysRevResearch.3.023079} {\bibfield  {journal} {\bibinfo  {journal} {Phys. Rev. Res.}\ }\textbf {\bibinfo {volume} {3}},\ \bibinfo {pages} {023079} (\bibinfo {year} {2021})}\BibitemShut {NoStop}%
\bibitem [{\citenamefont {Lamb}\ \emph {et~al.}(1987)\citenamefont {Lamb}, \citenamefont {Schlicher},\ and\ \citenamefont {Scully}}]{Lamb1987Sep}%
  \BibitemOpen
  \bibfield  {author} {\bibinfo {author} {\bibfnamefont {W.~E.}\ \bibnamefont {Lamb}}, \bibinfo {author} {\bibfnamefont {R.~R.}\ \bibnamefont {Schlicher}},\ and\ \bibinfo {author} {\bibfnamefont {M.~O.}\ \bibnamefont {Scully}},\ }\bibfield  {title} {\bibinfo {title} {{Matter-field interaction in atomic physics and quantum optics}},\ }\href {https://doi.org/10.1103/PhysRevA.36.2763} {\bibfield  {journal} {\bibinfo  {journal} {Phys. Rev. A}\ }\textbf {\bibinfo {volume} {36}},\ \bibinfo {pages} {2763} (\bibinfo {year} {1987})}\BibitemShut {NoStop}%
\bibitem [{\citenamefont {Milonni}\ \emph {et~al.}(1989)\citenamefont {Milonni}, \citenamefont {Cook},\ and\ \citenamefont {Ackerhalt}}]{Milonni1989Oct}%
  \BibitemOpen
  \bibfield  {author} {\bibinfo {author} {\bibfnamefont {P.~W.}\ \bibnamefont {Milonni}}, \bibinfo {author} {\bibfnamefont {R.~J.}\ \bibnamefont {Cook}},\ and\ \bibinfo {author} {\bibfnamefont {J.~R.}\ \bibnamefont {Ackerhalt}},\ }\bibfield  {title} {\bibinfo {title} {{Natural line shape}},\ }\href {https://doi.org/10.1103/PhysRevA.40.3764} {\bibfield  {journal} {\bibinfo  {journal} {Phys. Rev. A}\ }\textbf {\bibinfo {volume} {40}},\ \bibinfo {pages} {3764} (\bibinfo {year} {1989})}\BibitemShut {NoStop}%
\bibitem [{\citenamefont {De~Bernardis}\ \emph {et~al.}(2018)\citenamefont {De~Bernardis}, \citenamefont {Pilar}, \citenamefont {Jaako}, \citenamefont {De~Liberato},\ and\ \citenamefont {Rabl}}]{DeBernardis2018Nov}%
  \BibitemOpen
  \bibfield  {author} {\bibinfo {author} {\bibfnamefont {D.}~\bibnamefont {De~Bernardis}}, \bibinfo {author} {\bibfnamefont {P.}~\bibnamefont {Pilar}}, \bibinfo {author} {\bibfnamefont {T.}~\bibnamefont {Jaako}}, \bibinfo {author} {\bibfnamefont {S.}~\bibnamefont {De~Liberato}},\ and\ \bibinfo {author} {\bibfnamefont {P.}~\bibnamefont {Rabl}},\ }\bibfield  {title} {\bibinfo {title} {{Breakdown of gauge invariance in ultrastrong-coupling cavity QED}},\ }\href {https://doi.org/10.1103/PhysRevA.98.053819} {\bibfield  {journal} {\bibinfo  {journal} {Phys. Rev. A}\ }\textbf {\bibinfo {volume} {98}},\ \bibinfo {pages} {053819} (\bibinfo {year} {2018})}\BibitemShut {NoStop}%
\bibitem [{\citenamefont {Taylor}\ \emph {et~al.}(2020)\citenamefont {Taylor}, \citenamefont {Mandal}, \citenamefont {Zhou},\ and\ \citenamefont {Huo}}]{Taylor2020Sep}%
  \BibitemOpen
  \bibfield  {author} {\bibinfo {author} {\bibfnamefont {M.~A.~D.}\ \bibnamefont {Taylor}}, \bibinfo {author} {\bibfnamefont {A.}~\bibnamefont {Mandal}}, \bibinfo {author} {\bibfnamefont {W.}~\bibnamefont {Zhou}},\ and\ \bibinfo {author} {\bibfnamefont {P.}~\bibnamefont {Huo}},\ }\bibfield  {title} {\bibinfo {title} {{Resolution of Gauge Ambiguities in Molecular Cavity Quantum Electrodynamics}},\ }\href {https://doi.org/10.1103/PhysRevLett.125.123602} {\bibfield  {journal} {\bibinfo  {journal} {Phys. Rev. Lett.}\ }\textbf {\bibinfo {volume} {125}},\ \bibinfo {pages} {123602} (\bibinfo {year} {2020})}\BibitemShut {NoStop}%
\bibitem [{\citenamefont {Taylor}\ \emph {et~al.}(2022)\citenamefont {Taylor}, \citenamefont {Mandal}, \citenamefont {Huo},\ and\ \citenamefont {Huo}}]{Taylor2022Mar}%
  \BibitemOpen
  \bibfield  {author} {\bibinfo {author} {\bibfnamefont {M.~A.~D.}\ \bibnamefont {Taylor}}, \bibinfo {author} {\bibfnamefont {A.}~\bibnamefont {Mandal}}, \bibinfo {author} {\bibfnamefont {P.}~\bibnamefont {Huo}},\ and\ \bibinfo {author} {\bibfnamefont {P.}~\bibnamefont {Huo}},\ }\bibfield  {title} {\bibinfo {title} {{Resolving ambiguities of the mode truncation in cavity quantum electrodynamics}},\ }\href {https://doi.org/10.1364/OL.450228} {\bibfield  {journal} {\bibinfo  {journal} {Opt. Lett.}\ }\textbf {\bibinfo {volume} {47}},\ \bibinfo {pages} {1446} (\bibinfo {year} {2022})}\BibitemShut {NoStop}%
\bibitem [{\citenamefont {Wubs}\ \emph {et~al.}(2003)\citenamefont {Wubs}, \citenamefont {Suttorp},\ and\ \citenamefont {Lagendijk}}]{PhysRevA.68.013822}%
  \BibitemOpen
  \bibfield  {author} {\bibinfo {author} {\bibfnamefont {M.}~\bibnamefont {Wubs}}, \bibinfo {author} {\bibfnamefont {L.~G.}\ \bibnamefont {Suttorp}},\ and\ \bibinfo {author} {\bibfnamefont {A.}~\bibnamefont {Lagendijk}},\ }\bibfield  {title} {\bibinfo {title} {Multipole interaction between atoms and their photonic environment},\ }\href {https://doi.org/10.1103/PhysRevA.68.013822} {\bibfield  {journal} {\bibinfo  {journal} {Phys. Rev. A}\ }\textbf {\bibinfo {volume} {68}},\ \bibinfo {pages} {013822} (\bibinfo {year} {2003})}\BibitemShut {NoStop}%
\bibitem [{\citenamefont {Settineri}\ \emph {et~al.}(2018)\citenamefont {Settineri}, \citenamefont {Macr{\ifmmode\acute{\imath}\else\'{\i}\fi}}, \citenamefont {Ridolfo}, \citenamefont {Di~Stefano}, \citenamefont {Kockum}, \citenamefont {Nori},\ and\ \citenamefont {Savasta}}]{Settineri2018Nov}%
  \BibitemOpen
  \bibfield  {author} {\bibinfo {author} {\bibfnamefont {A.}~\bibnamefont {Settineri}}, \bibinfo {author} {\bibfnamefont {V.}~\bibnamefont {Macr{\ifmmode\acute{\imath}\else\'{\i}\fi}}}, \bibinfo {author} {\bibfnamefont {A.}~\bibnamefont {Ridolfo}}, \bibinfo {author} {\bibfnamefont {O.}~\bibnamefont {Di~Stefano}}, \bibinfo {author} {\bibfnamefont {A.~F.}\ \bibnamefont {Kockum}}, \bibinfo {author} {\bibfnamefont {F.}~\bibnamefont {Nori}},\ and\ \bibinfo {author} {\bibfnamefont {S.}~\bibnamefont {Savasta}},\ }\bibfield  {title} {\bibinfo {title} {{Dissipation and thermal noise in hybrid quantum systems in the ultrastrong-coupling regime}},\ }\href {https://doi.org/10.1103/PhysRevA.98.053834} {\bibfield  {journal} {\bibinfo  {journal} {Phys. Rev. A}\ }\textbf {\bibinfo {volume} {98}},\ \bibinfo {pages} {053834} (\bibinfo {year} {2018})}\BibitemShut {NoStop}%
\bibitem [{\citenamefont {Bamba}\ and\ \citenamefont {Ogawa}(2014)}]{Bamba2014Feb}%
  \BibitemOpen
  \bibfield  {author} {\bibinfo {author} {\bibfnamefont {M.}~\bibnamefont {Bamba}}\ and\ \bibinfo {author} {\bibfnamefont {T.}~\bibnamefont {Ogawa}},\ }\bibfield  {title} {\bibinfo {title} {{Recipe for the Hamiltonian of system-environment coupling applicable to the ultrastrong-light-matter-interaction regime}},\ }\href {https://doi.org/10.1103/PhysRevA.89.023817} {\bibfield  {journal} {\bibinfo  {journal} {Phys. Rev. A}\ }\textbf {\bibinfo {volume} {89}},\ \bibinfo {pages} {023817} (\bibinfo {year} {2014})}\BibitemShut {NoStop}%
\bibitem [{\citenamefont {Toyli}\ \emph {et~al.}(2016)\citenamefont {Toyli}, \citenamefont {Eddins}, \citenamefont {Boutin}, \citenamefont {Puri}, \citenamefont {Hover}, \citenamefont {Bolkhovsky}, \citenamefont {Oliver}, \citenamefont {Blais},\ and\ \citenamefont {Siddiqi}}]{PhysRevX.6.031004}%
  \BibitemOpen
  \bibfield  {author} {\bibinfo {author} {\bibfnamefont {D.~M.}\ \bibnamefont {Toyli}}, \bibinfo {author} {\bibfnamefont {A.~W.}\ \bibnamefont {Eddins}}, \bibinfo {author} {\bibfnamefont {S.}~\bibnamefont {Boutin}}, \bibinfo {author} {\bibfnamefont {S.}~\bibnamefont {Puri}}, \bibinfo {author} {\bibfnamefont {D.}~\bibnamefont {Hover}}, \bibinfo {author} {\bibfnamefont {V.}~\bibnamefont {Bolkhovsky}}, \bibinfo {author} {\bibfnamefont {W.~D.}\ \bibnamefont {Oliver}}, \bibinfo {author} {\bibfnamefont {A.}~\bibnamefont {Blais}},\ and\ \bibinfo {author} {\bibfnamefont {I.}~\bibnamefont {Siddiqi}},\ }\bibfield  {title} {\bibinfo {title} {{Resonance Fluorescence from an Artificial Atom in Squeezed Vacuum}},\ }\href {https://doi.org/10.1103/PhysRevX.6.031004} {\bibfield  {journal} {\bibinfo  {journal} {Phys. Rev. X}\ }\textbf {\bibinfo {volume} {6}},\ \bibinfo {pages} {031004} (\bibinfo {year} {2016})}\BibitemShut {NoStop}%
\bibitem [{\citenamefont {Carmichael}\ \emph {et~al.}(1987)\citenamefont {Carmichael}, \citenamefont {Lane},\ and\ \citenamefont {Walls}}]{PhysRevLett.58.2539}%
  \BibitemOpen
  \bibfield  {author} {\bibinfo {author} {\bibfnamefont {H.~J.}\ \bibnamefont {Carmichael}}, \bibinfo {author} {\bibfnamefont {A.~S.}\ \bibnamefont {Lane}},\ and\ \bibinfo {author} {\bibfnamefont {D.~F.}\ \bibnamefont {Walls}},\ }\bibfield  {title} {\bibinfo {title} {{Resonance Fluorescence from an Atom in a Squeezed Vacuum}},\ }\href {https://doi.org/10.1103/PhysRevLett.58.2539} {\bibfield  {journal} {\bibinfo  {journal} {Phys. Rev. Lett.}\ }\textbf {\bibinfo {volume} {58}},\ \bibinfo {pages} {2539} (\bibinfo {year} {1987})}\BibitemShut {NoStop}%
\bibitem [{\citenamefont {Mercurio}\ \emph {et~al.}(2022)\citenamefont {Mercurio}, \citenamefont {Macr{\ifmmode\grave{\imath}\else\`{\i}\fi}}, \citenamefont {Gustin}, \citenamefont {Hughes}, \citenamefont {Savasta},\ and\ \citenamefont {Nori}}]{Mercurio2022Apr}%
  \BibitemOpen
  \bibfield  {author} {\bibinfo {author} {\bibfnamefont {A.}~\bibnamefont {Mercurio}}, \bibinfo {author} {\bibfnamefont {V.}~\bibnamefont {Macr{\ifmmode\grave{\imath}\else\`{\i}\fi}}}, \bibinfo {author} {\bibfnamefont {C.}~\bibnamefont {Gustin}}, \bibinfo {author} {\bibfnamefont {S.}~\bibnamefont {Hughes}}, \bibinfo {author} {\bibfnamefont {S.}~\bibnamefont {Savasta}},\ and\ \bibinfo {author} {\bibfnamefont {F.}~\bibnamefont {Nori}},\ }\bibfield  {title} {\bibinfo {title} {{Regimes of cavity QED under incoherent excitation: From weak to deep strong coupling}},\ }\href {https://doi.org/10.1103/PhysRevResearch.4.023048} {\bibfield  {journal} {\bibinfo  {journal} {Phys. Rev. Res.}\ }\textbf {\bibinfo {volume} {4}},\ \bibinfo {pages} {023048} (\bibinfo {year} {2022})}\BibitemShut {NoStop}%
\bibitem [{\citenamefont {Gustin}(2025)}]{gustin2025gaugeinvariancenaturallineshape}%
  \BibitemOpen
  \bibfield  {author} {\bibinfo {author} {\bibfnamefont {C.}~\bibnamefont {Gustin}},\ }\href {https://arxiv.org/abs/2501.06594} {\bibinfo {title} {Gauge invariance of the natural lineshape and dissipative dynamics of a two-level atom}} (\bibinfo {year} {2025}),\ \Eprint {https://arxiv.org/abs/2501.06594} {arXiv:2501.06594} \BibitemShut {NoStop}%
\bibitem [{\citenamefont {Shahbazyan}(2021)}]{PhysRevB.103.045421}%
  \BibitemOpen
  \bibfield  {author} {\bibinfo {author} {\bibfnamefont {T.~V.}\ \bibnamefont {Shahbazyan}},\ }\bibfield  {title} {\bibinfo {title} {Interacting quantum plasmons in metal-dielectric structures},\ }\href {https://doi.org/10.1103/PhysRevB.103.045421} {\bibfield  {journal} {\bibinfo  {journal} {Phys. Rev. B}\ }\textbf {\bibinfo {volume} {103}},\ \bibinfo {pages} {045421} (\bibinfo {year} {2021})}\BibitemShut {NoStop}%
\bibitem [{\citenamefont {Shahbazyan}(2022)}]{PhysRevB.105.245411}%
  \BibitemOpen
  \bibfield  {author} {\bibinfo {author} {\bibfnamefont {T.~V.}\ \bibnamefont {Shahbazyan}},\ }\bibfield  {title} {\bibinfo {title} {{Non-Markovian effects for hybrid plasmonic systems in the strong coupling regime}},\ }\href {https://doi.org/10.1103/PhysRevB.105.245411} {\bibfield  {journal} {\bibinfo  {journal} {Phys. Rev. B}\ }\textbf {\bibinfo {volume} {105}},\ \bibinfo {pages} {245411} (\bibinfo {year} {2022})}\BibitemShut {NoStop}%
\bibitem [{\citenamefont {Kountouris}\ \emph {et~al.}(2022)\citenamefont {Kountouris}, \citenamefont {M{\o}rk}, \citenamefont {Denning},\ and\ \citenamefont {Kristensen}}]{Kountouris:22}%
  \BibitemOpen
  \bibfield  {author} {\bibinfo {author} {\bibfnamefont {G.}~\bibnamefont {Kountouris}}, \bibinfo {author} {\bibfnamefont {J.}~\bibnamefont {M{\o}rk}}, \bibinfo {author} {\bibfnamefont {E.~V.}\ \bibnamefont {Denning}},\ and\ \bibinfo {author} {\bibfnamefont {P.~T.}\ \bibnamefont {Kristensen}},\ }\bibfield  {title} {\bibinfo {title} {Modal properties of dielectric bowtie cavities with deep sub-wavelength confinement},\ }\href {https://doi.org/10.1364/OE.472793} {\bibfield  {journal} {\bibinfo  {journal} {Opt. Express}\ }\textbf {\bibinfo {volume} {30}},\ \bibinfo {pages} {40367} (\bibinfo {year} {2022})}\BibitemShut {NoStop}%
\bibitem [{\citenamefont {Kountouris}(2024)}]{Kountouris_thesis}%
  \BibitemOpen
  \bibfield  {author} {\bibinfo {author} {\bibfnamefont {G.}~\bibnamefont {Kountouris}},\ }\emph {\bibinfo {title} {Quantum optics of structures with extreme dielectric confinement}},\ \href@noop {} {Ph.D. thesis},\ \bibinfo  {school} {Technical University of Denmark} (\bibinfo {year} {2024})\BibitemShut {NoStop}%
\bibitem [{\citenamefont {Mercurio}\ \emph {et~al.}(2023)\citenamefont {Mercurio}, \citenamefont {Abo}, \citenamefont {Mauceri}, \citenamefont {Russo}, \citenamefont {Macr\`{\i}}, \citenamefont {Miranowicz}, \citenamefont {Savasta},\ and\ \citenamefont {Di~Stefano}}]{PhysRevLett.130.123601}%
  \BibitemOpen
  \bibfield  {author} {\bibinfo {author} {\bibfnamefont {A.}~\bibnamefont {Mercurio}}, \bibinfo {author} {\bibfnamefont {S.}~\bibnamefont {Abo}}, \bibinfo {author} {\bibfnamefont {F.}~\bibnamefont {Mauceri}}, \bibinfo {author} {\bibfnamefont {E.}~\bibnamefont {Russo}}, \bibinfo {author} {\bibfnamefont {V.}~\bibnamefont {Macr\`{\i}}}, \bibinfo {author} {\bibfnamefont {A.}~\bibnamefont {Miranowicz}}, \bibinfo {author} {\bibfnamefont {S.}~\bibnamefont {Savasta}},\ and\ \bibinfo {author} {\bibfnamefont {O.}~\bibnamefont {Di~Stefano}},\ }\bibfield  {title} {\bibinfo {title} {{Pure Dephasing of Light-Matter Systems in the Ultrastrong and Deep-Strong Coupling Regimes}},\ }\href {https://doi.org/10.1103/PhysRevLett.130.123601} {\bibfield  {journal} {\bibinfo  {journal} {Phys. Rev. Lett.}\ }\textbf {\bibinfo {volume} {130}},\ \bibinfo {pages} {123601} (\bibinfo {year} {2023})}\BibitemShut {NoStop}%
\bibitem [{\citenamefont {Reiter}\ \emph {et~al.}(2019)\citenamefont {Reiter}, \citenamefont {Kuhn},\ and\ \citenamefont {Axt}}]{reiter_distinctive_2019}%
  \BibitemOpen
  \bibfield  {author} {\bibinfo {author} {\bibfnamefont {D.~E.}\ \bibnamefont {Reiter}}, \bibinfo {author} {\bibfnamefont {T.}~\bibnamefont {Kuhn}},\ and\ \bibinfo {author} {\bibfnamefont {V.~M.}\ \bibnamefont {Axt}},\ }\bibfield  {title} {\bibinfo {title} {Distinctive characteristics of carrier-phonon interactions in optically driven semiconductor quantum dots},\ }\href {https://doi.org/10.1080/23746149.2019.1655478} {\bibfield  {journal} {\bibinfo  {journal} {Adv. Phys.: X}\ }\textbf {\bibinfo {volume} {4}},\ \bibinfo {pages} {1655478} (\bibinfo {year} {2019})}\BibitemShut {NoStop}%
\bibitem [{\citenamefont {Canales}\ \emph {et~al.}(2023)\citenamefont {Canales}, \citenamefont {Karmstrand}, \citenamefont {Baranov}, \citenamefont {Antosiewicz},\ and\ \citenamefont {Shegai}}]{canales_polaritonic_2023}%
  \BibitemOpen
  \bibfield  {author} {\bibinfo {author} {\bibfnamefont {A.}~\bibnamefont {Canales}}, \bibinfo {author} {\bibfnamefont {T.}~\bibnamefont {Karmstrand}}, \bibinfo {author} {\bibfnamefont {D.~G.}\ \bibnamefont {Baranov}}, \bibinfo {author} {\bibfnamefont {T.~J.}\ \bibnamefont {Antosiewicz}},\ and\ \bibinfo {author} {\bibfnamefont {T.~O.}\ \bibnamefont {Shegai}},\ }\bibfield  {title} {\bibinfo {title} {Polaritonic linewidth asymmetry in the strong and ultrastrong coupling regime},\ }\href {https://doi.org/10.1515/nanoph-2023-0492} {\bibfield  {journal} {\bibinfo  {journal} {Nanophotonics}\ }\textbf {\bibinfo {volume} {12}},\ \bibinfo {pages} {4073} (\bibinfo {year} {2023})}\BibitemShut {NoStop}%
\bibitem [{\citenamefont {Garziano}\ \emph {et~al.}(2020)\citenamefont {Garziano}, \citenamefont {Settineri}, \citenamefont {Di~Stefano}, \citenamefont {Savasta},\ and\ \citenamefont {Nori}}]{Garziano2020Aug}%
  \BibitemOpen
  \bibfield  {author} {\bibinfo {author} {\bibfnamefont {L.}~\bibnamefont {Garziano}}, \bibinfo {author} {\bibfnamefont {A.}~\bibnamefont {Settineri}}, \bibinfo {author} {\bibfnamefont {O.}~\bibnamefont {Di~Stefano}}, \bibinfo {author} {\bibfnamefont {S.}~\bibnamefont {Savasta}},\ and\ \bibinfo {author} {\bibfnamefont {F.}~\bibnamefont {Nori}},\ }\bibfield  {title} {\bibinfo {title} {{Gauge invariance of the Dicke and Hopfield models}},\ }\href {https://doi.org/10.1103/PhysRevA.102.023718} {\bibfield  {journal} {\bibinfo  {journal} {Phys. Rev. A}\ }\textbf {\bibinfo {volume} {102}},\ \bibinfo {pages} {023718} (\bibinfo {year} {2020})}\BibitemShut {NoStop}%
\bibitem [{\citenamefont {Rajabali}\ \emph {et~al.}(2021)\citenamefont {Rajabali}, \citenamefont {Cortese}, \citenamefont {Beck}, \citenamefont {De~Liberato}, \citenamefont {Faist},\ and\ \citenamefont {Scalari}}]{rajabali_polaritonic_2021}%
  \BibitemOpen
  \bibfield  {author} {\bibinfo {author} {\bibfnamefont {S.}~\bibnamefont {Rajabali}}, \bibinfo {author} {\bibfnamefont {E.}~\bibnamefont {Cortese}}, \bibinfo {author} {\bibfnamefont {M.}~\bibnamefont {Beck}}, \bibinfo {author} {\bibfnamefont {S.}~\bibnamefont {De~Liberato}}, \bibinfo {author} {\bibfnamefont {J.}~\bibnamefont {Faist}},\ and\ \bibinfo {author} {\bibfnamefont {G.}~\bibnamefont {Scalari}},\ }\bibfield  {title} {\bibinfo {title} {Polaritonic nonlocality in light–matter interaction},\ }\href {https://doi.org/10.1038/s41566-021-00854-3} {\bibfield  {journal} {\bibinfo  {journal} {Nat. Photon.}\ }\textbf {\bibinfo {volume} {15}},\ \bibinfo {pages} {690} (\bibinfo {year} {2021})}\BibitemShut {NoStop}%
\bibitem [{\citenamefont {Baranov}\ \emph {et~al.}(2020)\citenamefont {Baranov}, \citenamefont {Munkhbat}, \citenamefont {Zhukova}, \citenamefont {Bisht}, \citenamefont {Canales}, \citenamefont {Rousseaux}, \citenamefont {Johansson}, \citenamefont {Antosiewicz},\ and\ \citenamefont {Shegai}}]{baranov_ultrastrong_2020}%
  \BibitemOpen
  \bibfield  {author} {\bibinfo {author} {\bibfnamefont {D.~G.}\ \bibnamefont {Baranov}}, \bibinfo {author} {\bibfnamefont {B.}~\bibnamefont {Munkhbat}}, \bibinfo {author} {\bibfnamefont {E.}~\bibnamefont {Zhukova}}, \bibinfo {author} {\bibfnamefont {A.}~\bibnamefont {Bisht}}, \bibinfo {author} {\bibfnamefont {A.}~\bibnamefont {Canales}}, \bibinfo {author} {\bibfnamefont {B.}~\bibnamefont {Rousseaux}}, \bibinfo {author} {\bibfnamefont {G.}~\bibnamefont {Johansson}}, \bibinfo {author} {\bibfnamefont {T.~J.}\ \bibnamefont {Antosiewicz}},\ and\ \bibinfo {author} {\bibfnamefont {T.}~\bibnamefont {Shegai}},\ }\bibfield  {title} {\bibinfo {title} {Ultrastrong coupling between nanoparticle plasmons and cavity photons at ambient conditions},\ }\href {https://doi.org/10.1038/s41467-020-16524-x} {\bibfield  {journal} {\bibinfo  {journal} {Nat. Commun.}\ }\textbf {\bibinfo {volume} {11}},\ \bibinfo {pages} {2715} (\bibinfo {year} {2020})}\BibitemShut {NoStop}%
\bibitem [{\citenamefont {Holstein}\ and\ \citenamefont {Primakoff}(1940)}]{PhysRev.58.1098}%
  \BibitemOpen
  \bibfield  {author} {\bibinfo {author} {\bibfnamefont {T.}~\bibnamefont {Holstein}}\ and\ \bibinfo {author} {\bibfnamefont {H.}~\bibnamefont {Primakoff}},\ }\bibfield  {title} {\bibinfo {title} {{Field Dependence of the Intrinsic Domain Magnetization of a Ferromagnet}},\ }\href {https://doi.org/10.1103/PhysRev.58.1098} {\bibfield  {journal} {\bibinfo  {journal} {Phys. Rev.}\ }\textbf {\bibinfo {volume} {58}},\ \bibinfo {pages} {1098} (\bibinfo {year} {1940})}\BibitemShut {NoStop}%
\bibitem [{\citenamefont {Ressayre}\ and\ \citenamefont {Tallet}(1975)}]{PhysRevA.11.981}%
  \BibitemOpen
  \bibfield  {author} {\bibinfo {author} {\bibfnamefont {E.}~\bibnamefont {Ressayre}}\ and\ \bibinfo {author} {\bibfnamefont {A.}~\bibnamefont {Tallet}},\ }\bibfield  {title} {\bibinfo {title} {{Holstein-Primakoff transformation for the study of cooperative emission of radiation}},\ }\href {https://doi.org/10.1103/PhysRevA.11.981} {\bibfield  {journal} {\bibinfo  {journal} {Phys. Rev. A}\ }\textbf {\bibinfo {volume} {11}},\ \bibinfo {pages} {981} (\bibinfo {year} {1975})}\BibitemShut {NoStop}%
\bibitem [{\citenamefont {Lambert}\ \emph {et~al.}(2004)\citenamefont {Lambert}, \citenamefont {Emary},\ and\ \citenamefont {Brandes}}]{PhysRevLett.92.073602}%
  \BibitemOpen
  \bibfield  {author} {\bibinfo {author} {\bibfnamefont {N.}~\bibnamefont {Lambert}}, \bibinfo {author} {\bibfnamefont {C.}~\bibnamefont {Emary}},\ and\ \bibinfo {author} {\bibfnamefont {T.}~\bibnamefont {Brandes}},\ }\bibfield  {title} {\bibinfo {title} {{Entanglement and the Phase Transition in Single-Mode Superradiance}},\ }\href {https://doi.org/10.1103/PhysRevLett.92.073602} {\bibfield  {journal} {\bibinfo  {journal} {Phys. Rev. Lett.}\ }\textbf {\bibinfo {volume} {92}},\ \bibinfo {pages} {073602} (\bibinfo {year} {2004})}\BibitemShut {NoStop}%
\bibitem [{\citenamefont {Molesky}\ \emph {et~al.}(2018)\citenamefont {Molesky}, \citenamefont {Lin}, \citenamefont {Piggott}, \citenamefont {Jin}, \citenamefont {Vucković},\ and\ \citenamefont {Rodriguez}}]{molesky_inverse_2018}%
  \BibitemOpen
  \bibfield  {author} {\bibinfo {author} {\bibfnamefont {S.}~\bibnamefont {Molesky}}, \bibinfo {author} {\bibfnamefont {Z.}~\bibnamefont {Lin}}, \bibinfo {author} {\bibfnamefont {A.~Y.}\ \bibnamefont {Piggott}}, \bibinfo {author} {\bibfnamefont {W.}~\bibnamefont {Jin}}, \bibinfo {author} {\bibfnamefont {J.}~\bibnamefont {Vucković}},\ and\ \bibinfo {author} {\bibfnamefont {A.~W.}\ \bibnamefont {Rodriguez}},\ }\bibfield  {title} {\bibinfo {title} {Inverse design in nanophotonics},\ }\href {https://doi.org/10.1038/s41566-018-0246-9} {\bibfield  {journal} {\bibinfo  {journal} {Nat. Photon.}\ }\textbf {\bibinfo {volume} {12}},\ \bibinfo {pages} {659} (\bibinfo {year} {2018})}\BibitemShut {NoStop}%
\bibitem [{\citenamefont {Albrechtsen}\ \emph {et~al.}(2022)\citenamefont {Albrechtsen}, \citenamefont {Lahijani}, \citenamefont {Christiansen}, \citenamefont {Nguyen}, \citenamefont {Casses}, \citenamefont {Hansen}, \citenamefont {Stenger}, \citenamefont {Simgund}, \citenamefont {Jansen}, \citenamefont {Mork},\ and\ \citenamefont {Stobbe}}]{albrechtsen2022}%
  \BibitemOpen
  \bibfield  {author} {\bibinfo {author} {\bibfnamefont {M.}~\bibnamefont {Albrechtsen}}, \bibinfo {author} {\bibfnamefont {B.~V.}\ \bibnamefont {Lahijani}}, \bibinfo {author} {\bibfnamefont {R.~E.}\ \bibnamefont {Christiansen}}, \bibinfo {author} {\bibfnamefont {V.~T.~H.}\ \bibnamefont {Nguyen}}, \bibinfo {author} {\bibfnamefont {L.~N.}\ \bibnamefont {Casses}}, \bibinfo {author} {\bibfnamefont {S.~E.}\ \bibnamefont {Hansen}}, \bibinfo {author} {\bibfnamefont {N.}~\bibnamefont {Stenger}}, \bibinfo {author} {\bibfnamefont {O.}~\bibnamefont {Simgund}}, \bibinfo {author} {\bibfnamefont {H.}~\bibnamefont {Jansen}}, \bibinfo {author} {\bibfnamefont {J.}~\bibnamefont {Mork}},\ and\ \bibinfo {author} {\bibfnamefont {S.}~\bibnamefont {Stobbe}},\ }\bibfield  {title} {\bibinfo {title} {Nanometer-scale photon confinement in topology-optimized dielectric cavities},\ }\href {https://www.nature.com/articles/s41467-022-33874-w} {\bibfield  {journal} {\bibinfo  {journal} {Nat. Commun.}\ }\textbf {\bibinfo {volume} {13}},\
  \bibinfo {pages} {6281} (\bibinfo {year} {2022})}\BibitemShut {NoStop}%
\bibitem [{\citenamefont {Babar}\ \emph {et~al.}(2023)\citenamefont {Babar}, \citenamefont {Weis}, \citenamefont {Tsoukalas}, \citenamefont {Kadkhodazadeh}, \citenamefont {Arregui}, \citenamefont {Vosoughi~Lahijani},\ and\ \citenamefont {Stobbe}}]{babar_self-assembled_2023}%
  \BibitemOpen
  \bibfield  {author} {\bibinfo {author} {\bibfnamefont {A.~N.}\ \bibnamefont {Babar}}, \bibinfo {author} {\bibfnamefont {T.~A.~S.}\ \bibnamefont {Weis}}, \bibinfo {author} {\bibfnamefont {K.}~\bibnamefont {Tsoukalas}}, \bibinfo {author} {\bibfnamefont {S.}~\bibnamefont {Kadkhodazadeh}}, \bibinfo {author} {\bibfnamefont {G.}~\bibnamefont {Arregui}}, \bibinfo {author} {\bibfnamefont {B.}~\bibnamefont {Vosoughi~Lahijani}},\ and\ \bibinfo {author} {\bibfnamefont {S.}~\bibnamefont {Stobbe}},\ }\bibfield  {title} {\bibinfo {title} {Self-assembled photonic cavities with atomic-scale confinement},\ }\href {https://doi.org/10.1038/s41586-023-06736-8} {\bibfield  {journal} {\bibinfo  {journal} {Nature}\ }\textbf {\bibinfo {volume} {624}},\ \bibinfo {pages} {57} (\bibinfo {year} {2023})}\BibitemShut {NoStop}%
\bibitem [{\citenamefont {Xiong}\ \emph {et~al.}(2024)\citenamefont {Xiong}, \citenamefont {Christiansen}, \citenamefont {Schröder}, \citenamefont {Yu}, \citenamefont {Casses}, \citenamefont {Semenova}, \citenamefont {Yvind}, \citenamefont {Stenger}, \citenamefont {Sigmund},\ and\ \citenamefont {Mørk}}]{xiong_experimental_2024}%
  \BibitemOpen
  \bibfield  {author} {\bibinfo {author} {\bibfnamefont {M.}~\bibnamefont {Xiong}}, \bibinfo {author} {\bibfnamefont {R.~E.}\ \bibnamefont {Christiansen}}, \bibinfo {author} {\bibfnamefont {F.}~\bibnamefont {Schröder}}, \bibinfo {author} {\bibfnamefont {Y.}~\bibnamefont {Yu}}, \bibinfo {author} {\bibfnamefont {L.~N.}\ \bibnamefont {Casses}}, \bibinfo {author} {\bibfnamefont {E.}~\bibnamefont {Semenova}}, \bibinfo {author} {\bibfnamefont {K.}~\bibnamefont {Yvind}}, \bibinfo {author} {\bibfnamefont {N.}~\bibnamefont {Stenger}}, \bibinfo {author} {\bibfnamefont {O.}~\bibnamefont {Sigmund}},\ and\ \bibinfo {author} {\bibfnamefont {J.}~\bibnamefont {Mørk}},\ }\bibfield  {title} {\bibinfo {title} {Experimental realization of deep sub-wavelength confinement of light in a topology-optimized {InP} nanocavity},\ }\href {https://doi.org/10.1364/OME.513625} {\bibfield  {journal} {\bibinfo  {journal} {Opt. Mater. Express}\ }\textbf {\bibinfo {volume} {14}},\ \bibinfo {pages} {397} (\bibinfo {year} {2024})}\BibitemShut
  {NoStop}%
\bibitem [{\citenamefont {Nussbaum}\ \emph {et~al.}(2022)\citenamefont {Nussbaum}, \citenamefont {Rotenberg},\ and\ \citenamefont {Hughes}}]{PhysRevA.106.033514}%
  \BibitemOpen
  \bibfield  {author} {\bibinfo {author} {\bibfnamefont {E.}~\bibnamefont {Nussbaum}}, \bibinfo {author} {\bibfnamefont {N.}~\bibnamefont {Rotenberg}},\ and\ \bibinfo {author} {\bibfnamefont {S.}~\bibnamefont {Hughes}},\ }\bibfield  {title} {\bibinfo {title} {Optimizing the chiral {Purcell} factor for unidirectional single-photon emitters in topological photonic crystal waveguides using inverse design},\ }\href {https://doi.org/10.1103/PhysRevA.106.033514} {\bibfield  {journal} {\bibinfo  {journal} {Phys. Rev. A}\ }\textbf {\bibinfo {volume} {106}},\ \bibinfo {pages} {033514} (\bibinfo {year} {2022})}\BibitemShut {NoStop}%
\bibitem [{\citenamefont {Chikkaraddy}\ \emph {et~al.}(2016)\citenamefont {Chikkaraddy}, \citenamefont {De~Nijs}, \citenamefont {Benz}, \citenamefont {Barrow}, \citenamefont {Scherman}, \citenamefont {Rosta}, \citenamefont {Demetriadou}, \citenamefont {Fox}, \citenamefont {Hess},\ and\ \citenamefont {Baumberg}}]{chikkaraddy2016single}%
  \BibitemOpen
  \bibfield  {author} {\bibinfo {author} {\bibfnamefont {R.}~\bibnamefont {Chikkaraddy}}, \bibinfo {author} {\bibfnamefont {B.}~\bibnamefont {De~Nijs}}, \bibinfo {author} {\bibfnamefont {F.}~\bibnamefont {Benz}}, \bibinfo {author} {\bibfnamefont {S.~J.}\ \bibnamefont {Barrow}}, \bibinfo {author} {\bibfnamefont {O.~A.}\ \bibnamefont {Scherman}}, \bibinfo {author} {\bibfnamefont {E.}~\bibnamefont {Rosta}}, \bibinfo {author} {\bibfnamefont {A.}~\bibnamefont {Demetriadou}}, \bibinfo {author} {\bibfnamefont {P.}~\bibnamefont {Fox}}, \bibinfo {author} {\bibfnamefont {O.}~\bibnamefont {Hess}},\ and\ \bibinfo {author} {\bibfnamefont {J.~J.}\ \bibnamefont {Baumberg}},\ }\bibfield  {title} {\bibinfo {title} {Single-molecule strong coupling at room temperature in plasmonic nanocavities},\ }\href {https://www.nature.com/articles/nature17974} {\bibfield  {journal} {\bibinfo  {journal} {Nature}\ }\textbf {\bibinfo {volume} {535}},\ \bibinfo {pages} {127} (\bibinfo {year} {2016})}\BibitemShut {NoStop}%
\bibitem [{\citenamefont {Choi}\ \emph {et~al.}(2017)\citenamefont {Choi}, \citenamefont {Heuck},\ and\ \citenamefont {Englund}}]{PhysRevLett.118.223605}%
  \BibitemOpen
  \bibfield  {author} {\bibinfo {author} {\bibfnamefont {H.}~\bibnamefont {Choi}}, \bibinfo {author} {\bibfnamefont {M.}~\bibnamefont {Heuck}},\ and\ \bibinfo {author} {\bibfnamefont {D.}~\bibnamefont {Englund}},\ }\bibfield  {title} {\bibinfo {title} {{Self-Similar Nanocavity Design with Ultrasmall Mode Volume for Single-Photon Nonlinearities}},\ }\href {https://doi.org/10.1103/PhysRevLett.118.223605} {\bibfield  {journal} {\bibinfo  {journal} {Phys. Rev. Lett.}\ }\textbf {\bibinfo {volume} {118}},\ \bibinfo {pages} {223605} (\bibinfo {year} {2017})}\BibitemShut {NoStop}%
\bibitem [{\citenamefont {Li}\ \emph {et~al.}(2022)\citenamefont {Li}, \citenamefont {Li}, \citenamefont {Liu}, \citenamefont {Zhong}, \citenamefont {Liu}, \citenamefont {Chen},\ and\ \citenamefont {Wang}}]{li_room-temperature_2022}%
  \BibitemOpen
  \bibfield  {author} {\bibinfo {author} {\bibfnamefont {J.-Y.}\ \bibnamefont {Li}}, \bibinfo {author} {\bibfnamefont {W.}~\bibnamefont {Li}}, \bibinfo {author} {\bibfnamefont {J.}~\bibnamefont {Liu}}, \bibinfo {author} {\bibfnamefont {J.}~\bibnamefont {Zhong}}, \bibinfo {author} {\bibfnamefont {R.}~\bibnamefont {Liu}}, \bibinfo {author} {\bibfnamefont {H.}~\bibnamefont {Chen}},\ and\ \bibinfo {author} {\bibfnamefont {X.-H.}\ \bibnamefont {Wang}},\ }\bibfield  {title} {\bibinfo {title} {Room-{Temperature} {Strong} {Coupling} {Between} a {Single} {Quantum} {Dot} and a {Single} {Plasmonic} {Nanoparticle}},\ }\href {https://doi.org/10.1021/acs.nanolett.2c00606} {\bibfield  {journal} {\bibinfo  {journal} {Nano Lett.}\ }\textbf {\bibinfo {volume} {22}},\ \bibinfo {pages} {4686} (\bibinfo {year} {2022})}\BibitemShut {NoStop}%
\bibitem [{\citenamefont {Kuisma}\ \emph {et~al.}(2022)\citenamefont {Kuisma}, \citenamefont {Rousseaux}, \citenamefont {Czajkowski}, \citenamefont {Rossi}, \citenamefont {Shegai}, \citenamefont {Erhart},\ and\ \citenamefont {Antosiewicz}}]{kuisma_ultrastrong_2022}%
  \BibitemOpen
  \bibfield  {author} {\bibinfo {author} {\bibfnamefont {M.}~\bibnamefont {Kuisma}}, \bibinfo {author} {\bibfnamefont {B.}~\bibnamefont {Rousseaux}}, \bibinfo {author} {\bibfnamefont {K.~M.}\ \bibnamefont {Czajkowski}}, \bibinfo {author} {\bibfnamefont {T.~P.}\ \bibnamefont {Rossi}}, \bibinfo {author} {\bibfnamefont {T.}~\bibnamefont {Shegai}}, \bibinfo {author} {\bibfnamefont {P.}~\bibnamefont {Erhart}},\ and\ \bibinfo {author} {\bibfnamefont {T.~J.}\ \bibnamefont {Antosiewicz}},\ }\bibfield  {title} {\bibinfo {title} {Ultrastrong {Coupling} of a {Single} {Molecule} to a {Plasmonic} {Nanocavity}: {A} {First}-{Principles} {Study}},\ }\href {https://doi.org/10.1021/acsphotonics.2c00066} {\bibfield  {journal} {\bibinfo  {journal} {ACS Photonics}\ }\textbf {\bibinfo {volume} {9}},\ \bibinfo {pages} {1065} (\bibinfo {year} {2022})}\BibitemShut {NoStop}%
\bibitem [{\citenamefont {S\'aez-Bl\'azquez}\ \emph {et~al.}(2023)\citenamefont {S\'aez-Bl\'azquez}, \citenamefont {de~Bernardis}, \citenamefont {Feist},\ and\ \citenamefont {Rabl}}]{PhysRevLett.131.013602}%
  \BibitemOpen
  \bibfield  {author} {\bibinfo {author} {\bibfnamefont {R.}~\bibnamefont {S\'aez-Bl\'azquez}}, \bibinfo {author} {\bibfnamefont {D.}~\bibnamefont {de~Bernardis}}, \bibinfo {author} {\bibfnamefont {J.}~\bibnamefont {Feist}},\ and\ \bibinfo {author} {\bibfnamefont {P.}~\bibnamefont {Rabl}},\ }\bibfield  {title} {\bibinfo {title} {{Can We Observe Nonperturbative Vacuum Shifts in Cavity QED?}},\ }\href {https://doi.org/10.1103/PhysRevLett.131.013602} {\bibfield  {journal} {\bibinfo  {journal} {Phys. Rev. Lett.}\ }\textbf {\bibinfo {volume} {131}},\ \bibinfo {pages} {013602} (\bibinfo {year} {2023})}\BibitemShut {NoStop}%
\bibitem [{\citenamefont {Franke}(2020)}]{Franke2020thesis}%
  \BibitemOpen
  \bibfield  {author} {\bibinfo {author} {\bibfnamefont {S.~R.}\ \bibnamefont {Franke}},\ }\emph {\bibinfo {title} {{Quantization of quasinormal modes in dissipative media}}},\ \href {https://depositonce.tu-berlin.de/items/9211ef06-67ba-49d8-ace4-7bf41fcdf49d} {Ph.D. thesis},\ \bibinfo  {school} {Technical University of Berlin} (\bibinfo {year} {2020})\BibitemShut {NoStop}%
\bibitem [{\citenamefont {Kamandar~Dezfouli}\ \emph {et~al.}(2017)\citenamefont {Kamandar~Dezfouli}, \citenamefont {Gordon},\ and\ \citenamefont {Hughes}}]{PhysRevA.95.013846}%
  \BibitemOpen
  \bibfield  {author} {\bibinfo {author} {\bibfnamefont {M.}~\bibnamefont {Kamandar~Dezfouli}}, \bibinfo {author} {\bibfnamefont {R.}~\bibnamefont {Gordon}},\ and\ \bibinfo {author} {\bibfnamefont {S.}~\bibnamefont {Hughes}},\ }\bibfield  {title} {\bibinfo {title} {Modal theory of modified spontaneous emission of a quantum emitter in a hybrid plasmonic photonic-crystal cavity system},\ }\href {https://doi.org/10.1103/PhysRevA.95.013846} {\bibfield  {journal} {\bibinfo  {journal} {Phys. Rev. A}\ }\textbf {\bibinfo {volume} {95}},\ \bibinfo {pages} {013846} (\bibinfo {year} {2017})}\BibitemShut {NoStop}%
\end{thebibliography}%
\end{document}